\documentclass[twocolumn]{aastex631}

\usepackage{xcolor}

\newcommand{\Msol}{M{$_{\odot}$}}

\newcommand{\CO}{{$^{12}$CO}}
\newcommand{\HtwoCO}{{H$_{2}$CO}}
\newcommand{\Htwo}{{H$_{2}$}}
\newcommand{\tco}{{$^{13}$CO}}
\newcommand{\co}{C{$^{18}$O}}

\newcommand{\Feii}{Fe{\sc ii}}
\newcommand{\Fei}{Fe{\sc i}}
\newcommand{\Sii}{S{\sc ii}}
\newcommand{\Si}{S{\sc i}}
\newcommand{\Oiii}{O{\sc iii}}

\newcommand{\kms}{km~s{$^{-1}$}}
\newcommand{\Ha}{H$\alpha$}

\newcommand{\cmq }{cm$^{-3}$}

\shorttitle{The HH 24 Star Forming Complex}
\shortauthors{Reipurth et al.}

\begin{document}

\title{The HH 24 Complex:\\ 
Jets, Multiple Star Formation, and Orphaned Protostars}

\author[0000-0001-8174-1932]{Bo Reipurth}      
\affiliation{Institute for Astronomy, University of Hawaii, 640 North A'Ohoku Place, Hilo, HI 96720, USA}

\author[0000-0001-8135-6612]{J. Bally}
\affiliation{Center for Astrophysics and Space Astronomy, 
          University of Colorado, Boulder, CO 80309, USA}

\author{Hsi-Wei Yen}
\affiliation{Academia Sinica Institute of Astronomy and Astrophysics, 11F of Astro-Math Bldg, 1, Sec. 4, Roosevelt Rd, Taipei 10617, Taiwan}

\author[0000-0001-5653-7817]{H.G. Arce}
\affiliation{Department of Astronomy, Yale University, P.O. Box 208101, New Haven, CT 06520-8101, USA}

\author{L.-F. Rodr{\'\i}guez}        
\affiliation{Instituto de Radioastronom\'\i a y Astrof\'\i sica, 
Universidad Nacional Aut\'onoma de M\'exico, Apdo. Postal 3-72 (Xangari), 58089 Morelia, Michoac\'an, M\'exico and
Mesoamerican Center for Theoretical Physics, Universidad
Aut\'onoma de Chiapas, Carretera Emiliano Zapata km. 4,
Real del Bosque (Ter\'an). 29050 Tuxtla Guti\'errez, Chiapas,
M\'exico}

\author[0000-0002-0835-1126]{A.C. Raga}
\affiliation{Instituto de Ciencias Nucleares, Universidad Nacional Aut\'onoma de M\'exico, Ap. 70-543, 04510 D.F., M\'exico}

\author[0000-0003-2824-3875]{T.R. Geballe} 
\affiliation{Gemini Observatory/NSF's NOIRLab, 670 N. Aohoku Place, Hilo, HI 96720, USA}

\author{R. Rao}       
\affiliation{Submillimeter Array, Academia Sinica Institute of Astronomy and Astrophysics, 645 N. A'ohoku Place, Hilo, HI 96720, USA}

\author[0000-0002-7838-2606]{F. Comer\'on}        
\affiliation{European Southern Observatory, Karl-Schwarzschild-Strasse 2, D-85748 Garching bei M\"unchen, Germany}

\author[0000-0003-1448-8767]{S. Mikkola}       
\affiliation{University of Turku, Dept. of Physics and Astronomy, Vesilinnantie 5, FIN-20014, Finland }

\author[0000-0001-6601-8906]{C.A. Aspin}       
\affiliation{Institute for Astronomy, University of Hawaii, 640 North A'Ohoku Place, Hilo, HI 96720, USA}

\author[0000-0002-6092-8295]{J. Walawender}
\affiliation{W. M. Keck Observatory, 65-1120 Mamalahoa Hwy, Kamuela, HI 96743, USA}

\vspace{2cm}



\begin{abstract}
 
The HH 24 complex harbors five collimated jets emanating from a small
protostellar multiple system. We have carried out a multi-wavelength
study of the jets, their driving sources, and the cloud core hosting
the embedded stellar system, based on data from the HST, Gemini,
Subaru, APO 3.5m, VLA, and ALMA telescopes. The data show that the
multiple system, SSV~63, contains at least 7 sources, ranging in mass
from the hydrogen-burning limit to proto-Herbig Ae stars. The stars
are in an unstable non-hierarchical configuration, and one member, a
borderline brown dwarf, is moving away from the protostellar system
with 25~\kms, after being ejected $\sim$5,800~yr ago as an orphaned
protostar. Five of the embedded sources are surrounded by small,
possibly truncated, disks resolved at 1.3 mm with ALMA. Proper motions
and radial velocities imply jet speeds of 200-300~\kms. The two main
HH~24 jets, E and C, form a bipolar jet system which traces the
innermost portions of parsec-scale chains of Herbig-Haro and H$_2$
shocks with a total extent of at least 3 parsec.  H$_2$CO and
C$^{18}$O observations show that the core has been churned and
continuously fed by an infalling streamer. $^{13}$CO and $^{12}$CO
trace compact, low-velocity, cavity walls carved by the jets and an
ultra-compact molecular outflow from the most embedded object.
Chaotic N-body dynamics likely will eject several more of these
objects.  The ejection of stars from their feeding zones sets their
masses. Dynamical decay of non-hierarchical systems can thus be a
major contributor to establishing the initial mass function.

\end{abstract}


\keywords{
Herbig-Haro objects (722) ---
Multiple stars (1081) ---
Young stellar objects (1834) ---
Circumstellar disks (235) ---
Protostars (1302) ---
Herbig Ae/Be stars (723) ---
Star formation (1569)
}

\section{INTRODUCTION}\label{sec:introduction}


Evidence is mounting that stars rarely form in isolation as single
objects, but rather as binaries or small multiple systems (e.g.,
Duch\^ene \& Kraus 2013). Small multiple systems are produced through
fragmentation of prestellar cores, as first studied by Hoyle (1953)
and Larson (1972). In modern terms, the two principal pathways for
fragmentation are turbulent fragmentation, which tends to operate on
larger scales (e.g., Lee et al. 2019), and disk fragmentation, which
operates on small scales in massive protostellar disks (e.g, Kratter
\& Matzner 2006).  Most multiple systems form in non-hierarchical
configurations, but soon undergo dynamical interactions. Over about a
hundred crossing times such systems tend to rearrange into
hierarchical configurations consisting of compact binaries and members
that either are ejected into a distant bound orbit, or escape (e.g.,
Anosova 1986, Delgado-Donate et al. 2004).  Half of all such escapes
occur during the embedded phase, leading to the ejection and exposure
of {\em orphaned protostars}, some of which did not have time to gain
enough mass to become hydrogen burning stars (Reipurth \& Clarke 2001,
Reipurth et al. 2010).  This competition between accretion and
ejection was shown by Bate \& Bonnell (2005) to be the key driver in
shaping the initial mass function at all masses.

The reconfiguring of a non-hierarchical triple system occurs after a
close triple approach, when three bodies can exchange energy and
momentum. After an ejection the remaining binary has a high
eccentricity, leading to disk-disk interactions during periastron
passages, and a gradual inspiral of the binary. The periastron
passages lead to disk disturbances and accretion events, with ensuing
outflow. The stellar magnetohydrodynamic jet engines are thus
force-fed, resulting in spectacular giant Herbig-Haro (HH) flows
(Reipurth 2000).

Large-scale numerical simulations have offered insight into the
formation of multiple systems and their dynamical interactions (Bate
2009, 2012). Such dynamical interactions can help to bind components
into tighter binaries, but to produce the observed frequency of close
binaries, dissipative interactions are needed, during which the
presence of gas serves to transport angular momentum and dissipate
energy in star-disk and disk-disk interactions. While any
non-hierarchical system will eventually always evolve into a
hierarchical configuration on dynamical grounds alone, the presence of
gas plays an important role in the subsequent orbital evolution of the
binary and its mass-ratio (Bate et al. 2002).

Evidently significant dynamical evolution is expected to occur during
early stellar evolution, as borne out by observations. Early optical
surveys of T~Tauri stars showed an excess of companions relative to
field stars (e.g., Reipurth \& Zinnecker 1993, Leinert et al. 1993).
This was further demonstrated with near-infrared observations of
Class~I protostars, which revealed not only an excess of companions,
but also a bimodal distribution of the separation distribution
function with a second peak at several thousand AU (Connelley et
al. 2008a,b). This population of distant companions decreases for the
more evolved Class~I sources, suggesting that the companions
dynamically evolve and become unbound. Most recently, ALMA and VLA
observations of Class~0 and Class~I sources have yielded insights into
the high multiplicity of the youngest protostars (Tobin et al. 2016,
2018, 2022) and have confirmed the existence of the bimodal binary
separation distribution.

For reviews of multiple systems and their dynamical
evolution, see Reipurth et al. (2014) and Offner et al. (2022).

\begin{figure*} 
\centerline{\includegraphics[angle=0,width=14cm]{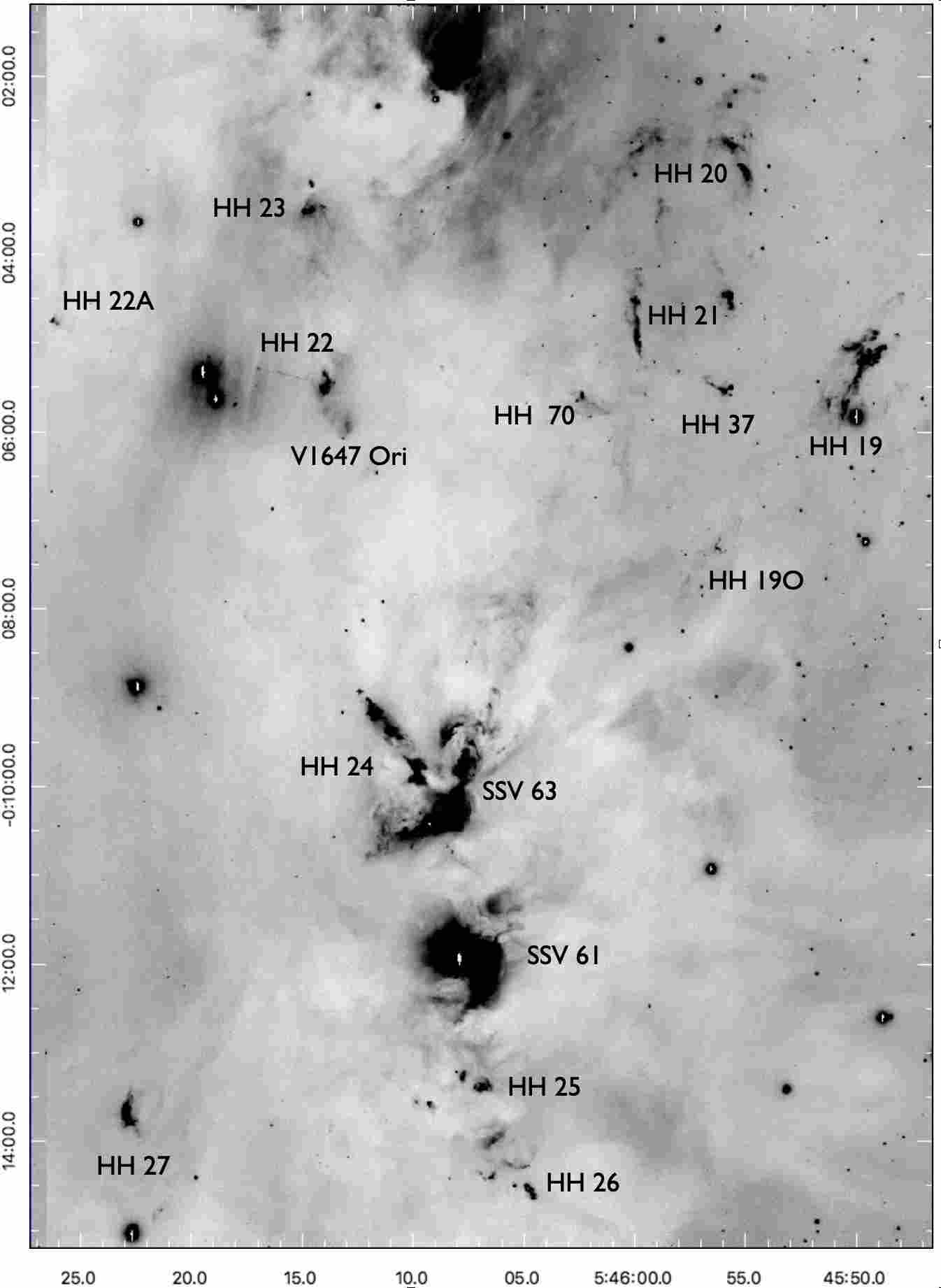}}
\caption{Deep H$\alpha$+[\Sii] image obtained at the
Subaru 8m telescope shows the central part of L1630 with
identifications of objects discussed in the text. HH~24 is the cluster
of jets emanating from the multiple system SSV~63, while HH~19, 20,
21, 27, 37, and 70 are distant bow shocks related to the HH~24
jets. HH~22 is driven by an embedded source and HH~23 possibly by
V1647~Ori.  HH~25 and 26 are driven by embedded sources south of
SSV~61. The
height of the figure corresponds to approximately 1.5~pc. Star
formation occurs along a narrow ridge oriented N-S with a length of
$\sim$1~pc and a total mass around 230~M$_\odot$. Coordinates are
equinox 2000.
 \label{overview}} 
\end{figure*}

In this paper we present a detailed study of the HH~24 jet complex and
the compact multiple system that drives these jets. This is a complex
region of star formation, in which a small multiple system has formed
within a cloud core and through dynamical interactions has triggered
disk disturbances that have lead to massive accretion events and
ensuing outflow activity. This has resulted in the highest
concentration of finely collimated HH jets known. An overview of the
region is shown in Figure~\ref{overview}  and some of the
general properties of the outflows are given in
Table~\ref{table:overview}.

The paper is organized as follows: In Section~\ref{sec:previous} we
present a summary of key results obtained in previous studies, and in
Section~\ref{sec:observations} a description of the observations
obtained for this study. This is followed by an overview of the HH~24
complex in Section~\ref{sec:hh24complex}, and a summary of the
properties of the multiple system in
Section~\ref{sec:sources}. Section~\ref{sec:jets} contains a
discussion of the individual jets and shocks, and
Section~\ref{sec:hh24mms} presents an analysis of the neighboring
protostar HH~24~MMS. The discovery of a low-mass runaway borderline
brown dwarf that was ejected 5,800~yr ago from the multiple system is
discussed in Section~\ref{sec:halo}. After that the star formation
efficiency is derived in Section~\ref{sec:efficiency}. Details of our
ALMA observations are presented in Section~\ref{sec:ALMA-I} and
Section~\ref{sec:ALMA-II}, where the individual disks and the
large-scale cloud structures, respectively, are studied.  Finally
Section~\ref{sec:discussion} and Section~\ref{sec:conclusions} contain
a detailed discussion and a summary of our results.

\begin{deluxetable}{lcccc}
 
\tablecaption{Overview of the HH 24 Complex\label{table:overview}}
\tablewidth{0pt}
\tablehead{
    \colhead{Jet} &
    \colhead{PA} & 
    \colhead{Orient.} &			   
    \colhead{Giant Bow-Shocks$^a$} &
    \colhead{Source} 
}      
\startdata
C  & 333$^\circ$ & Blue     & HH 20/21/37/70 & Ea         \\
E  & 149$^\circ$ & Red      &                & Ea         \\
A  &  -          & Red      &                & Ea/HOPS317 \\
G  &  39$^\circ$ & Blue     &                & NE         \\
J  & 311$^\circ$ & Blue$^b$ & HH 19/27       & Wb         \\
L  &  38$^\circ$ &  -       &                & HOPS317    \\
X  & 143$^\circ$ &  -       &                & S(?)       \\
B  &  -          & Blue     &                & Wa         \\
\enddata
\tablecomments{a: Additional very distant bow shocks exist. 
b: Deduced from the blue-shift of HH 19.}
\end{deluxetable}


\clearpage

\vspace{0.3cm}

\section{PREVIOUS WORK}\label{sec:previous}

HH~24 is located in the L1630 cloud (aka Orion~B), in a dense core
that is part of a chain of north-south oriented cores detected in both
millimeter line emission and sub-millimeter continuum (e.g., Gibb \&
Heaton 1993, Lis et al. 1999, Mitchell et al. 2001, Kirk et al. 2016a,
Hsieh et al. 2021). The driving source of HH~24 was detected in a
near-infrared survey by Strom et al. (1976). This source, SSV~63, was
later found to be a multiple protostellar system. We here assume HH 24
and the L1630 cloud to be at a distance of $\sim$400~pc (e.g.,
Anthony-Twarog 1982), a distance supported by the more recent studies
of Lombardi et al. (2011) [398$\pm$12~pc], Kounkel et al. (2017)
[388$\pm$10~pc], and Zucker et al. (2019) [423$\pm$21~pc]. For an
overview of star formation in L1630, see the review by Gibb (2008).

The HH~24 complex was discovered by Herbig \& Kuhi (1963) in their
search for H$\alpha$ emission stars in L1630\footnote{The first
mention of HH~24 is in a letter from George Herbig to Jesse Greenstein
dated August 9, 1952 in which Herbig speculates that the faint
nebulous emission-line objects he found on his photographic plates of
the HH~24 region might be similar to the recently discovered objects
HH~1 and 2.}. Subsequently HH~24 has been the subject of numerous
studies, a selection of which are listed here. Schmidt \& Miller
(1979) and Scarrott et al. (1987) used polarimetric observations to
infer that the HH~24 nebulosity is a mixture of emission from shocks
and reflected light from embedded sources. HH~24 has been imaged
optically by Herbig (1974), Strom et al. (1974a), Jones et al. (1987)
and Mundt et al.  (1991). Two of the knots in HH~24 were detected in
H$_2$ 2.122~$\mu$m emission by Davis et al.  (1997).  Optical or
ultraviolet spectroscopy of various components in HH~24 has been
presented by Strom et al. (1974), Brugel et al. (1981), Jones et al.
(1987), Solf (1987), and B\"ohm et al.  (1992). Some of the HH~24 jets
are associated with distant bow shocks, as noted by Jones et
al. (1987) and Eisloeffel \& Mundt (1997).

For the following detailed discussion of the HH~24 complex, it is
important to have clear definitions of the nomenclature of the
multitude of shocks in the region. Unfortunately, the existing knot
designations were developed over a number of years by many different
researchers, and along the way a number of mistakes occurred, so that
it is difficult to compare various studies. HH~24 was discovered by
George Herbig, but besides the brief mention in Herbig
\& Kuhi (1964), he did not provide further information until his HH
catalog appeared, in which he identified four components A,B,C,D
(Herbig 1974).  Simultaneously Strom et al. (1974a,b) labeled five
knots A-E, but used E for knot D in Herbig's notation, a knot that was
later shown to be an H$\alpha$-strong reflection nebula. Schmidt \&
Miller (1979) adopted the Strom et al. (1974a,b) nomenclature. Solf
(1987) added the label F, which simultaneously was labeled E by Jones
et al. (1987), who also introduced more detailed designations of
knots. In this paper we follow and extend the consistent designations
by Herbig (1974), Jones et al. (1987), Mundt et al. (1991), and
Eisloeffel \& Mundt (1997).

\vspace{0.3cm}

\section{OBSERVATIONS}\label{sec:observations}

\subsection{HST WFC3}

The HH 24 complex was observed with HST under program GO-13485 (PI:
Reipurth) in an H$\alpha$ (F656N) filter on UT 2014-03-10 with a total
exposure time of 5578 sec, in a [\Sii] (F673N) filter on UT 2014-02-26
for 5578 sec, in a [\Feii] (F164N) filter on UT 2014-02-18 for two
exposures of 3596 sec and 1798 sec. Parallel observations of HH~19
were made with ACS in H$\alpha$ on UT 2014-03-10 for 5165 sec. Two
years later, on UT 2016-02-03, a second-epoch [\Feii] image of HH~24 was
obtained under program GO-14344 with an exposure time of 5395 sec.

\subsection{Subaru SuprimeCam images}

The Subaru 8m telescope was used to observe HH 24 with SuprimeCam
(field of view 34$'$ $\times$ 27$'$ and scale 0\farcs20/pxl) on UT
2006-01-05 using a [\Sii] filter (N-A-L671, FWHM 130 \AA, transmission
88\%) with 5$\times$12 min dithered exposures; the sky was clear and
seeing varied between 0.51 and 0.70 arcsec. On UT 2006-01-06 HH 24 was
observed using an H$\alpha$ filter (N-A-L659, FWHM 99 \AA,
transmission 88\%) with 5$\times$12 min dithered exposures through
intermittent light cirrus and seeing between 0.57 and 0.67
arcsec. The pixel scale was 0.20 arcsec/pxl.  Second-epoch
observations with 5$\times$6 min were similarly acquired on UT
2015-12-17 in a [\Sii] filter in seeing of $\sim$0.8-0.9~arcsec.

\subsection{Gemini observations}

Several observing runs were carried out at the Gemini-North Frederick
C. Gillett 8m telescope.  GMOS was used on 2010-03-13 and 2010-03-16
under program GN-2010A-Q-10 to obtain g, r, i, H$\alpha$, and [\Sii]
images and multi-slit spectra of the SSV~63 region. At the time of
these observations GMOS had a ~5.5'$\times$7.4' field of view with
0.0727\arcsec \/ pixels. Three exposures of 60~sec were obtained through
the broadband filters and three 5~min exposures in the narrowband
filters. The R400 grating with a dispersion of 0.0673~nm/pxl was used
for 6 exposures of 20~min using slitless spectroscopy. NIRI was used
on 2009-12-26 and 2010-02-09 to obtain near-infrared images in the J, H,
K', H$_2$, and [\Feii] filters. Eighteen 30~sec exposures were obtained
in the two narrowband filters and in nearby continuum filters, 9
$\times$ 25~sec exposures were obtained in the J-filter and 9 $\times$
10~sec exposures in H and K'.  Near-infrared spectroscopy was obtained
with GNIRS under program GN-2013B-Q-77 in cross-dispersed SXD mode
using the 32 l/mm grating and a 0.3 arcsec slit. Source Wb was
observed on 2014-03-19 for 2400~sec in 0.87$''$ seeing, and Ea on
2014-03-20 for 1200~sec in 0.62$''$ seeing. Subsequently near-infrared
imaging of SSV~63 using NIRI and Gemini's adaptive optics module ALTAIR with a laser guide star was performed on
2013-12-15 in J, H, K' filters.

\subsection{Apache Point Observatory} 

Radial velocities of various knots and features in the HH 24 field
were measured using the ARCES echelle spectrograph on the APO 3.5
meter telescope on UT 2018-11-19 and on UT 2021-02-27.
ARCES 
captures the entire spectrum between 3200-10000~\AA\ 
with a resolution (2.5 pixels) of about R$\sim$32,000. 
The ARCES entrance aperture is a small slit 1.6$''$ by
3.2$''$ in extent on the sky. 
A one pixel interval near the H$\alpha$ and red [\Sii] doublet lines
corresponds to a Doppler shift of $\sim$4 km s$^{-1}$ per pixel.
The ARCES spectrograph was also used to obtain spectra of the new knot
in HH24 jet C on UT 2022-01-26.  A set of five 300 second exposures
was combined for the final spectrum.

All ARCES velocities reported here are referenced to the mean H$\alpha$
radial velocity of the Orion Nebula in the vicinity of the Trapezium
cluster which is assumed to have a heliocentric radial velocity of 21
km s$^{-1}$, corresponding to V$_{lsr}$ = +2 km s$^{-1}$. This
reference frame is within a few km s$^{-1}$ of the radial velocity of
the Orion B cloud in which HH 24 is embedded. The Orion Nebula is
located within 5$^\circ$ of HH 24, making the relative correction
between the observatory reference frame and heliocentric (or LSR)
reference frame smaller than the errors in radial velocity
determinations.  The measurement errors in the spectral line profiles
are dominated by the large observed line-widths and low
signal-to-noise ratios and are estimated to be between 5 to 10 km
s$^{-1}$.

[\Sii] images of the HH 24 outflow were obtained with a new [\Sii]
filter having a passband of 78 Angstroms and providing full
illumination of the 8\arcmin\ field of view of the ARCTIC CCD camera
on UT 2021-12-01 with the APO 3.5 meter reflector.  A dithered set
of three to six 300 second exposures were acquired at four different
pointings to cover the entire HH 24 outflow complex.


Near-infrared observations were obtained with the NICFPS camera on
the APO 3.5 meter telescope on UT 2018-11-19, 
2018-12-23, 2022-01-25, and 2022-01-27. 
The pixel scale of this instrument is 0.273$''$ per pixel with a field
of view 4.58$'$ on a side. Dithered images with 300 second exposures were
obtained in the 2.122 $\mu$m S(1) line of H$_2$ using a narrow-band
filter (FWHM=0.4\% of the central wavelength) plus identical separate
sky frames.
Atmospheric seeing produced 1.2$''$ FWHM stellar images.





\subsection{VLT}

An unpublished data set of images of SSV 63 in the Ks and L' band
obtained with NACO, the adaptive optics-assisted infrared imager and
spectrograph at the Very Large Telescope (Lenzen et al., 2003, Rousset
et al., 2003), was retrieved from the ESO Science Archive Facility
together with its relevant calibration frames. The data set consists
of 33 individual frames through the Ks filter obtained on the night of
20/21 December 2007, with a total exposure time of 30 minutes, and 82
images through the L' filter obtained on the night of 31 December 2007
/ 1 January 2008, with total exposure time of 41 minutes. The Ks- and
L'-band images were flux calibrated using respectively the standard
stars S252-D (Persson et al. 1998) and S842-E (Leggett et
al. 2003). Data reduction was carried out using IRAF-based scripts.







\begin{figure*}
\centerline{\includegraphics[angle=0,width=15cm]{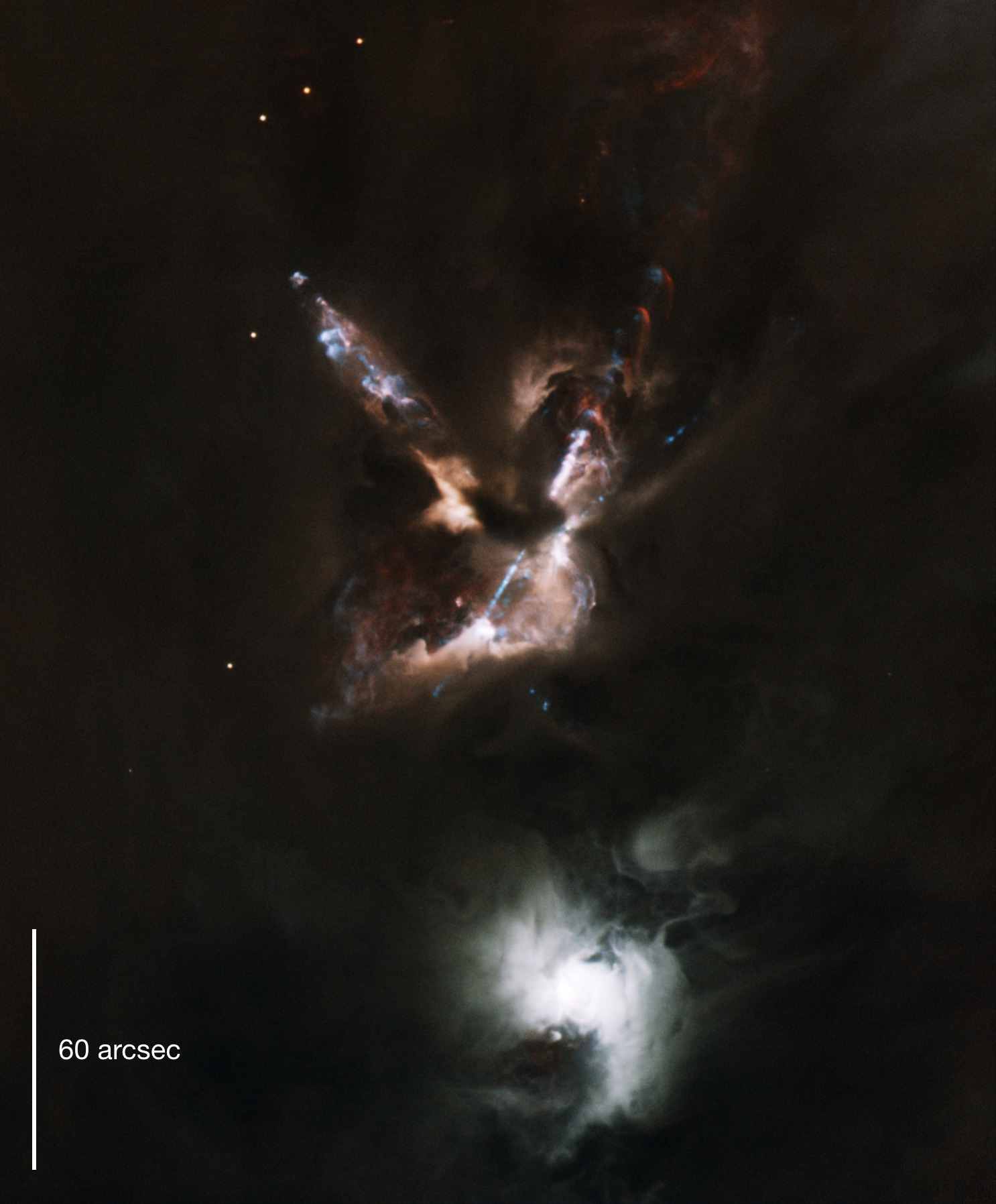}}
\caption{The HH 24 complex (top) and SSV~61 reflection nebula (bottom) 
as seen in a color mosaic from Gemini composed of  
$g'$ (blue), $r'$ (cyan), $i'$ (orange), H$\alpha$ (red) and [\Sii] (blue).
Color figure prepared by Travis Rector. The figure is $\sim$4$\times$5~arcminutes, corresponding to about 1/2~pc wide. North is up and east is left. 
  \label{gemini}}
\end{figure*}

\subsection{ALMA} 

The Atacama Large Millimeter Array (ALMA) was used to observe
molecular line and dust thermal continuum emission from the HH 24
region in the 1.3 mm region of the spectrum (ALMA Band 6). The
observations, part of the Cycle 6 project 2018.1.01194 (PI:
Reipurth), included one spectral configuration that allowed 
simultaneously observations of a 1.875 GHz-wide band of continuum emission,
centered at 232.6 GHz, and the following spectral lines:
$^{12}$CO(2-1), $^{13}$CO(2-1), C$^{18}$O(2-1),
H$_2$CO(3$_{0,3}$-$2_{0,2}$), and SiO(5-4).  The (single) pointing of
the ALMA 12 m array observations was centered at 05:46:08.35,
-00:10:01.5 (2000), which was chosen to be able to cover, well within the
25\arcsec \/ Half-Power Beam Width (HPBW) of the primary beam at the
observed frequency, the circumstellar environment of the previously
known protostars in the HH 24 region.

The goal of the observations was to study the link between the small
scale structure (i.e., disks), with sizes of about 50 to 100 AU, and the
larger structures with scales of $\sim$1000 AU (e.g., circumstellar
envelopes, outflows). As such, a range of baselines was needed to be
sensitive to this range of scales and therefore the observations were
done using two array configurations (named C43-3 and C43-6). The more
compact configuration (C43-3) consisted of baselines ranging from
about 15 to 500 m, while the more extended configuration (C43-6)
contained baselines of up to approximately 3070 m. The angular
resolution and maximum recoverable scale of the compact configuration
was about 0.7\arcsec\ and 7.5\arcsec, while for the extended
configuration these were 0.12\arcsec\ and 2\arcsec, respectively.  The
data from the C43-3 configuration were taken with three execution
blocks, obtained in December 2018 and April 2019, while the three
execution blocks with the C43-6 configuration were observed in September
2019.


\begin{deluxetable*}{cccccc}
\tablecolumns{6}
\tablecaption{ALMA Observations \label{table:almamaps}}
\tablehead{
\colhead{Map} & \colhead{Configurations\textsuperscript{a}} & \colhead{Beam Size} & \colhead{Beam P.A.} & \colhead{$\Delta V$\textsuperscript{b}} & \colhead{rms\textsuperscript{c}} \\
\colhead{} &   \colhead{ } & \colhead{[arcsec]} & \colhead{[deg E of N]} & \colhead{[km s\textsuperscript{-1}]} & \colhead{[mJy beam\textsuperscript{-1}]}
}
\startdata
Continuum & C43-3 + C43-6 & $0.13 \times 0.08$ & -87 & --- & 0.038\\
C$^{18}$O(2-1) & C43-3 + C43-6 & $0.24 \times 0.21$ & -70 & 0.2 & 1.4\\
$^{12}$CO(2-1) &  C43-3 & $0.78 \times 0.52$ & 86 & 0.16 & 4.0\\
$^{13}$CO(2-1) &  C43-3  & $0.81 \times 0.54$ & 87 & 0.08 & 5.5\\
C$^{18}$O(2-1) &  C43-3   & $0.82 \times 0.54$ & 87 & 0.08 & 4.5\\
H$_2$CO(3$_{0,3}$-$2_{0,2}$) & C43-3 & $0.83 \times 0.53$ & 87 & 0.17 & 2.8\\
\hline
\enddata
\tablecomments{
\textsuperscript{a}ALMA configurations used to make map.  \textsuperscript{b}Velocity resolution of molecular line maps. \textsuperscript{c}rms per velocity channel at the quoted velocity resolution.
}
\end{deluxetable*}

The Common Astronomy Software Application Package (CASA, McMullin et
al.~2007) was used to reduce the data. Version 5.4 of the CASA
pipeline was used to calibrate the raw visibility data taken in
configuration C43-3, while version 5.6 was used for data taken in
configuration C43-6.  We combined the calibrated data from both
configurations to study the dust continuum and C$^{18}$O emission at
small (disk) scales and used CASA version 5.7 for self-calibration of
the continuum data and imaging. We iteratively performed phase-only
self-calibration with a minimum solution interval of 10~s, and then
applied the solution to both the continuum and C$^{18}$O data. These
were subsequently imaged using the {\it tclean} task in CASA, with the
multi-scale deconvolver with scales of 0, 0\farcs3, 0\farcs7, and
2\farcs1 for the continuum image and 0, 0\farcs5, 1\farcs1, 2\farcs3,
and 5\farcs2 for the C$^{18}$O line map (which approximately
correspond to 0, 2, 5, and 10--20 times the beam sizes), and using
Briggs weighting with robust parameters of $-1$ and 0.5, respectively.

In order to study the gas structure and kinematics at larger scales ($\sim$1000~AU) we used the $^{12}$CO, $^{13}$CO, C$^{18}$O and
H$_2$CO line maps obtained with the C43-3 configuration. These were
all imaged with version 5.4 of the CASA pipeline. Imaging of the
visibility data was done using the {\it tclean} task in CASA with a
robust parameter of -0.5.  The continuum was subtracted from all the
molecular line maps using the CASA task {\it uvcontsub}. Primary beam
correction was applied to all maps, except for the high-resolution
C$^{18}$O map.  The synthesized beam and rms noise of the resulting
images are shown in Table~\ref{table:almamaps}.



\subsection{VLA}

The observations were part of our VLA project 19A-012, made with the
NSF's Karl G. Jansky Very Large Array (VLA) of NRAO\footnote{The National
Radio Astronomy Observatory is a facility of the National Science
Foundation operated under cooperative agreement by Associated
Universities, Inc.}. The observations were obtained in the A
configuration, those at 44.0 GHz (Q band) on UT 2019-8-19 and those at
10.0 GHz (X band) on UT 2019-8-24. These are the deepest observations
     of the HH~24 region obtained to date in those bands. The flux and
     bandpass calibrator was J0542+4951 (=3C147) 
and the phase calibrator was J0552+0313.  The digital correlator of
the VLA was configured in spectral windows of 128 MHz width, each
divided in 64 channels of spectral resolution of 2 MHz. The total
bandwidths were 4.0 and 8.0 GHz for the X band and Q band
observations, respectively.  The data were processed and analyzed in
the standard manner using the CASA package of NRAO and the pipeline
provided for
VLA\footnote{https://science.nrao.edu/facilities/vla/data-processing/pipeline}
observations. Maps were made using a robust weighting (Briggs 1995) of
2 in order to optimize the sensitivity at the expense of losing some
angular resolution.

 


\begin{figure} 
\centerline{\includegraphics[angle=0,width=8.3cm]{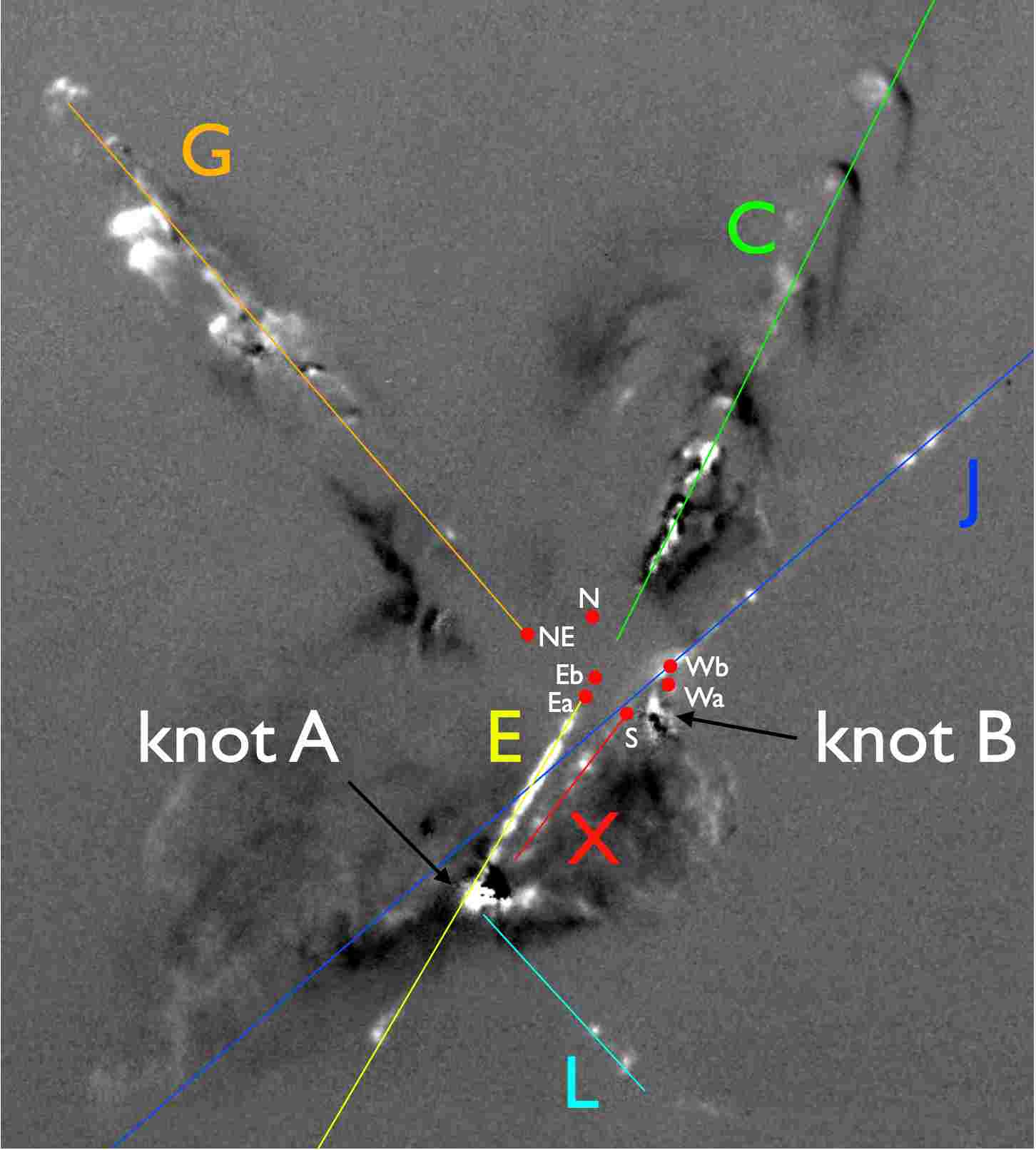}}
\caption{Annotated H$\alpha$--[\Sii] image obtained at the Subaru telescope 
showing the individual jets from SSV~63. The labels preserve and expand existing nomenclature. The multiple system is shown as red 
circles. White is [\Sii]-strong, black is H$\alpha$-strong. 
North is up and east is left. 
\label{definitions}} 
\end{figure}

\begin{figure*}
\centerline{\includegraphics[angle=0,width=15cm]{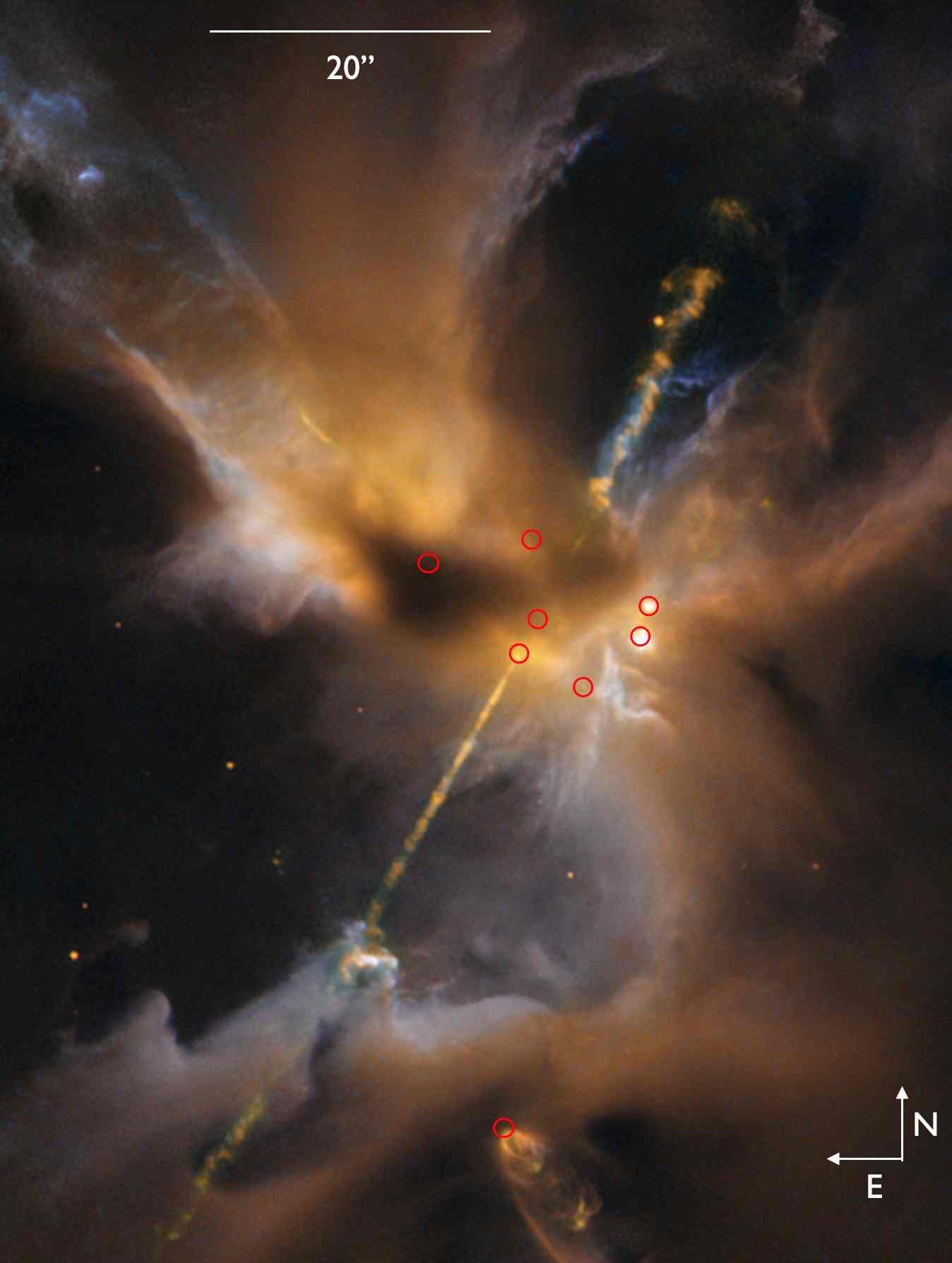}}
\caption{HST multi-filter image with the protostellar components of
the SSV~63 multiple system superposed (red circles).   The
red circle at the bottom of the figure marks the location of the
embedded source HOPS~317, which illuminates an outflow cavity and
drives the HH~24L flow. The filters used are: F814W (I-band) as blue,
F814W+F160W as green, F160W (H-band) as orange, and F164N ([\Feii]) as
red. This mixture of narrowband and wideband images render jets,
clouds, and outflow cavities particularly well. Color image courtesy
Judy Schmidt/NASA/ESA. The F814W image is from HST programs 9160 (PI
D. Padgett) and the F160W image from program 11548 (PI
S.T. Megeath). The [\Feii] image is from this paper.
\label{pressrelease}}
\end{figure*}

\begin{figure*}[tbp]
\centerline{\includegraphics[angle=0,width=18cm]{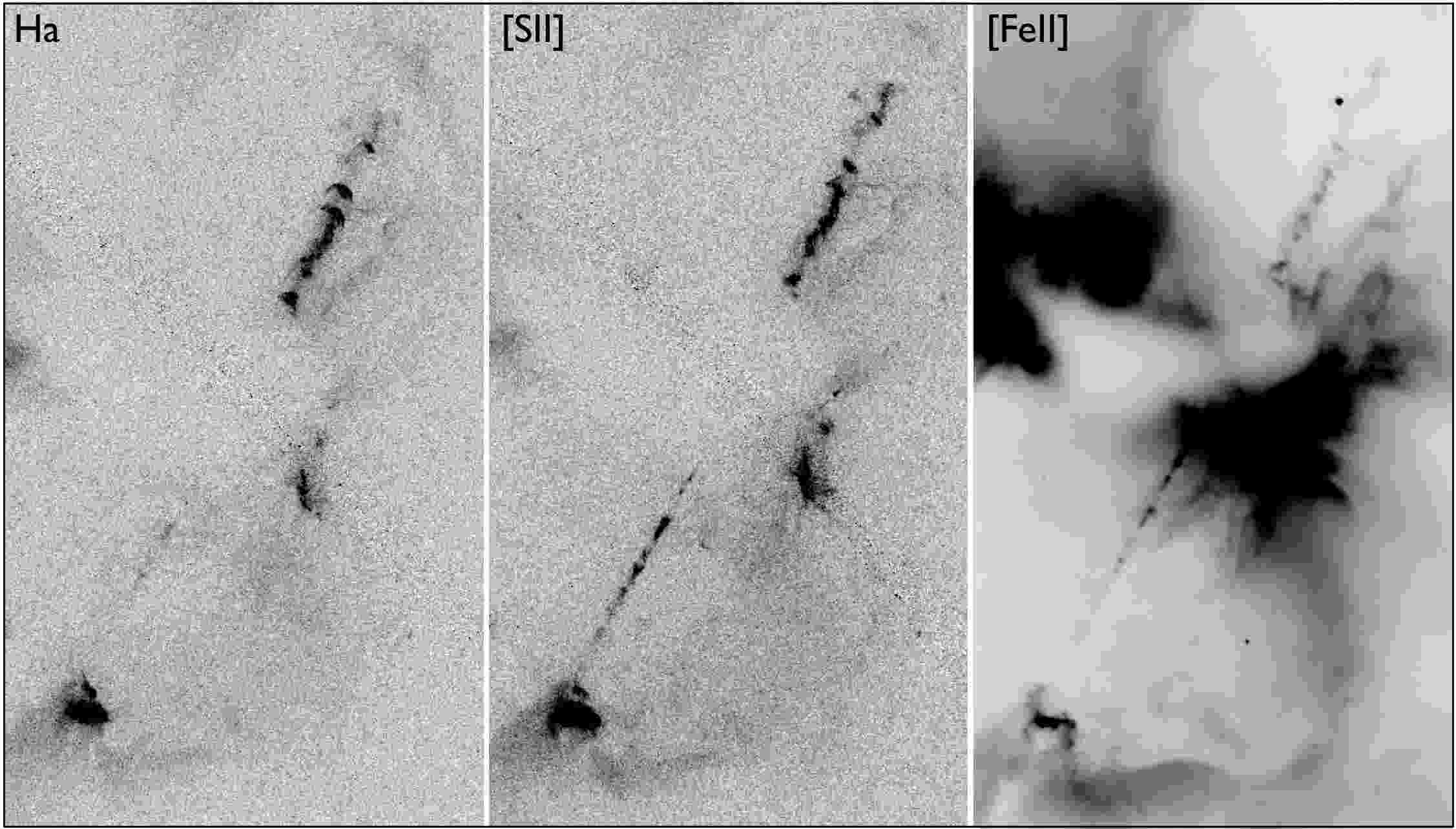}}
\caption{Triptych of WFC3 HST images showing the HH~24 jet~E and jet~C in
the H$\alpha$ 6563~\AA, [\Sii] 6717/31~\AA, and [\Feii] 1.644~\AA\
lines. North is up and east is left.
\label{triptych}}
\end{figure*}

\begin{figure}
\centerline{\includegraphics[angle=0,width=6cm]{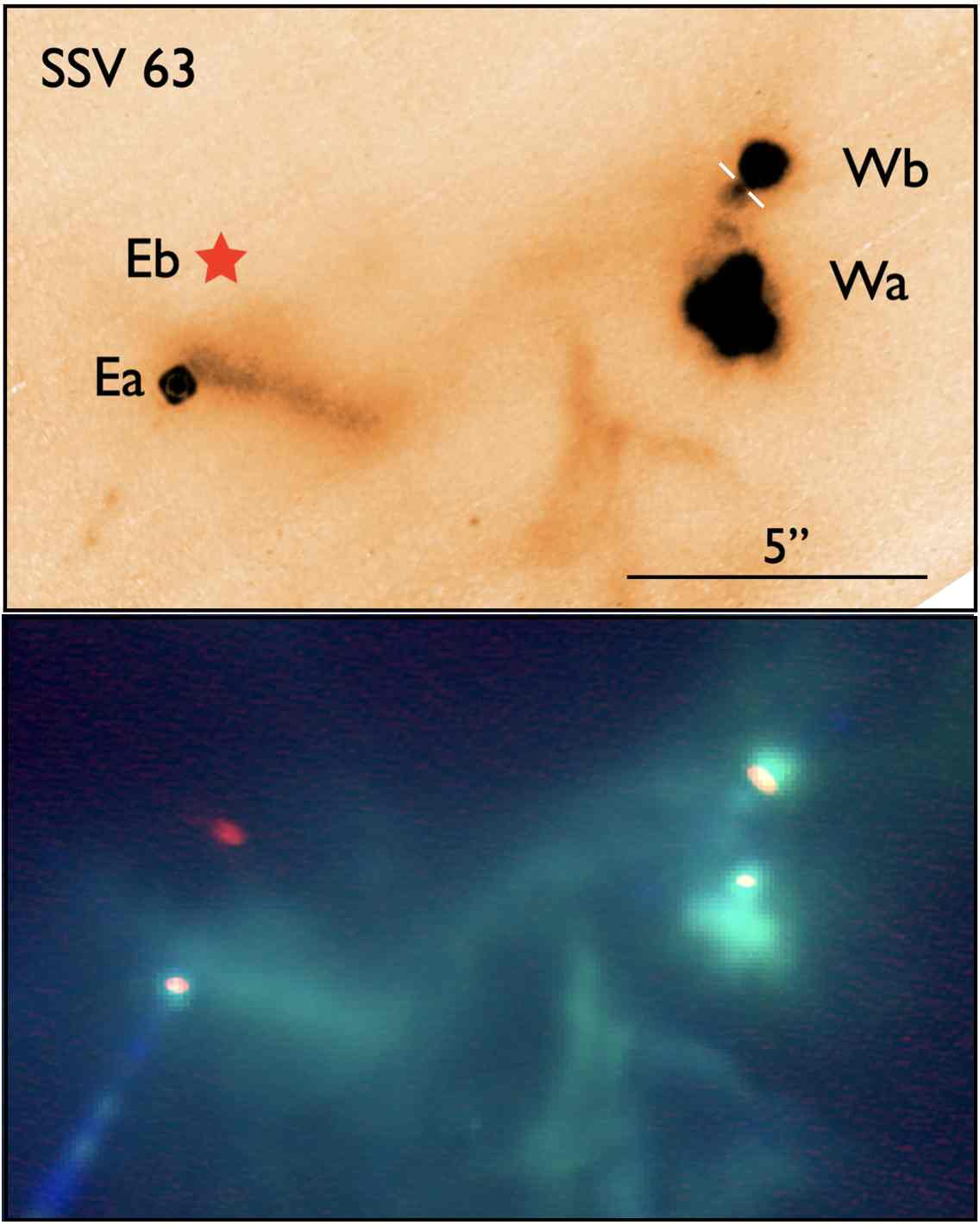}}
\caption{(top) H$_2$ HST image of the SSV 63 multiple system. The
deeply embedded mid-infrared source Eb is not detectable at
2.1~$\mu$m, but is marked with an asterisk. Archival image obtained
with NICMOS (Program 11205, PI Muzerolle). (bottom) A color composite
of an HST [\Feii] image, an HST H-band image (Program 11548, PI
Megeath), and an ALMA 1.3mm continuum image. The circumstellar disks
are clearly resolved, and it is seen that the near-infrared source Wb
is not a star, but the compact NW lobe of a bipolar reflection nebula
on either side of a silhouette disk (marked with white lines). North
is up and east is left.
\label{ssv63-h2}}
\end{figure}

\begin{figure}
\centerline{\includegraphics[angle=0,width=6cm]{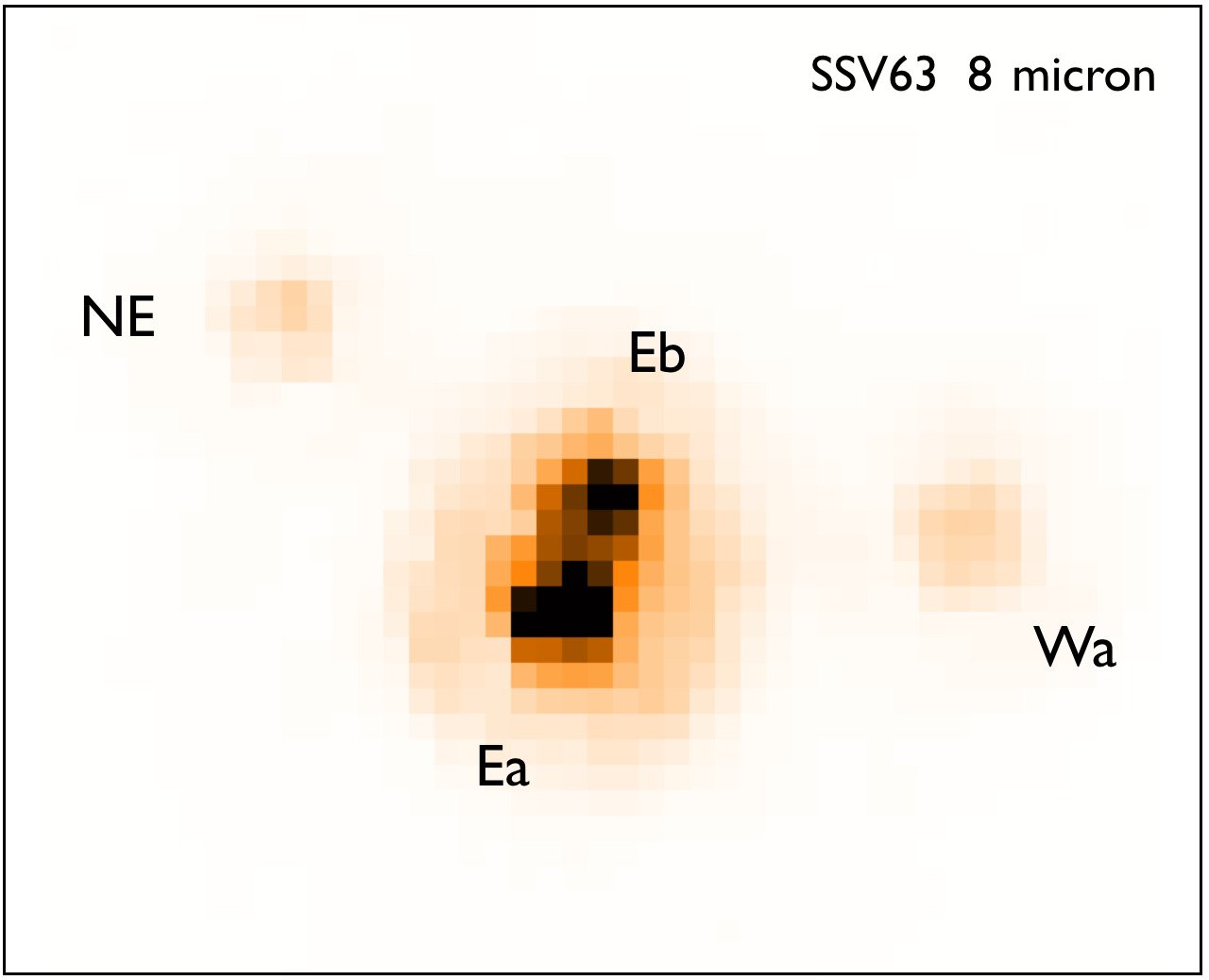}}
\caption{Spitzer 8~$\mu$m image of the SSV 63 multiple system.
The source Eb is clearly resolved from Ea. 
The figure is about 40 arcsec across. North is up and east is left. 
\label{ssv63-8micron}}
\end{figure}

\begin{figure} 
\centerline{\includegraphics[angle=0,width=8.3cm]{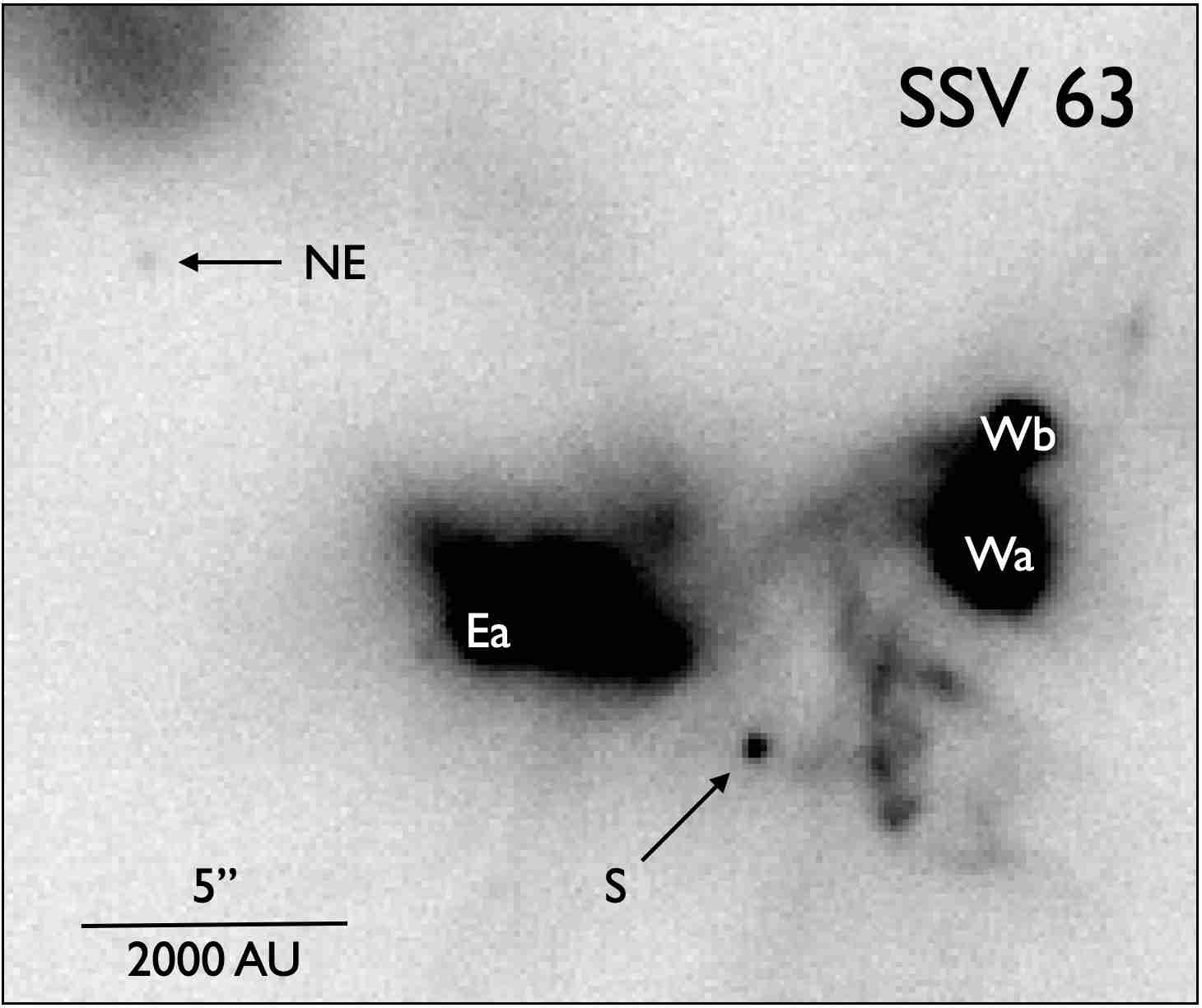}}
\caption{ An H$_2$ image obtained at the Gemini-N telescope showing
the faint source S just south of the E and W binaries together with a
weak detection of the embedded source NE. Together with the
ALMA-detected source N, SSV~63 thus constitutes at least a septuple
stellar system. A complex of H$_2$ knots is seen between knots Wa and
Ea. North is up and east is left. 
\label{embedded-S}} 
\end{figure}

\begin{figure}
\centerline{\includegraphics[angle=0,width=8.25cm]{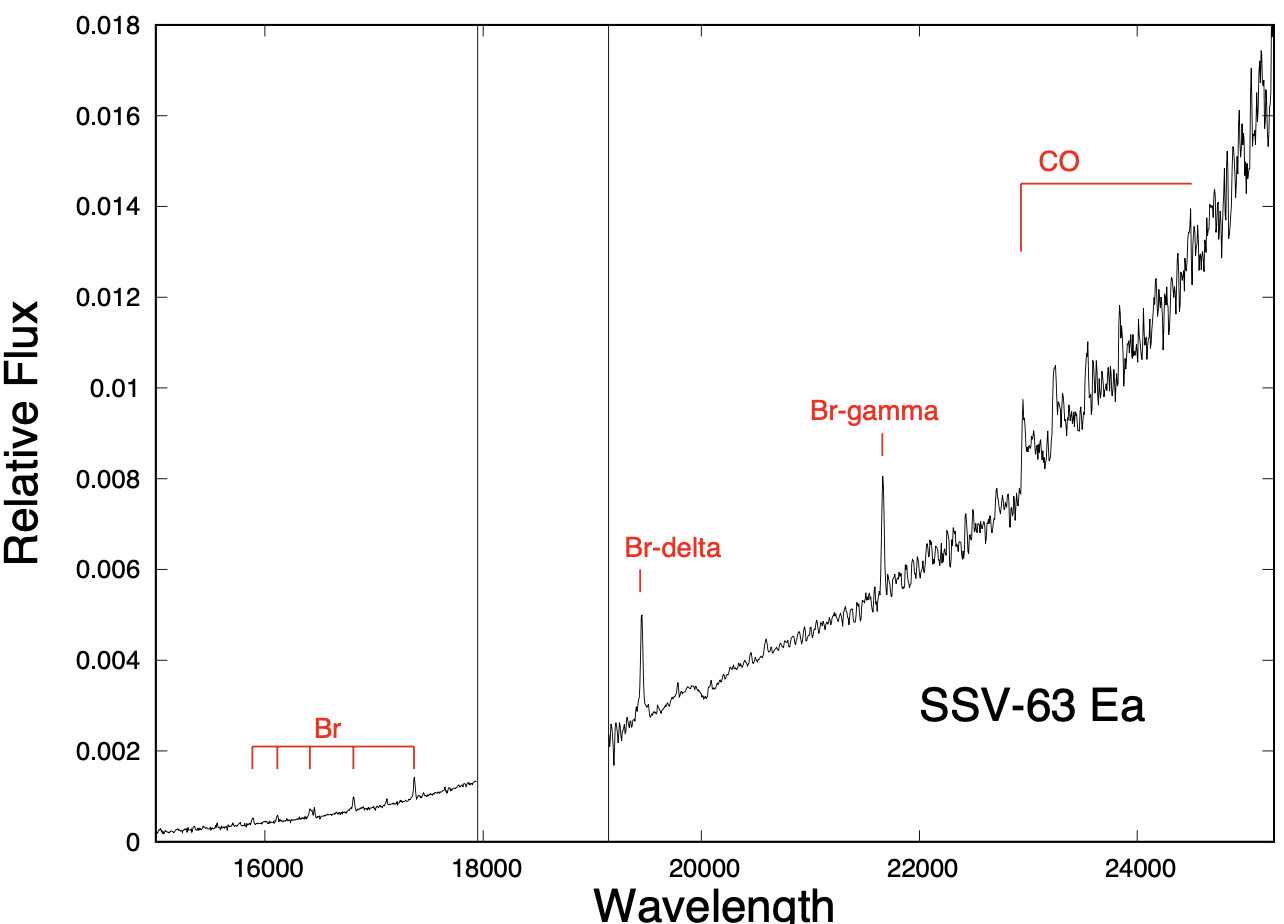}}
\centerline{\includegraphics[angle=0,width=8.3cm]{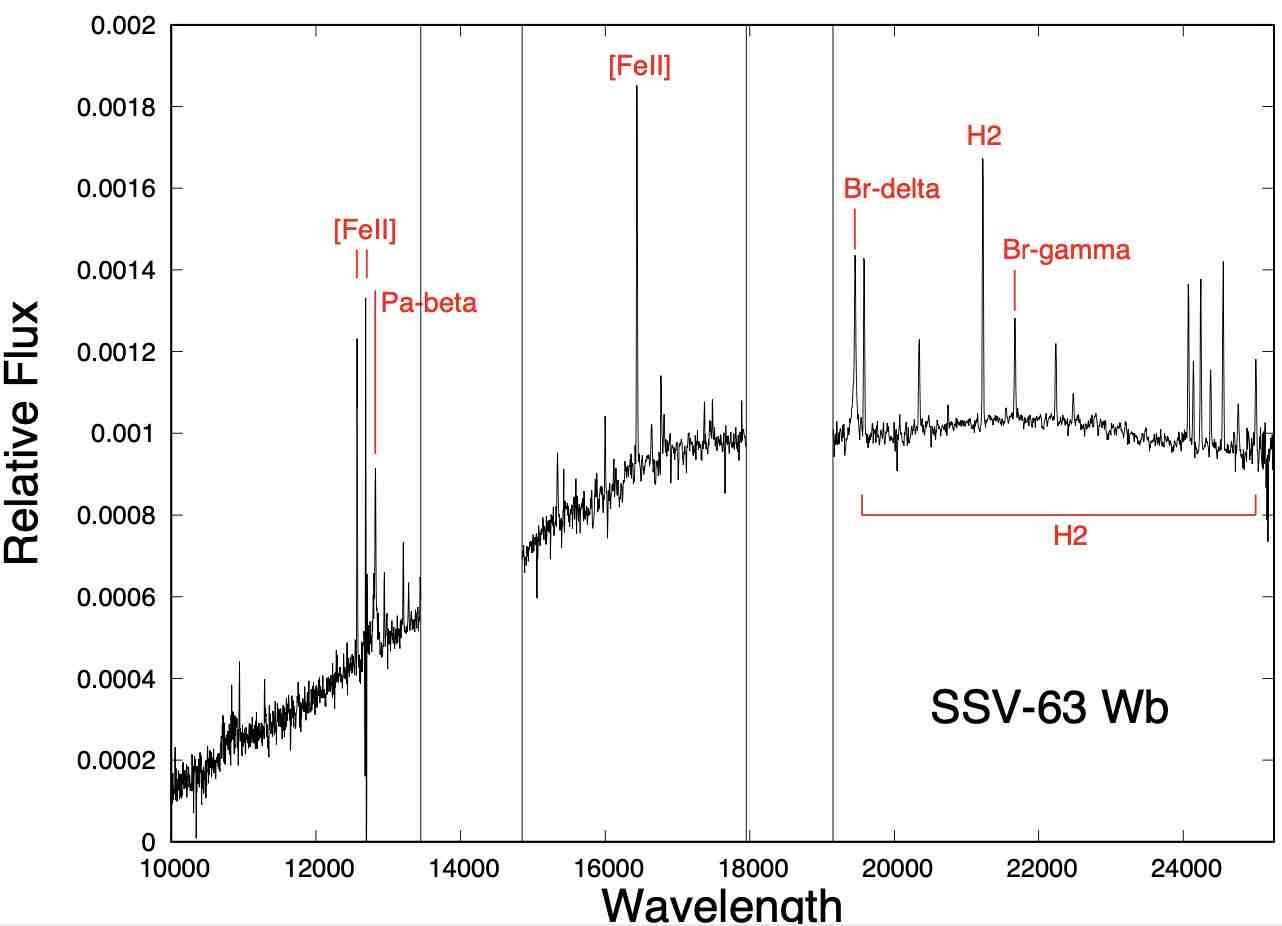}}
\caption{GNIRS spectra of SSV63 Ea and Wb. Source Ea shows a heavily reddened continuum with a few emission lines and the CO-bands in emission and no absorption features. In contrast, source Wb shows little reddening but a forest of molecular and atomic hydrogen lines, together with [\Feii], indicative of shocked outflow. Spectral regions with poor atmospheric transmission are omitted.
\label{gnirs}}
\end{figure}

\vspace{0.3cm}

\section{THE HH 24 JET COMPLEX}\label{sec:hh24complex}

In the following we study in detail the complex structure of the
HH~24 jet group, based on Gemini, Subaru, and HST images. We discuss
all the individual jets in the HH~24 complex based on new ultradeep
high spatial resolution groundbased images. These reveal numerous new
previously unseen or unresolved knots, which allow a better
understanding of the multiple flow structures in the HH~24 complex.
We introduce a new flow, HH~24X, and extend current knot nomenclature
for the principal jets C and E, see Section~\ref{sec:observations}.

The environment of HH~24 in a $\sim$6$'$$\times$10$'$ field is shown
in Figure~\ref{overview}, which is the sum of deep (1 hour) exposures
in H$\alpha$ and [\Sii] obtained with SuprimeCam at the Subaru 8m
telescope. HH 24 is located in a highly structured N-S oriented cloud
filament studied at mm-wavelengths by,
e.g., Lada et al. (1991), and in the sub-mm by, e.g., Kirk et
al. (2016). Figure~\ref{gemini} shows more detail of the jets in an
optical color-figure based on the broadband and narrowband Gemini
images. The figure shows how the group of jets that constitute HH~24
is emanating from a dense cloud core and in the process is tearing
apart the cloud environment. Figure~\ref{definitions} shows a
difference image between H$\alpha$ and [\Sii] displayed such that
H$\alpha$ dominant regions are black and [\Sii]-dominant regions are
white. The figure is annotated with designations for the individual
jets.

We have also obtained HST images using WFC3 with H$\alpha$, [\Sii],
and [\Feii] filters, see Section~\ref{sec:observations} for full
details. Figure~\ref{pressrelease} shows a color image based on our
narrow-band filter HST images and archival broadband HST images, which
provides a more detailed overview of the region. The individual
narrow-band images of the E- and C-jets are shown in
Figure~\ref{triptych}. These images do not have the same field-of-view
as the Subaru and Gemini images, but offer higher resolution. In
Section~\ref{sec:jets} we discuss the properties of the HH 24 jet
complex based on these and other data sets.


\vspace{0.3cm}

\section{THE SSV 63 MULTIPLE SYSTEM}\label{sec:sources}

In this section we consider the multiple system that drives the
cluster of jets discussed above, and we attempt to associate specific
jets with individual sources.

Strom et al. (1976) detected a near-infrared source associated with
HH~24 in a survey of L1630. It was subsequently detected in the 6~cm
radio continuum (Bieging et al. 1984) and later at mid- and
far-infrared wavelengths as IRAS 05436-0011 (Cohen \& Schwartz
1987) and with Herschel as HOPS~387 (Furlan et al. 2016).
SSV~63 was resolved as a binary source with $\sim$10$''$ separation by
Zealey et al. (1992) and Moneti \& Reipurth (1995) and in the radio
continuum by Bontemps et al. (1995). Subsequently, Davis et al. (1997)
found that SSV~63W is itself a binary with a separation we measure as
1.95$''$.  Reipurth et al. (2002) found yet another source, SSV~63NE,
further to the north-east at 3.6~cm, which was detected
at mid-infrared wavelengths by Huelamo et al. (2007). In the same
study, Huelamo et al. found a new source at mid-infrared wavelengths,
labeled Eb, located about 2.6\arcsec\ NNW of source E, henceforth Ea.  
Source~Eb was also detected by Tobin et al. (2020) in their
large-scale sub-mm and radio continuum survey of Orion protostars.

Figure~\ref{ssv63-h2} shows an archival H$_2$ image obtained with
NICMOS on HST (PI Muzerolle, Program 11205) which demonstrates that
SSV~63 is a non-hierarchical quadruple system. Such systems are
unstable and will eventually break apart. This is further discussed in
Section~7.  Source~Eb appears prominently in a Spitzer 8~$\mu$m image
where it is well separated from Ea (Figure~\ref{ssv63-8micron}).


Properties of these and other sources are
listed in Table~\ref{table:coordinates}. 
Additional photometry with adaptive optics is listed in Table~\ref{table:vlt}.


\begin{deluxetable*}{lrrrrrrrrrrrrrr}
\tabletypesize{\scriptsize}
\tablecaption{Coordinates and 2MASS-, SPITZER-, and WISE-Photometry of HH~24 Sources and  H$\alpha$ emission stars \label{table:coordinates}}
\tablewidth{0pt}
\tablehead{
\colhead{Object} & $\alpha(2000)^a$ &$\delta(2000)^a$ & J & H & K & W1 & I1 & I2 & W2 & I3 & I4 & W3 & W4 & M1 }
\startdata 
               &           &             &1.25 &1.65 &2.2  &3.4    &   3.6 & 4.5   & 4.6   & 5.8   &  8    &  12   &   22  &  24 \\ 
\cline{1-15}\\
IRS~1$^b$           &05:46:07.77&--00:09:38.3 &     &     &     & 13.12 & 12.60 & 11.59 & 11.09 & 10.76 &  9.81 &  8.33 &  4.47 &  -- \\ 
               &           &             &     &     &     &  0.04 &  0.01 &  0.01 &  0.03 &  0.01 &  0.01 &  0.05 &  0.07 &  -- \\
HH24-Wb        &05:46:07.84&--00:09:59.3 &     &     &     &       &       &       &       &       &       &       &       &     \\ 
               &           &             &     &     &     &       &       &       &       &       &       &       &       &     \\
HH24-Wa        &05:46:07.86&--00:10:01.2 &15.20&13.46&11.94&  9.32 &  9.93 &  8.60 &  7.52 &  7.63 &  6.62 &  4.45 &  0.73 & 2.12\\ 
               &            &            & 0.12& 0.13& 0.08&  0.05 &  0.01 &  0.01 &  0.04 &  0.01 &  0.01 &  0.02 &  0.02 & 0.03\\
HH24-S         &05:46:08.16&--00:10:05.3 &     &     &     &       &       &       &       &       &       &       &       &     \\ 
               &           &             &     &     &     &       &       &       &       &       &       &       &       &     \\
HH24-Eb        &05:46:08.40&--00:10:00.6 &     &     &     &       &       &       &       &       &       &       &       &     \\ 
               &           &             &     &     &     &       &       &       &       &       &       &       &       &     \\
HH24-Ea        &05:46:08.49&--00:10:03.0 &15.86&14.16&11.16&  8.40 &  8.10 &  6.69 &  5.49 &  5.57 &  4.39 &  2.86 &--0.11 & 0.00\\ 
               &           &             & --  &  .12&  .05&  0.02 &  0.01 &  0.01 &  0.03 &  0.01 &  0.01 &  0.01 &  0.01 & 0.01\\
HOPS 317          &05:46:08.53&--00:10:39.1 &17.79&16.79&15.13& 12.80 & 12.31 & 10.65 & 10.20 &  9.39 &  8.31 &  7.19 &  2.57 & 3.59\\ 
               &           &             & --  & --  &  .13&  0.03 &  0.01 &  0.01 &  0.02 &  0.01 &  0.01 &  0.02 &  0.02 &     \\
HH24-N$^c$     &05:46:08.46&--00:09:54.8 &     &     &     &       &       &       &       &       &       &       &       &     \\ 
               &           &             &     &     &     &       &       &       &       &       &       &       &       &     \\
HH24-NE        &05:46:08.92&--00:09:56.1 &     &     &     &       & 11.63 &  9.33 &       &  7.87 &  6.97 &       &       & 3.40\\ 
               &            &            &     &     &     &       &  0.04 &  0.01 &       &  0.01 &  0.01 &       &       & 0.07\\
HH24-H$\alpha$1 &05:46:11.34&--00:07:55.1&17.61&15.86&15.15&       & 13.85 & 13.10 &       & 12.72 & 11.83 &       &       & 8.06\\ 
               &           &             &  .28&  .14&  .15&       &  0.01 &  0.01 &       & 0.03  &  0.04 &       &       & 0.06\\
HH24-H$\alpha$2 &05:46:12.27&--00:08:07.8&14.63&13.94&13.51& 12.70 & 12.72 & 12.32 & 12.07 & 11.91 & 11.10 &  9.06 &  6.83 & 8.21\\ 
               &           &             &  .03&  .02&  .04&  0.03 &  0.01 &  0.01 &  0.03 &  0.02 & 0.02  &  0.05 &  0.02 & 0.07\\
HH24-H$\alpha$3 &05:46:12.99&--00:08:14.8&16.55&15.63&14.98& 14.45 & 14.32 & 13.59 & 13.63 & 12.73 & 10.90 &  8.83 &  6.35 & 6.32\\ 
               &           &             &  .11&  .11&  .12&  0.04 &  0.01 &  0.01 &  0.05 &  0.03 & 0.02  &  0.04 &  0.13 & 0.02\\
HH24-H$\alpha$4 &05:46:13.17&--00:09:10.0&16.34&15.39&15.34& 15.09 &       &       & 14.74 &       &       & 12.36 &  7.81 &     \\
               &           &             &  .10&  .09&  .16&  0.04 &       &       &  0.06 &       &       &  --   &  --   &     \\ 
IRS~2           &05:46:13.47&--00:08:56.2 &18.68&15.95&13.54& 12.25 & 11.71 & 11.04 & 10.94 & 10.55 &  9.53 &  8.49 &  6.00 & 5.72\\ 
               &           &             & --  &  .15&  .04&  0.02 &  0.01 &  0.01 &  0.02 &  0.01 &  0.01 &  0.03 &  0.06 & 0.02\\
HH24-H$\alpha$5 &05:46:13.58&--00:10:34.0&16.66&16.08&15.69& 15.09 &       &       & 14.88 &       &       & 12.38 &  8.86 &     \\
               &           &             &  .13&  .20&  .23&  0.04 &       &       &  0.07 &       &       &  --   &  --   &     \\
\enddata

\tablecomments{
\textsuperscript{a}
Coordinates for HH24-Ea, -Wa, -NE, and MMS-VLA1 are
3.6~cm VLA astrometry from Reipurth et al. (2002), for HH24-N from
ALMA (this paper), for MMS-HOPS317 from 2MASS, for IRS~1 from WISE, for
IRS~2 from Spitzer I1-image, for SSV63-Eb from
Spitzer I4-image, and for the rest they are from 2MASS. The Spitzer
photometry is from Megeath et al. (2012). Note that a few sources that
are close to brighter sources or surrounded by bright reflection
nebulae can be seen in Spitzer images, but meaningful photometry
cannot be extracted.
\textsuperscript{b}
IRS~2 is not in the 2MASS catalog, even though it is optically visible, 
presumably due to confusion from its proximity to the knots in the C-jet.
\textsuperscript{c} 
HH24-N is a submm source only detected by ALMA. 
}
\end{deluxetable*}\label{sources}


\begin{deluxetable}{llrl}
\tablecaption{VLT Photometry of SSV 63 Components\label{table:vlt}}
\tablecolumns{3}
\tablewidth{0pt}
\tablehead{
   \colhead{Star} &
   \colhead{K$_s$}  &
   \colhead{L'} &
  }
\startdata
Wb & $>$16.5        & 12.34$\pm$0.04 \\
Wa & 12.79$\pm$0.03 &  9.64$\pm$0.03 \\
Eb & $>$16.5        &  9.19$\pm$0.03 \\
Ea & 12.70$\pm$0.03 &  8.32$\pm$0.03 \\ 
S  & 16.16$\pm$0.22 & 13.67$\pm$0.06 \\
NE & $>$16.5        & 11.21$\pm$0.03 \\
\enddata  
\tablecomments{ These adaptive optics data from the ESO archive 
were obtained with NACO at the ESO VLT.}  
\end{deluxetable}


\subsection{Near-IR Imaging and Spectroscopy}\label{subsec:imaging}

None of the three sources Wa, Wb, and Ea are visible at optical
wavelengths, and at near-infrared wavelengths the dominant source is
Wa. 
At longer wavelengths, the Ea and Eb sources are dominant. From their
energy distributions, all the five main components of SSV~63 are
likely Class~I sources as determined by Furlan et al. (2016), who used
near-, mid-, and far-infrared data to study the sources (under the
designations HOPS 386 and HOPS 387). We note that Eb is highly
obscured and detectable only at mid-infrared and longer wavelengths,
so it is likely a borderline Class~0 source.

One additional source is found in the region on a deep K-band image
from the Gemini-N 8m telescope. Figure~\ref{embedded-S} shows this
image, with the new very faint source, marked S, identified. The
source is midway between and slightly to the south of the prominent Wa
and Ea sources. It is faint, with K$\sim$16.2, and it is
not detected at shorter wavelengths, most likely due to extinction. In
the L'-band it is much brighter, L$\sim$13.7, see
Table~\ref{table:vlt}.  Since the source is not seen in Spitzer
images it is unlikely to be as luminous as the other sources, nor to
be a background red giant. Given its location towards a dense cloud
core, we assume that the source is a deeply embedded very low-mass
star or brown dwarf.

Figure~\ref{gnirs} shows the Gemini/GNIRS spectra of SSV~63 Ea and Wb.
Source Ea shows a steeply rising continuum devoid of absorption lines
 with the CO bands as well as the Bracket hydrogen series in
emission.  In contrast, source Wb is much less reddened but
sufficiently veiled to wash out absorption features. Its spectrum
displays prominently a forest of molecular and atomic lines, as well
as lines of [\Feii], indicative of a shocked outflow. A planned spectrum
of Wa was weathered out, but Simon et al. (2004) have presented a K-band
spectrum of this source which shows a red continuum with a prominent
Br$\gamma$ emission line and some weaker H$_2$ lines.

\subsection{Spitzer Imaging}\label{subsec:spitzer}

Spitzer observed the L1630 cloud and Megeath et al. (2012)
compiled a catalog of all young stellar objects in the region. SSV~63
Wa, Ea, and NE are detected in all bands, whereas Wb is only weakly seen at
3.6~$\mu$m. As mentioned earlier, the Spitzer images reveal a new
source, SSV~63~Eb, located just 2.8~arcsec ($\sim$1100~AU)
NNW of what is now labeled Ea. At 3.6~$\mu$m, Eb is seen as an
extension to Ea, increasing in brightness at longer wavelengths, and
at 8~$\mu$m it is nearly as bright as Ea. At 24~$\mu$m the pair is
blended, but it appears that Eb has become the dominant source.

\begin{figure} 
\centerline{\includegraphics[angle=0,width=6.9cm]{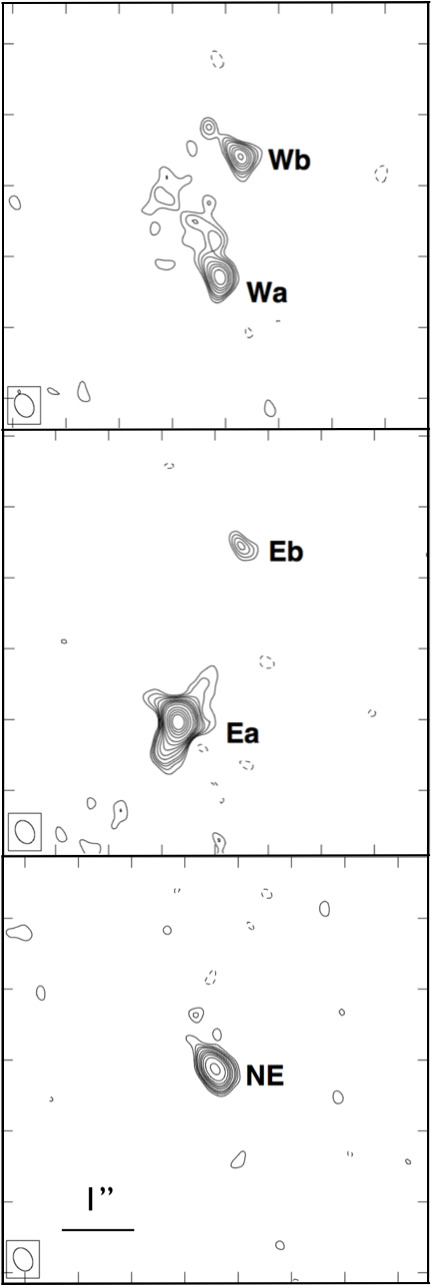}}
\caption{ VLA X-band maps of the 5 main
sources of SSV 63. A possible companion to Wb is seen, as well as
filamentary structure linked to Wa. The Ea source shows a radio jet
along the axis of the E/C jet pair, and an orthogonal stubby bipolar
structure. Positions and flux densities are given in
Table~\ref{table:vla-parameters}. North is up and east is left.
\label{vla-composite}} 
\end{figure}

\subsection{VLA Observations}\label{subsec:vla}

SSV~63 was detected in the 6~cm radio continuum by Bieging et
al. (1984) and at 3.6~cm by Bontemps et al. (1995), who resolved the
SSV~63 E-W binary.  Reipurth et al. (2002) carried out a 3.6 cm study
in the A-configuration which detected a new source, labeled SSV~63~NE.
Most recently, Tobin et al. (2020) observed the SSV~63 region as part
of the large VANDAM protostellar survey.\footnote{Tobin et al. (2020)
use the following nomenclature for the 5 main sources in SSV~63: Ea =
HOPS~386A, Eb = HOPS~386B, NE = HOPS~386C, Wb = HOPS~387A, Wa =
HOPS~387B.}


\begin{deluxetable*}{lccccccc}
\tabletypesize{\scriptsize}
\tablecaption{Parameters of the Radio Sources in the SSV 63 Region\label{table:vla-parameters}}
\tablehead{
\colhead{}  & \multicolumn{3}{c}{X Band$^a$} &  \multicolumn{3}{c}{Q band$^b$} & Spectral\\
\colhead{Source} & \colhead{$\alpha(2000)^c$} & \colhead{$\delta(2000)^c$}
& \colhead{S($\mu$Jy)$^d$} & \colhead{$\alpha(2000)^c$} & \colhead{$\delta(2000)^c$}
& \colhead{S($\mu$Jy)$^d$} & Index}
\startdata
Wb &  $07\rlap.{^s}836\pm0\rlap.{^s}001$ &  $09'~59\rlap.{''}59\pm0\rlap.{''}02$ & 51$\pm$6 &  $07\rlap.{^s}837\pm0\rlap.{^s}001$ &  $09'~59\rlap.{''}60\pm0\rlap.{''}02$ & 1578$\pm$60 & 2.3$\pm$0.1 \\
Wa &  $07\rlap.{^s}855\pm0\rlap.{^s}001$ &  $10'~01\rlap.{''}29\pm0\rlap.{''}03$ & 119$\pm$11 &  $07\rlap.{^s}855\pm0\rlap.{^s}001$ &  $10'~01\rlap.{''}30\pm0\rlap.{''}01$ & 376$\pm$60 & 0.8$\pm$0.1 \\
Ea &  $08\rlap.{^s}485\pm0\rlap.{^s}001$ &  $10'~03\rlap.{''}04\pm0\rlap.{''}01$ & 203$\pm$9 &  $08\rlap.{^s}485\pm0\rlap.{^s}001$ &  $10'~03\rlap.{''}04\pm0\rlap.{''}01$ & 1181$\pm$90 & 1.2$\pm$0.1\\
Eb & $08\rlap.{^s}426\pm0\rlap.{^s}001$ &  $10'~00\rlap.{''}54\pm0\rlap.{''}02$ &    15$\pm$2 &  ... & ...  & $\leq$50  & $\leq$0.8  \\
NE &  $08\rlap.{^s}922\pm0\rlap.{^s}001$ &  $09'~56\rlap.{''}12\pm0\rlap.{''}01$ & 83$\pm$6 &  $08\rlap.{^s}922\pm0\rlap.{^s}001$ &  $09'~56\rlap.{''}11\pm0\rlap.{''}02$ & 315$\pm$80 & 0.9$\pm$0.2 \\
HH~24 MMS &  $08\rlap.{^s}380\pm0\rlap.{^s}004$ &  $10'~43\rlap.{''}71\pm0\rlap.{''}05$ & 141$\pm$15 &  $08\rlap.{^s}381\pm0\rlap.{^s}002$ &  $10'~43\rlap.{''}70\pm0\rlap.{''}02$ & 10750$\pm$120 & 2.9$\pm$0.1 \\
\enddata
\tablenotetext{a}{10.0 GHz} 
\tablenotetext{b}{44.0 GHz}
\tablenotetext{c}{$\alpha(2000) = 05^h~46^m$; $\delta(2000) = -00^\circ$.}
\tablenotetext{d}{Total flux density in $\mu$Jy.}
\end{deluxetable*}


We have carried out a deep high-resolution study of SSV~63 with the
JVLA in the X-band ($\sim$3~cm, see Section~2 for details of the
observations). The five dominant sources in the SSV~63 multiple
system, Ea, Eb, Wa, Wb, and NE, are detected, and
Table~\ref{table:vla-parameters} lists the VLA coordinates and total
flux density for each YSO. Source~Ea is by far the brightest in the
radio continuum. Extended structure is seen around the sources, see
Figure~\ref{vla-composite}. Noteworthy is what appears to be a faint
companion to Wb at a separation of 0.6~arcsec and a position angle of
43$^\circ$ ($\alpha_{2000}$ = 5:46:07.866, $\delta_{2000}$ = 
--00:09:59.18). However, more observations are needed to confirm its
stellar nature. Source Wa exhibits what appears to be an almost 2
arcsec long wiggling outflow towards the NNE. 
Alternatively the extended emission may be thermal emission from a
ridge of dust. Source Ea displays a prominent bipolar radio continuum
jet along the axis of jet~E, with evidence for another weaker outflow
perpendicular to the first, suggesting that source Ea is a close
binary. A similar quadrupolar structure is seen around the prominent
jet source HH~111 VLA-1 (Reipurth et al. 1999). There is also a weak
extension from Source NE towards the HH~24~G flow, although it should
be noted that the source extension in that direction almost coincides
with the direction of the slightly elongated beam profile. Perhaps the
more surprising result is that source Eb, which is so prominent in the
mid-infrared, is the weakest of the sources.

\subsection{ALMA Observations}\label{subsec:sourceN}

We have observed the SSV~63 multiple system with ALMA in the 1.3~mm
continuum, see Section~\ref{sec:observations}. The sources Ea, Eb, Wa,
Wb, and NE were all detected, and additionally a new source, here
labeled N, was detected. Source~S discussed in
Section~\ref{subsec:imaging} was not detected. The ALMA observations of
these sources are discussed in detail in Section~\ref{sec:ALMA-I}.

\subsection{X-ray Observations}\label{subsec:xray}

SSV 63 has been observed several times at X-ray wavelengths. Ozawa et
al. (1999) obtained a 30~ks exposure with ASCA, but were not able to fully
resolve SSV~63 from the bright X-ray source SSV~61 (HBC~502) to the
south (see Figure~\ref{gemini}). Simon et al. (2004) used Chandra to
resolve SSV~63 into Ea, Wa, and NE. The companion Wb was not detected.
All three components have hard X-ray spectral indices. Spectral
modeling of the brightest X-ray source, Wa, suggested a visible
extinction of roughly 48 mag. However, they found that the depth of
the 3.08~$\mu$m ice band indicated only 10-20 mag of extinction.
Principe et al (2014) did a very deep X-ray study of the L1630 region
and also detected these three sources. In none of these X-ray studies
was HH~24~MMS detected.

\begin{figure}
\centerline{\includegraphics[angle=0,width=8.3cm]{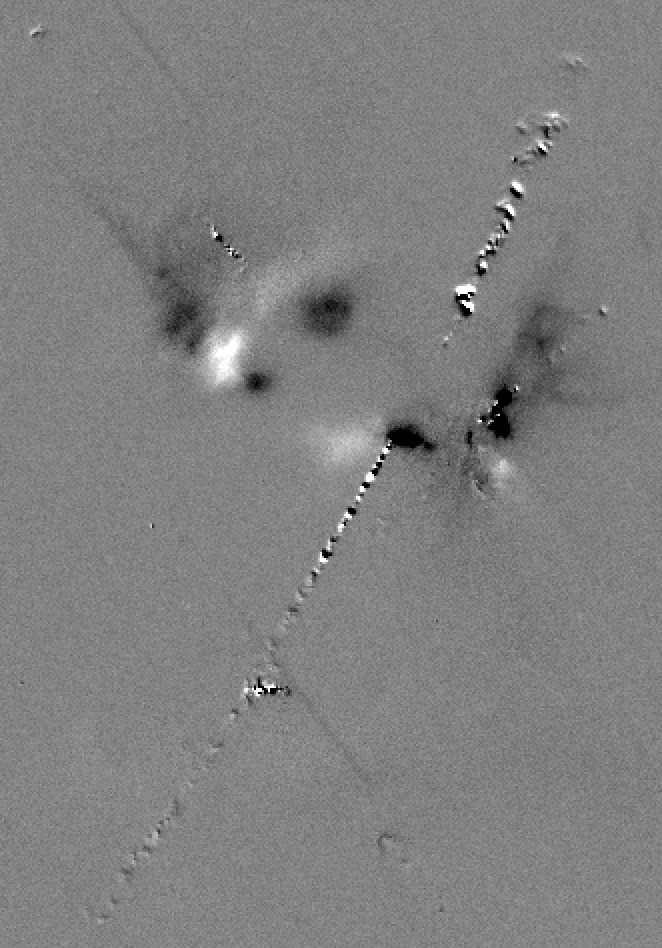}}
\caption{Difference between the 2014 (black) and 2016 (white) HST [\Feii] 
images, showing the motion of the jet knots, seen especially clearly
in the E and C jets. Note the $\sim$5$^\circ$ change in position angle
of the southeastern portion of the E-jet. Substantial variability in
the reflection nebulae appears as black and white pairs of
nebulosity. Three parallel line-segments running from upper left to
lower right are artifacts. North is up and east is left.  The figure is about 55\arcsec \/ wide.
\label{difference}}
\end{figure}

\subsection{Reflection Nebulae}\label{subsec:reflection}

The HH 24 complex contains several bright reflection nebulae. The
early polarization studies by Strom et al. (1974b), Schmidt \& Miller
(1979) and Scarrott et al. (1987) demonstrated that the source of
illumination is associated with SSV~63, but the angular resolution was
too low to identify any specific source. The principal reflection
nebulosity, labeled knot~D by Herbig (1974), is seen towards the base
of the G-jet, see e.g., Figure~\ref{gemini}. It is likely, at least in part, to originate
from the NE source, which is obscured by a dense core of gas and dust
(Figure~\ref{pressrelease}). This is corroborated by comparing the
optical H$\alpha$ and [\Sii] images with an infrared image, see
Section~\ref{subsec:jetG}.

These reflection nebulae are variable, as can be seen when comparing
images from the two epochs of HST observations
(Figure~\ref{difference}). Such variability of reflection nebulosity
around a young star was first seen by Hubble
(1917) and Knox-Shaw (1917) and can be caused by light escaping 
from a partly embedded source (e.g., Reipurth \& Bally 1986, Dahm \&
Hillenbrand 2017). Such variations are shadowing effects from
material moving close to the illuminating star (Graham \& Phillips
1987). Additional compact reflection nebulae are located around the
sources Ea/b and Wa/b (Figure~\ref{embedded-S}).

\subsection{Association of Jets and Sources}\label{subsec:association}

As discussed above, there are at least five sources in the SSV~63
multiple system, and together with source~S and the additional companions
suggested by the VLA observations as well as yet another component
(source~N) detected by ALMA (see Section~\ref{sec:ALMA-I}),
the system contains at least 7 components. We here attempt to
sort out the connection between the multiple jets and the individual
sources.

The most eye-catching of the many jets in HH 24 is the E/C pair. Jet E
is evidently launched by source Ea, as clearly seen in the HST and VLA
images (Figures~\ref{pressrelease} and \ref{vla-composite}). Jet~C
lies within just a few degrees of a line through jet~E, and it is
blueshifted whereas jet~E is redshifted, and hence it would be
reasonable to assume that they form one bipolar pair. However, the two
jets have rather different morphologies, with jet~E being perfectly
collimated whereas jet~C has an irregular and wobbling
appearance. Also, with the discovery of the embedded source Eb on the
line connecting jets~E and C, there is a potential different source to
drive jet~C. However, our ALMA observations
(Section~\ref{sec:ALMA-II}) show that there is almost no high-velocity
emission associated with Eb, and the little there is forms a stubby
bipolar outflow along an axis inclined by roughly 20$^\circ$ to the
axis of jet~E. Moreover, the southeastern lobe of this microflow is
blueshifted and the northwestern is redshifted, opposite to that of
jets~E and C. We conclude that source Eb is not related to the
C-jet. This leaves open the question of why the E and C jets have such
different morphologies. One possibility is that the C jet is forcing
its way through the dense core in which the two sources Ea and Eb have
formed, and through internal deflections in the core is losing an
initial high collimation.

Jet G has an unusual structure, as discussed in
Section~\ref{subsec:jetG}. Despite its morphology it does have a well
defined axis, and SSV63~NE lies precisely along this axis. Our VLA
observations show that the source is elongated along this axis.

\begin{figure}
\centerline{\includegraphics[angle=0,width=7cm]{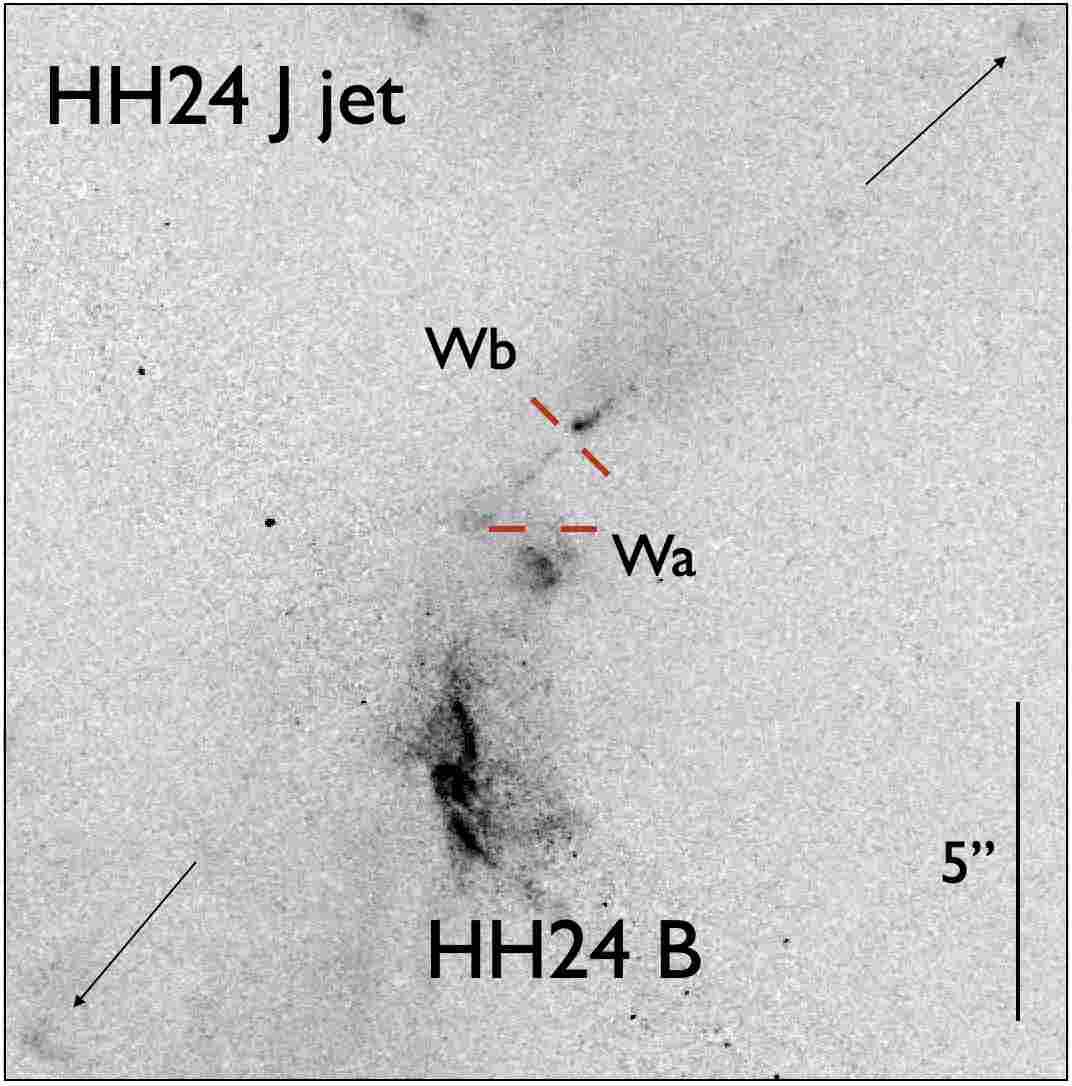}}
\caption{Jet J in a [\Sii] image taken with HST and
WFC3. The jet emanates from the source Wb which is deeply embedded and 
only detected at mm and cm wavelengths, the object seen at optical and
infrared light is a combination of shocks and reflected light. The
source drives a very faint but highly collimated jet towards the NW
and pointing to the large HH~19 bow shock. The SE lobe is bent
slightly southwards, and points to the bright HH~27 bow shock. The
location of the embedded sources Wb and Wa are marked with red lines.
North is up and east is left.
\label{jetJ}}
\end{figure}

Jet J consists of a series of [\Sii]-dominated knots located on a very
well defined line that passes directly through the Wb source, which is
likely the driving source. This alignment shows that jet~J is {\em
not} driven by the nearby bright source Wa.  VLA observations suggest
that Wb may be a binary with 0.6$''$ separation, and either of the two
sources could be driving the jet. There is a bit of emission just to
the SE of Wb, the rest of the jet is only seen in the NW lobe. HH~19
is a distant bow shock driven by source Wb
(Section~\ref{subsec:giant-flows}). Figure~\ref{jetJ} shows the inner
region of jet~J around the driving source. The precise location of the
source derived from ALMA data reveals that the optical knot is not the
driving source, but a compact reflection nebula mixed with shocked
emission (see also Figure~\ref{ssv63-h2}).

Jet X is an inconspicuous slightly wobbly chain of faint
[\Sii]-dominant knots (Figure~\ref{definitions}). It points directly
away from the very faint source~S, which is likely a brown dwarf seen
through significant extinction (see Section~\ref{subsec:imaging}). An
increasing number of outflows have been found from very young brown
dwarfs, e.g., Riaz et al. (2017), Riaz \& Bally (2021).

Jet L is not driven by any of the sources in the SSV~63 multiple
system, but by the nearby source HOPS 317 or by the embedded Class~0
source HH~24~MMS further to the south. This is discussed in detail in
Section~\ref{sec:hh24mms}.

In summary, the SSV~63 multiple system is found to consist of at least
7 sources: Ea, Eb, Wa, Wb, NE, S, and N within an ellipse of roughly
10\arcsec $\times$ 20\arcsec (4000~AU $\times$ 8000~AU).  Additionally
the VLA observations suggest that Ea is an unresolved binary driving a
quadrupolar jet, and Wb appears to have a faint companion. These
sources are likely Class~I sources, but the lack of near-infrared
emission and X-ray emission from Eb and Wb suggest that they could be
Class~0 sources. However, blending at longer wavelengths precludes a
more precise classification. The very low luminosity of sources S and
N suggest that they may be very low-mass stars or brown dwarfs.

\begin{figure}
\centerline{\includegraphics[angle=90,width=5.2cm]{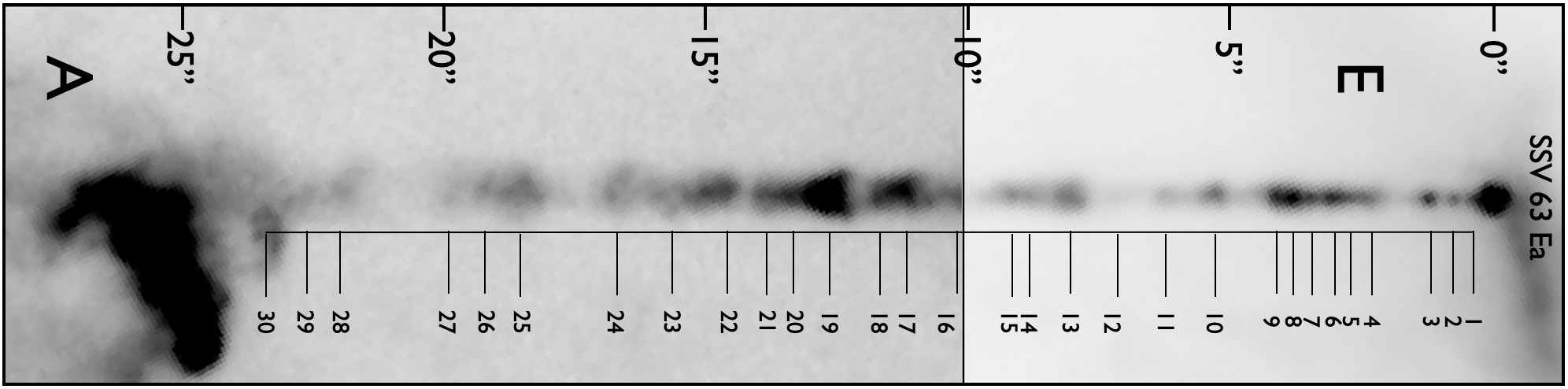}}
\caption{Structure of the E jet, based on the HST [\Feii] image in two cuts. 5 arcsec corresponds to 2000~AU.  The panel shows jet~E emanating from the Ea~source. Knot~A is the large bright knot at the bottom. Individual knots in the E jet are numbered, see text for details. 
\label{feii-E-jet}}
\end{figure}

\begin{figure}
\centerline{\includegraphics[angle=0,width=8.3cm]{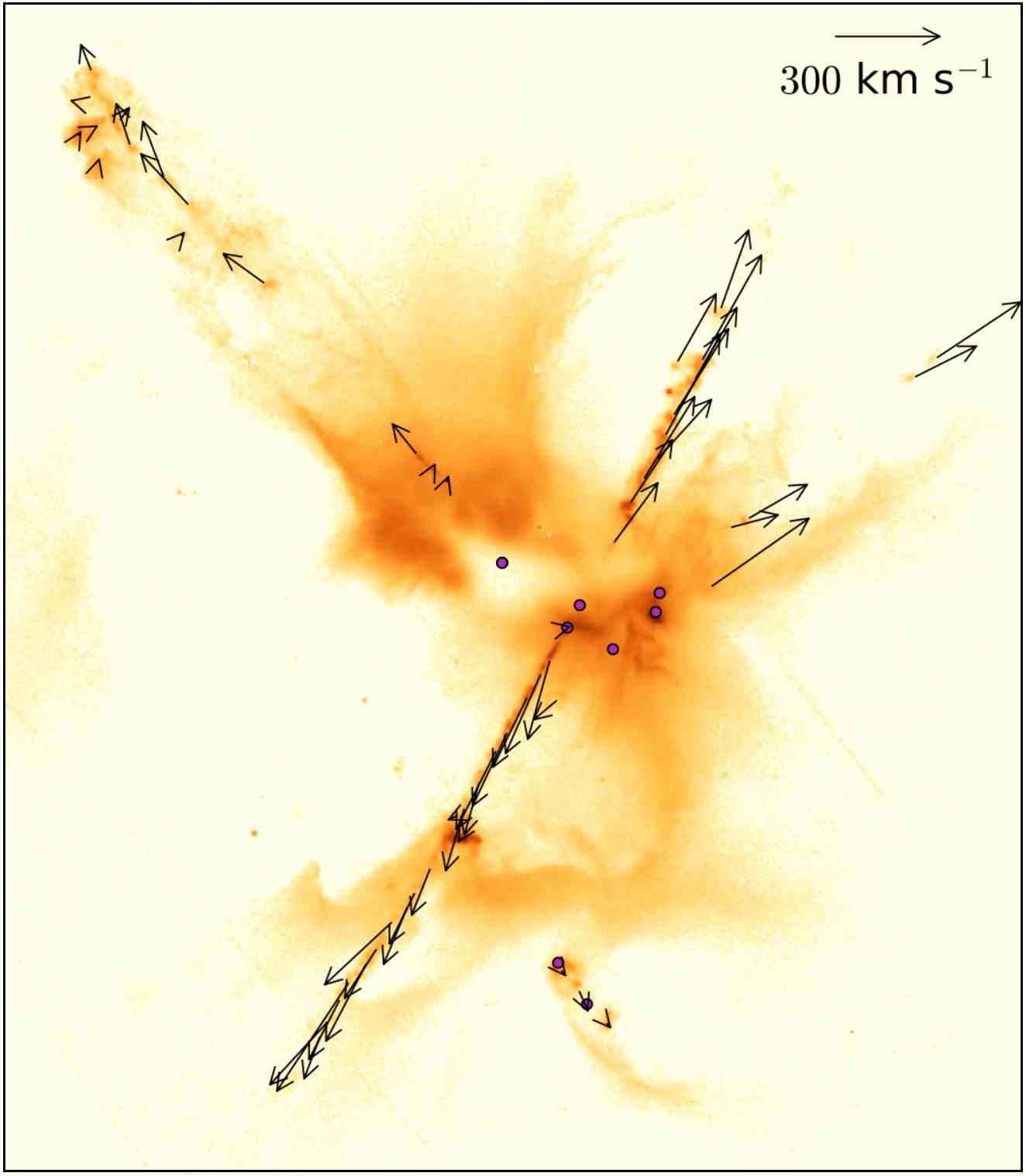}}
\caption{Proper motions based on two epochs of HST images superposed
on an H$\alpha$ HST image.  The 300 \kms \/ velocity
vector is about 11\arcsec \/ long and shows the motion in about
75~yr.
\label{propermotion}}
\end{figure}

\begin{figure}
\centerline{\includegraphics[angle=0,width=8.3cm]{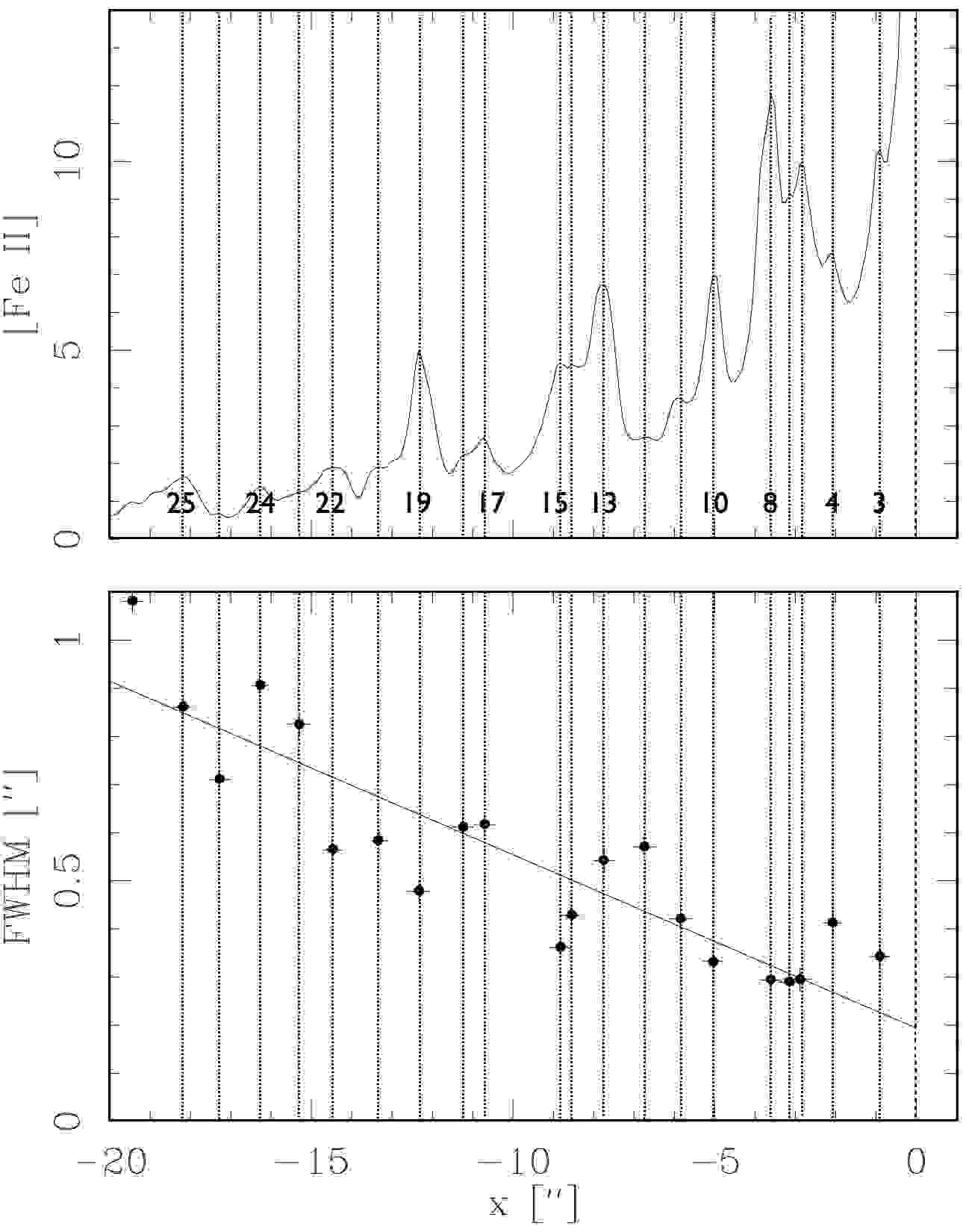}}
\caption{(top) A tracing of the HH~24E-jet from the [\Feii] HST image. Knots are identified with the nomenclature defined in Figure~\ref{feii-E-jet}.
(bottom) The FWHM of the individual knots of the E-jet within the first 20~arcsec (8000~AU) of the source were calculated by subtracting in quadrature the point-spread function. 
\label{jet-expansion}}
\end{figure}

\vspace{0.3cm}

\section{INDIVIDUAL JETS AND SHOCKS}\label{sec:jets}


\subsection{HH 24 Jet E}\label{subsec: jetE}

As is evident in Figure~\ref{pressrelease}, jet~E is the most
prominent of the multiple jets in the HH~24 complex, and is remarkable for its
highly collimated appearance.  It is very weak in H$\alpha$ and
strong in [\Sii], indicating a series of very weak shocks. The near-infrared [\Feii] and H$_2$ images at the Gemini-N
telescope reveal that jet~E is very bright in [\Feii]. In
contrast, jet~E is not emitting in H$_2$.


\subsubsection{Structure and Excitation}\label{subsubsec:structure-E}


The perfect collimation of jet~E is seen well in the new HST images,
and is particularly evident in the [\Feii] image in
Figure~\ref{feii-E-jet}. However, beyond the large shock~A, the jet
slightly shifts course towards the southeast, as if it was deflected
by an angle of $\sim$5$^\circ$. The nature of shock A is further
discussed in Section~\ref{subsec:HH24A}.

Figure~\ref{triptych} shows that the E-jet has a different appearance
in the three filters transmitting H$\alpha$, [\Sii], and
[\Feii]. Since H$\alpha$ and [\Sii] have similar wavelengths, they are
affected similarly by extinction. Hence the ratio between the two
relates to intrinsic properties of the shocks. H$\alpha$ is much
weaker, and it follows that jet~E is a very low-excitation flow, and
hence has low-velocity shocks. In contrast, the [\Feii]/[\Sii] ratio
is heavily affected by extinction. Because the [\Feii] 1.64~$\mu$m and
[\Sii] 0.67~$\mu$m lines have similar energies of 1.7 and 1.8~eV above
ground, and \Fei\ and \Si\ atoms have comparable ionization potentials
of 7.87 and 10.36 eV, respectively, it follows that the intensity
ratio of the two transitions is a good indicator of
extinction. Whereas jet~E can be traced all the way to the source in
[\Feii], the first knot that is (barely) visible in the [\Sii] image
is E6. In projection this is 1000~AU from source Ea. But there is
still some extinction out to a projected distance of about 3000~AU
from Ea. From the bright knot~E13 and outwards, the [\Feii]/[\Sii]
ratio is essentially constant, indicating that the jet has broken out
of the cloud core. This situation is very similar to the case of the
HH~1 jet, which undergoes two abrupt steps in extinction at 1400 and
3000~AU (Reipurth et al. 2000). We discuss the cloud core in more
detail in Section~\ref{sec:ALMA-II}, and interpret the [\Feii]/[\Sii]
ratio in Section~\ref{subsec:Ea-flow}.

\subsubsection{Proper Motions and Radial Velocities}\label{subsubsec:propermotion-E}

Our two HST images of the HH~24 complex in the [\Feii] 1.644~$\mu$m
line are separated by 744 days. As is evident in
Figure~\ref{difference} the motion of the jets is readily visible,
allowing us to measure the proper motions of the shocked outflows. We
have used a code that convolves the images with wavelet functions of
chosen width, see Raga et al. (2016b) for details. Jet~E shows
pronounced motion, as illustrated in Figure~\ref{propermotion}. The
slight deviations of some vectors from the well defined direction of
the jet are likely due to slight changes in the structure of the
knots. Especially near the source, such deviations can have
significant impact on the angles. The mean tangential velocity of the
knots between the source and HH~24A is about 250~\kms. This is
comparable to other HH jets,
e.g., the HH~1 jet has a proper motion of $\sim$280~\kms\ (Bally et
al. 2002) and the HH~34 jet $\sim$190~\kms\ (Reipurth et al. 2002a).

In our medium-resolution spectroscopy of jet~E with the Apache Point
3.5m telescope, the [\Sii] 6717/6731 lines are the brightest and have 
heliocentric velocities from about +170 to 200~\kms\, with a peak
around +170 to 180~\kms. If the bulk radial motion is about +175~\kms,
and we adopt a proper motion of 250~\kms, then it follows that jet~E
moves away from the observer at an angle of roughly 35$^\circ$ to the
plane of the sky with a total space velocity of $\sim$300~\kms.



\subsubsection{Ejection Variability}\label{subsubsec:ejection-E}

The [\Feii] emission along the HH~24E jet is divided into three main
groups of peaks: one at distances $x=2'' \to 5''$, the second
$5'' \to 10''$ and the third $10'' \to 15''$ from the outflow
source.  Figure~\ref{feii-E-jet} provides a detailed view of the
jet, with individual knots numbered.

Selecting the points of highest intensity within each of the three
groups, we obtain a mean separation between the groups of knots
$<\Delta x>_1=7.7''\pm 3.6''$. Together with a mean proper motion
velocity $v_{pm}=250$~km~s$^{-1}$, this gives a timescale
$\tau_1=<\Delta x>_1/v_{pm}=(33\pm 16)$~yr.

Similarly, if we take all of the intensity peaks in the top frame of
Figure 13, we obtain a mean knot separation $<\Delta x>_2=0.93''\pm
0.36''$, which for $v_{pm}=250$~km~s$^{-1}$ gives a timescale
$\tau_2=(7.1\pm 2.8)$~yr.

Conceivably, $\tau_1$ and $\tau_2$ could correspond to two modes of a
quasi-periodic, time-dependent ejection variability. Also, the
ejections could be non-periodic with a characteristic timescale of
$\sim 7$~yr (corresponding to the timescale deduced including the
fainter intensity peaks along the jet, see above), and with the
brighter knots corresponding to mergers of the fainter knots. There is
at least partial evidence that such knot mergers occur in the HH~34
jet (see Raga \& Noriega-Crespo 2013), for which more detailed
observations have been made.




\subsubsection{Jet Expansion}\label{subsubsec:expansion-E}

It has been found in several well collimated HH jets that the knots
widen as they move away from the source, e.g., the HH ~1 and HH~34
jets (Reipurth et al. 2000, 2002a). It is clear from
Figure~\ref{feii-E-jet} that this is also the case for the
HH~24E jet. The knots are well resolved in the HST images, and
Figure~\ref{jet-expansion} shows a gradual expansion of 0.7~arcsec in
total width along the first 20~arcsec until it enters the complex
region around the bright knot~A. This corresponds to a full opening
angle of the jet of 2.6$^\circ$, which is comparable to the opening
angles measured for other jets (e.g., Erkal et al. 2021). 
A jet velocity of 300 \kms\ implies that a half-angle of 1.3\arcdeg\
corresponds to knots spreading orthogonally to the jet axis with a
velocity of 7 \kms, comparable to the sound speed expected in the
post-shock cooling layers where [\Sii] emission originates. If the
plasma is fully ionized ($\mu \approx$ 0.6), the temperature of this
region is about 3,500 K. For mostly neutral gas ($\mu \approx$ 1.3)
the temperature is $\sim$8000~K.


\begin{figure}
\centerline{\includegraphics[angle=0,width=8cm]{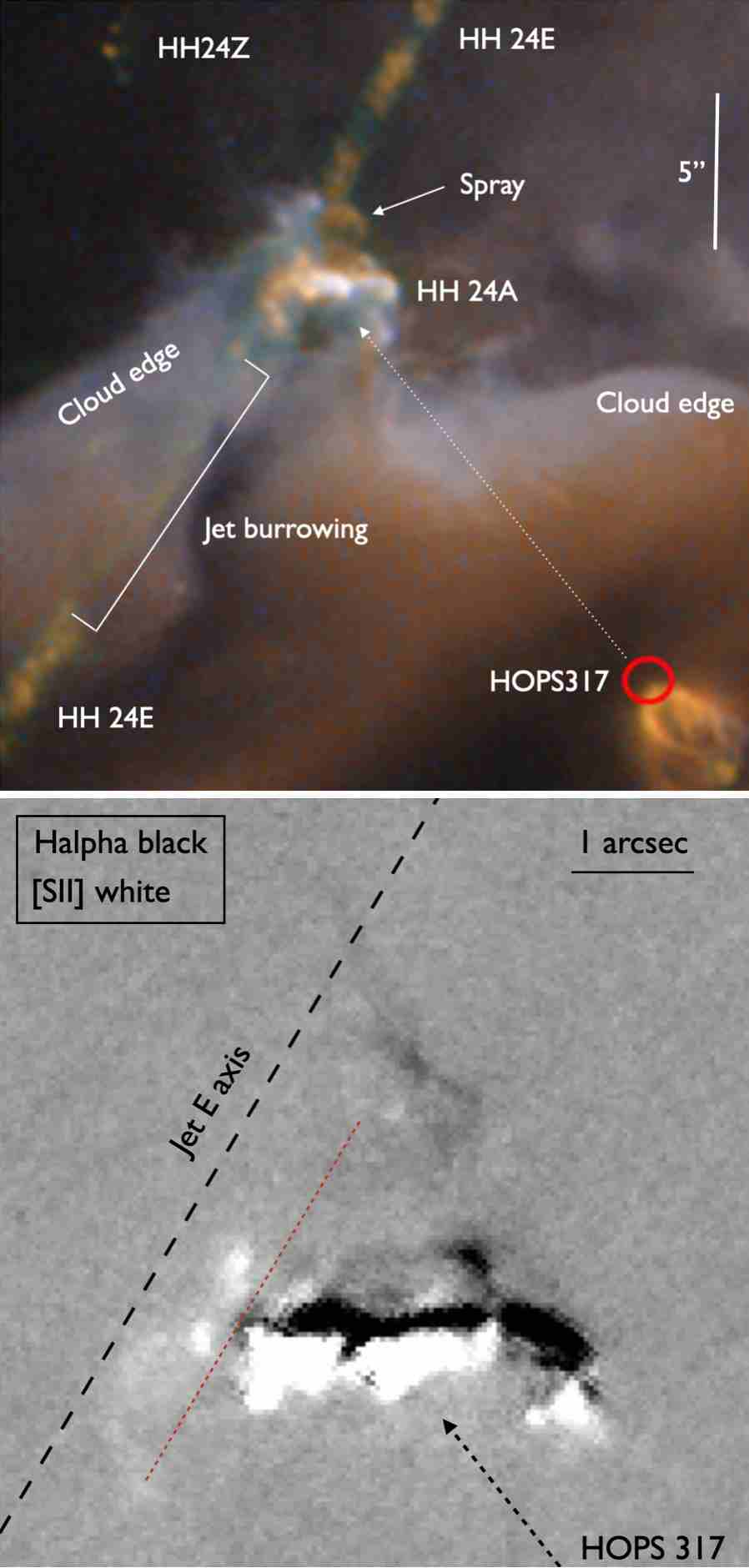}}
\caption{(top) Detail of Figure~\ref{pressrelease} showing the bright 
bow shock HH~24A, located at the
intersection of two flows originating from the embedded sources Ea and
HOPS~317. The well collimated jet HH~24E launched from the Class~I source
Ea impacts a cloud edge (seen well in Figure~\ref{pressrelease}) and
partly burrows through the cloud to re-emerge further down in a
slightly different direction.   
(bottom) An H$\alpha$-[\Sii] difference
image of HH~24A, with H$\alpha$ black and [\Sii] white. The little
group of faint [SII]-bright knots to the left of the red dashed line move
approximately along the jet-E axis towards the SSE with about 40-50
~\kms\ and evidently form part of this outflow. The bright central
region of HH~24A is stationary, while the western extension is either
stationary or has at most a slight motion towards the west.
The dotted arrow indicates the direction from HOPS~317. North is up and east is left.
\label{HH24A}}
\end{figure}

\subsection{The HH 24A Shock}\label{subsec:HH24A} 

The two jets HH 24E and C are located in the interior of a pair of
low-extinction cavities, north and south of the SSV~63 core, that are
rendered visible in the near-infrared by scattered light
(Figure~\ref{pressrelease}).  These cavities may have been excavated
by the long-term action of the SSV~63 jets and outflows.  Two pillars
facing the SSV~63 region are located along the south wall of the
southern cavity. 
The HH~24A shock is located 25\arcsec\ (10,000 AU) south of SSV~63~Ea
and about 2\arcsec\ south of the tip of the largest pillar in the
cloud wall at the southern end of jet~E.
It is the brightest shock in the HH~24 complex, and has long been
assumed to be a working surface for the HH~24E jet, possibly
interacting with the cloud.


Spectra of the brightest part of the HH~24A shock show peak velocities
ranging from +30 to +40~\kms, much less than the radial velocity of
the HH~24E jet, and thus supporting the above picture that HH~24A is a
shock driven into a stationary cloud. The [\Sii] 6717/6731 ratio is
$\sim$0.68, indicating an electron density of 2400~cm$^{-3}$ for a
temperature of 10,000~K (or 1800 at 5,000~K). Jones et al. (1987)
present low-resolution spectra in which [\Oiii] is detected, thus
showing that at least some part of HH~24A has a high excitation, very
different from the very low excitation of the HH~24E jet.

Jet E disappears at the pillar tip near the HH~24A shock, but
re-appears about 8\arcsec\ farther south, bent towards the east by
about 5\arcdeg .  One possible interpretation is that jet E impacts
the back-side of the pillar, and is deflected towards the east by the
interaction.\footnote{It should be noted that the little jet~X
associated with source~S (see Section~\ref{subsec:association}) is
pointing straight towards the deflected part of the HH~24E jet so, at
least in principle, it cannot be excluded that this deflected part of
the jet could have an origin different from source~Ea.}  At right
angles to jet E, HH~24A extends about 2\arcsec\ farther west than the
western edge of the jet (Figure~\ref{HH24A}-top).
Figure~\ref{HH24A}-bottom shows an HST H$\alpha$-[\Sii] difference
image of HH~24A, which reveals a two-shock structure of the main body
of HH~24A, with an H$\alpha$-strong part facing north and a southern
side that is [\Sii]-bright. 

\begin{figure} 
\centerline{\includegraphics[angle=0,width=8.3cm]{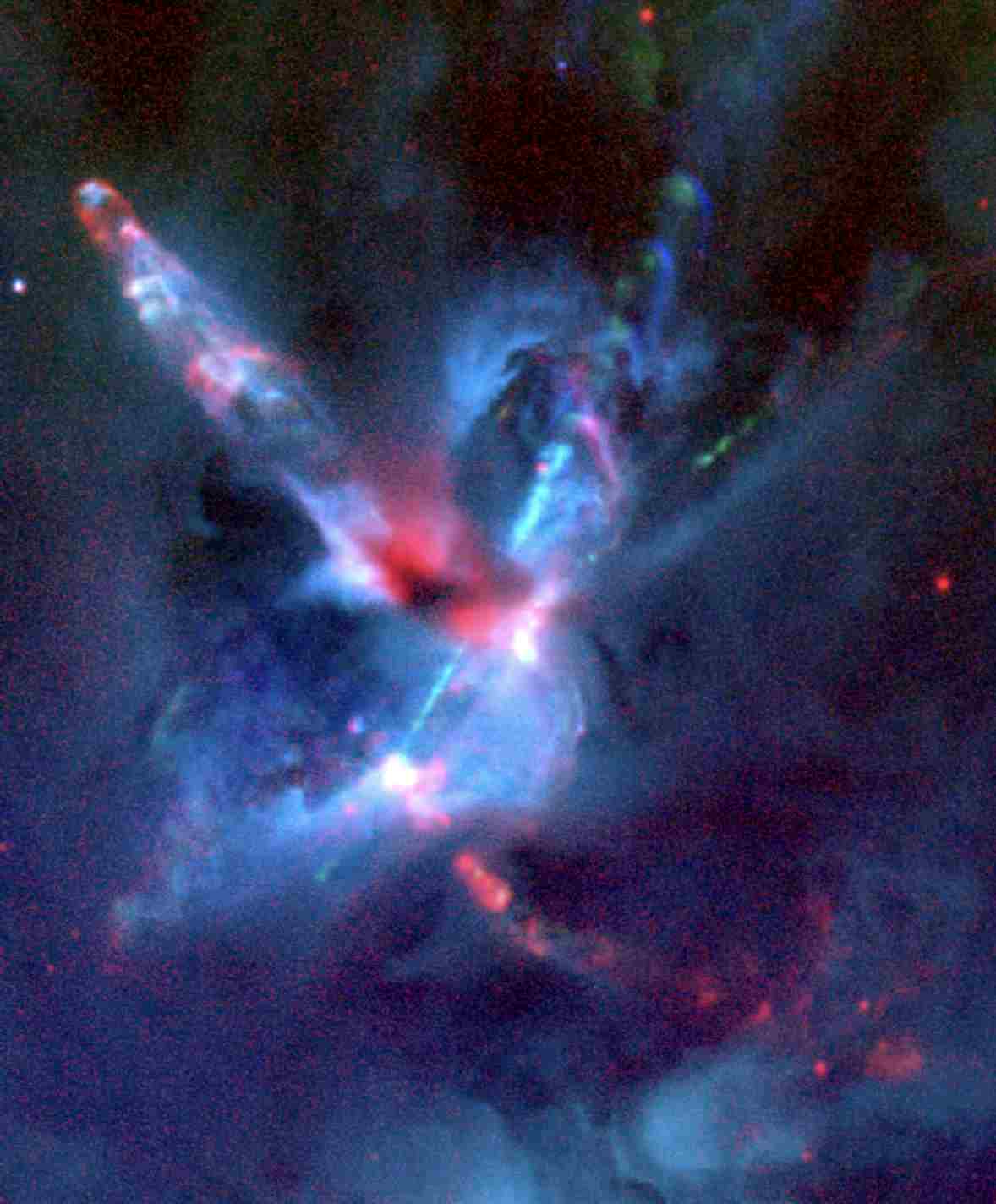}}
\caption{ A superposition of an H$\alpha$ (blue), a [\Sii] (green), and a 2.12~$\mu$m molecular hydrogen (red) image of the HH 24 complex obtained at the APO 3.5m telescope. {\color{red} The figure is 2.5' wide.} 
\label{opt-ir-mosaic}} 
\end{figure}

HH~24A is located only about 22$''$ from the Class~0 source
HH~24~MMS (see Section~\ref{sec:hh24mms}), and Bontemps et al. (1996)
suggested that HH~24A may be a separate shock from a flow originating
in this embedded source. HH~24~MMS is located just outside the lower
right corner of Figure~\ref{HH24A}-top. That image shows a conical
outflow cavity of another source, the Class~0 source HOPS~317, which
is located even closer, only 17$\arcsec$, to HH~24A.  This reflection
nebula is opening up towards the southwest, suggesting that the
blueshifted lobe of outflow L is located southwest of this YSO.
HH~24A, which is redshifted, is located along the expected counterflow
direction of outflow L.  A line from HOPS~317 to HH~24A is aligned
with the outflow cavity of HOPS~317 as well as the molecular hydrogen
outflow (HH~24L) extending SW from HOPS~317 (see
Section~\ref{sec:hh24mms}).  Figure~\ref{opt-ir-mosaic} shows that
this lobe of the L-counterflow also contains shock-excited 2.12 $\mu$m
\Htwo\ emission connecting HOPS~317 to HH~24A.
In addition, to the NE of HH~24A a new faint shock, here called
HH~24Z, is found (Figure~\ref{HH24A}-top), which could be part of the
outflow driven by HOPS~317.  
It thus appears, on morphological grounds, that HH~24A might be a bow
shock powered by HOPS~317.

Ideally, proper motions should resolve the issue of the origin of the
HH~24A bow shock. Unfortunately, the 2-yr time interval between our
two epochs of [\Feii] HST images are not sufficient to show any motion
reliably, but adding an archival wideband image including the [\Feii]
1.64~$\mu$m line does show some rather slow motions. 

Figure~\ref{HH24A}-bottom shows two areas of HH~24A separated by a red
dashed line. The [SII]-bright knots to the left of the line have
motions towards the SSE with about 40-50~\kms, roughly along the
direction defined by jet~E. They are slightly displaced from the axis
of jet~E, either because the jet has been disturbed by
burrowing through the cloud, similar to jet~C, or they may be shocks
from a wider angle wind interacting with a flow cavity. 

The central part of HH~24A to the right of the red line is essentially
stationary, indicating that the shock is ramming into the cloud. The
western wing may have a slow tangential motion of 20-30$\pm$15~\kms\
approximately due west. HH~24A shows a classical two-shock structure,
with an H$\alpha$-strong and a [\Sii]-strong component. The dashed
arrow shows the direction from the HOPS~317 source.

The data available do not allow a definite conclusion on the origin of
HH~24A, it could originate from either source Ea or HOPS~317. If the
gentle westward motion of the wesstern wing is real it would in both
cases represent gas squirting sideways along the wing of the bow
shock. If the bright part of the HH~24A shock comes from HOPS~317 then
both flows from Ea and HOPS~317 interact with the pillar, but not
necessarily with each other.  The high-surface brightness of HH~24A
and detection of [\Oiii] suggests that it is interacting with the
front side of the pillar while the cloud interaction with jet~E occurs
mainly within or on the back side of the pillar.


\subsection{HH 24 Jet C}\label{subsec: jetC}

\subsubsection{Structure and Excitation}\label{subsubsec:structure-HST-C}

\begin{figure}
\centerline{\includegraphics[angle=0,width=5cm]{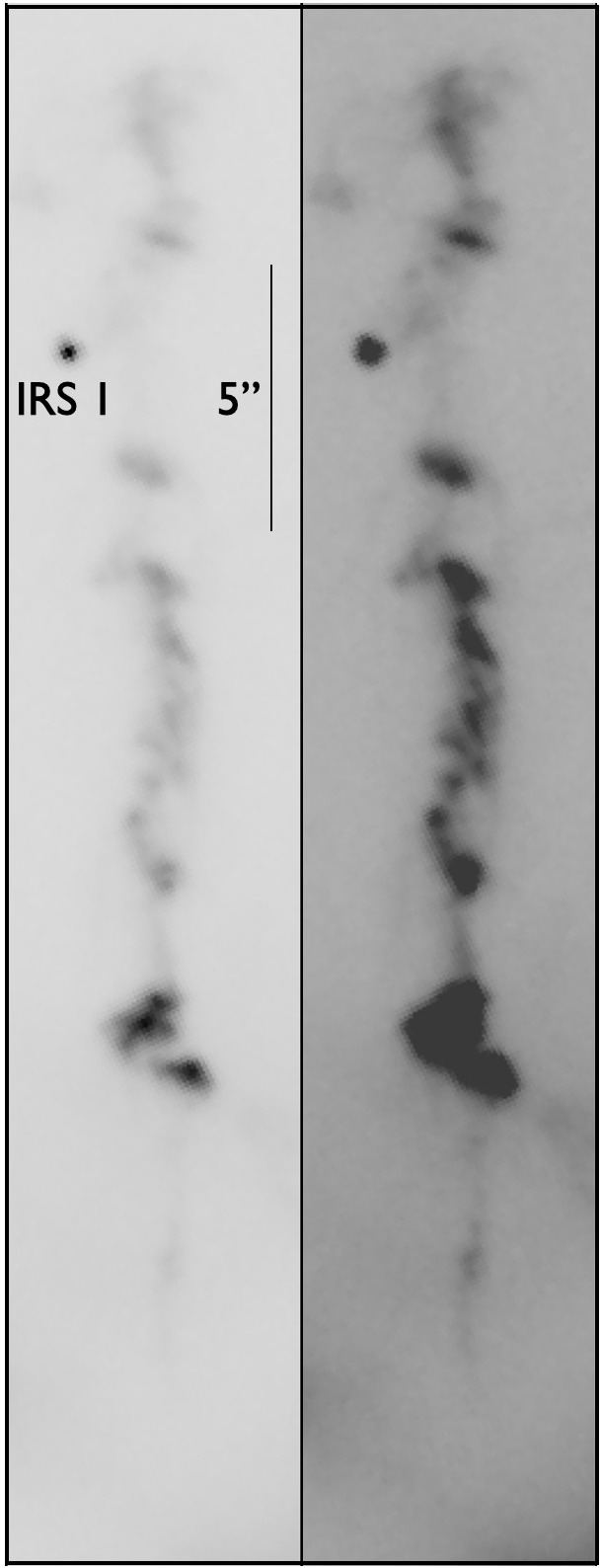}}
\caption{Structure of the C jet, based on the HST [\Feii] image in two cuts. 5 arcsec corresponds to 2000~AU.  
\label{feii-C-jet}}
\end{figure}

Figure~\ref{feii-C-jet} shows the detailed structure of jet~C as seen
in the HST [\Feii] image. Although it appears to be a counter-jet to
jet~E, it does not share the perfect collimation of jet~E. Another
puzzling fact is that while jet~E can be traced directly back to the
source even though it is red-shifted, in contrast jet~C only becomes
visible (in the 1.644~$\mu$m [\Feii] line) about 8.5~arcsec north of
the source Ea (see below).

\begin{figure}
\centerline{\includegraphics[angle=0,width=5cm]{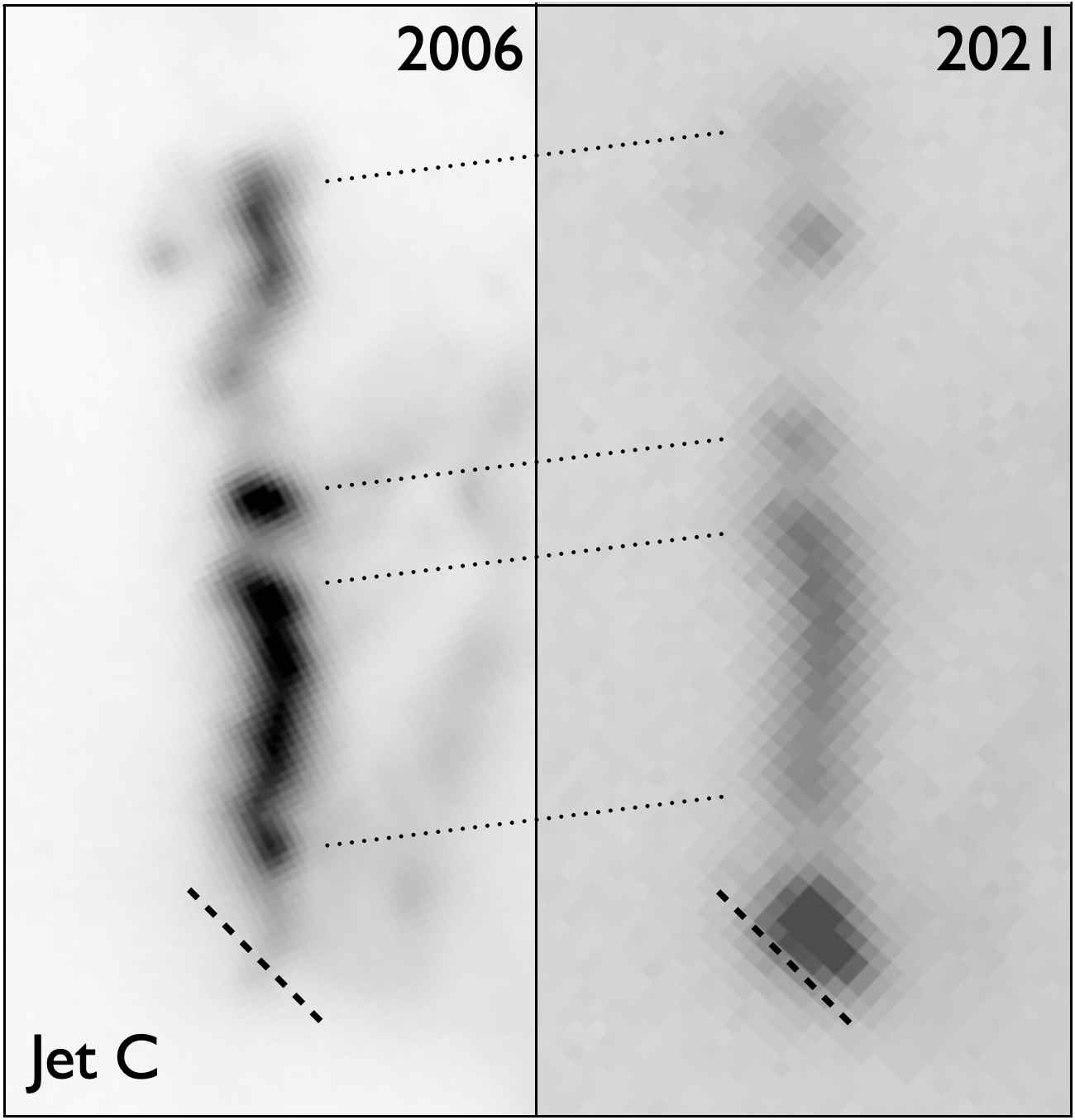}}
\caption{The C jet in two groundbased [\Sii] images taken in 2006 at Subaru and in 2021 at Apache Point Observatory. The emergence of two bright knots from behind a cloud edge, indicated by the dotted line, is clearly seen.
The motion of the jet during the 15 years between exposures is
indicated.  
\label{jetC-2006-2021}}
\end{figure}

Figure~\ref{definitions} shows that near the outflow source, jet~C is
strong in [\Sii] and is surrounded by H$\alpha$-strong shocks sitting
on the 'shoulders' of the individual knots that protrude to either
side of the main jet axis, in a very similar fashion to what is seen
in the equally wiggling HH~46/47 jet (Heathcote et al. 1996).
These H$\alpha$ arcs are deflection shocks or spur
shocks, caused by knots glancing off the side of a mostly-evacuated
cavity, and they are seen in several other prominent jets, like HH~1,
34, and 47 (Heathcote et al. 1996, Hartigan et al. 2005, 2011).  
Further out along the flow axis a series of bow shocks are seen, which
show a clear double-shock structure, with an inner [\Sii]-strong shock
and an outer H$\alpha$-strong envelope. This is as expected from a
heavy jet pushing through a tenuous ambient medium, either stationary
or co-moving, which will produce a double-shock working surface, with
a weak jet-shock and a stronger bow shock (Hartigan
1989, Reipurth \& Heathcote 1992).

\subsubsection{Proper Motions and Radial Velocities}\label{subsubsec:kinematics-C}

Jones et al. (1987) measured the proper motion of part of the C-jet
and derived a tangential velocity of about 320~\kms\ to the NNW away
from SSV~63. Our proper motion study concurs with this result,
indicating motion of about 300~\kms\ away from the Ea/Eb pair
(Figure~\ref{propermotion}). Our spectra along the C-jet show
blue-shifted emission across the velocity range -180 to -230 \kms\
with a peak radial velocity of -200~\kms. If we adopt these two
numbers, then we find that jet~C is moving towards us at an angle to
the plane of the sky of roughly 34$^\circ$. This is comparable to the
angle of $\sim$35$^\circ$ for the redshifted jet~E, and although these
angles have uncertainties of at least several degrees, their
similarity supports the interpretation that the two jets form one
bipolar outflow.

Figure~\ref{jetC-2006-2021} shows two groundbased [\Sii] images, 
one taken in 2006 at the Subaru telescope and a new taken in 2021 at
the Apache Point Observatory. A new knot has appeared, emerging from
behind a dense cloud edge. This new knot is also seen in the H$\alpha$
and [\Sii] images obtained with HST in 2014 (Figure~\ref{triptych}),
narrowing the interval during which it appeared to between 2006 and
2014. Our spectra yield a [\Sii] ratio 6717/6731 of 0.63, 
indicating an electron density of 2300 - 3000 (T=10,000~K). The knot is
blueshifted.

\subsubsection{Origin of Jet~C}\label{subsubsec:structure-C}

Given that jets C and E are almost perfectly aligned with each other,
and the fact that jet~C is blueshifted while jet~E is redshifted (Solf
1987) with the same angle to the plane of the sky, it appears evident
that they form parts of a single bipolar outflow. However, as was
discussed in Section~\ref{sec:sources}, there are two sources between
the two jets, Ea and Eb, so in principle the jets could arise from
separate sources, which would make it easier to understand the curious
difference in morphology of jets C and E. However, if jets C and E
were driven by two separate sources, then we would expect that each
source would also have a counterjet. Given the limited size of the
SSV~63 cloud core, such counterjets should be readily visible. Also,
the VLA observations of Ea clearly show a bipolar radio continuum jet
along the common E/C jet axis. It thus seems well established that
jets C and E form opposite sides of a bipolar outflow from
source~Ea. The slight mis-alignment of the C and E jets could be
explained if source Ea moves towards the southwest through the SSV~63
cloud core with a speed of $\sim$2 \kms , consistent with the expected
motion of stars within the gravitational potential of the core.  In
this scenario, source Eb does not drive any jet. Our ALMA data shows
that source~Eb does drive an ultra-compact arcsecond-scale molecular
flow along a northwest-southeast axis with the redshifted lobe
oriented to the northwest (Section~\ref{subsec:MO1}).

\subsubsection{Wiggling of Jet~C}\label{subsubsec:wiggling-C}

As already mentioned, jet~C shows pronounced wiggling, which might
suggest that the source is either a binary or the jet is anchored in a
precessing disk. Raga et al. (2009a) have made models of precessing
accretion disks around a star in a binary system, and find that it
leads to a reflection-symmetric spiraling outflow on small scales from
the orbital motion together with a reflection-symmetric spiral on
large scales due to the precessing disk.


On closer examination, however, this interpretation runs into difficulties.
If we estimate the ratio between a typical extent of one of the wiggles
(d$\sim$10$''$ or $\sim$4000 AU) and its sideways displacement (h$\sim$1$''$),
together with the measured jet velocity  v$_j$ = 250 km/s
the orbiting jet model then  yields an estimate for the orbital velocity:

v$_o$ =v$_j$ $\times$ h/d = 25 km/s   

\noindent Also, the orbital period is:

t$_o$ = d/v$_j$= 76 yr 

\noindent corresponding to an orbital radius

r$_o$ = v$_o$ t$_o$ / (2 $\pi$) = 64 AU 

\noindent which is uncomfortably large.

For a binary with two stars of equal mass M in circular orbits, the mass of
one of the two stars can be obtained as:

M = 2 v$_o$${^2}$ r$_o$/G = 180 M$_\odot$

which clearly is unrealistic. No matter what tweaks are made to the
above numbers the resulting mass is far too large. There are HH jets
with a wiggling that is convincingly interpreted as the result of binary
motion, but the wiggles here are more irregular and are spread over
longer distances along the jet.  This may be due to
unknown density perturbations that accompany the velocity differences
along the jet, and when material 
runs into itself new knots will come
and go.




Another possible explanation for the difference in morphology of the E/C
jet pair could be that, while source Ea is located at the edge of the
cloud core and launching jet~E unhindered, jet~C is burrowing through
the cloud core. Shear might excite Kelvin-Helmholtz instabilities
along the cavity walls, and the jet could be slighly deflected by
these ripples in a quasi-periodic fashion.


\subsection{HH 24 Jet G}\label{subsec:jetG}

\begin{figure}
\centerline{\includegraphics[angle=0,width=8.3cm]{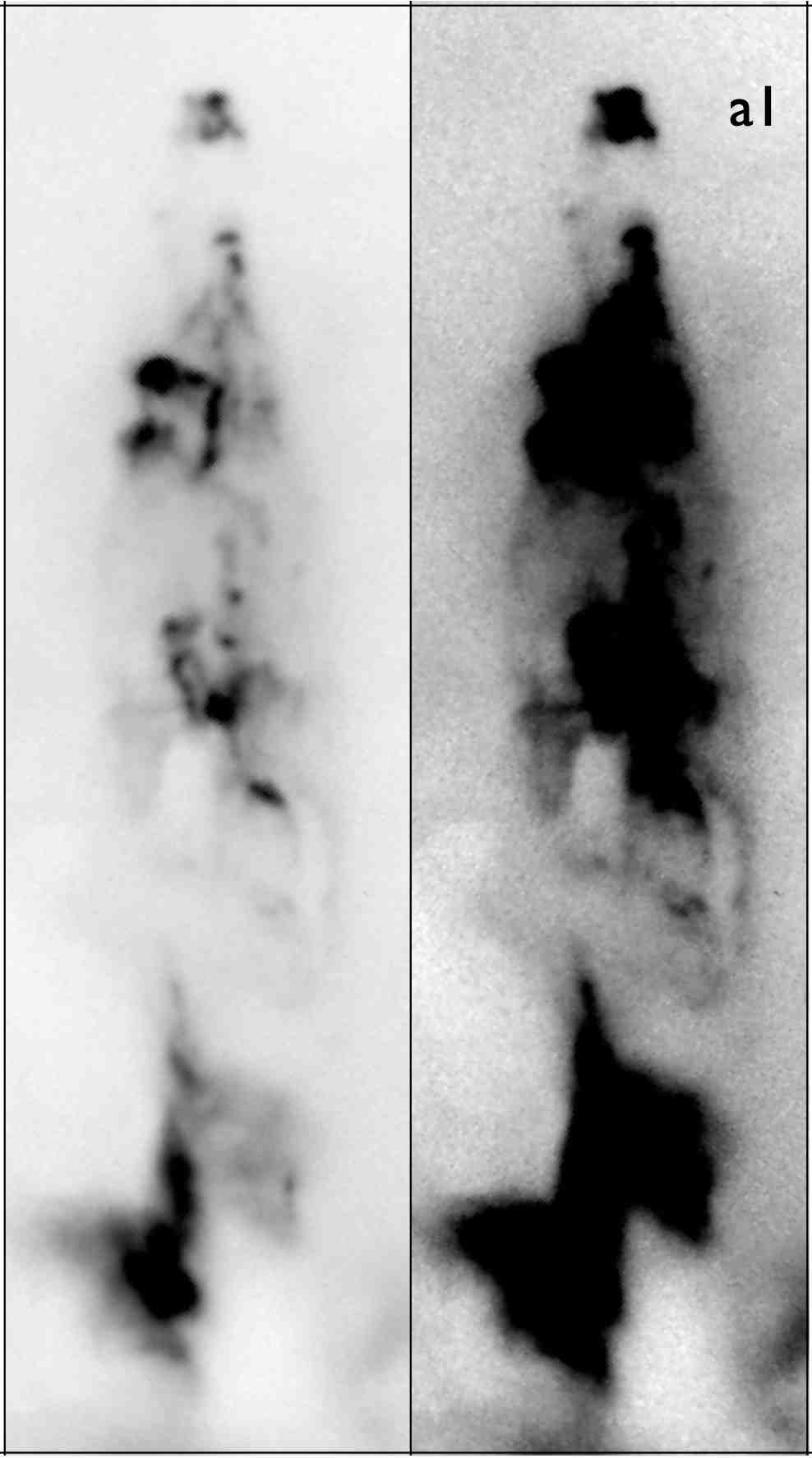}}
\caption{Jet~G in the optical in two different cuts, illustrating the
brighter interior and fainter exterior structures. Each figure is the
sum of deep H$\alpha$ and [\Sii] images obtained at the Subaru 8m
telescope. The vertical dimension is 84 arcsec
corresponding to 0.16~pc. The apex of the jet is denoted a1, and
more features are labeled in Figure~\ref{G-hst}. 
\label{G-subaru}}
\end{figure}

\begin{figure}
\centerline{\includegraphics[angle=0,width=5cm]{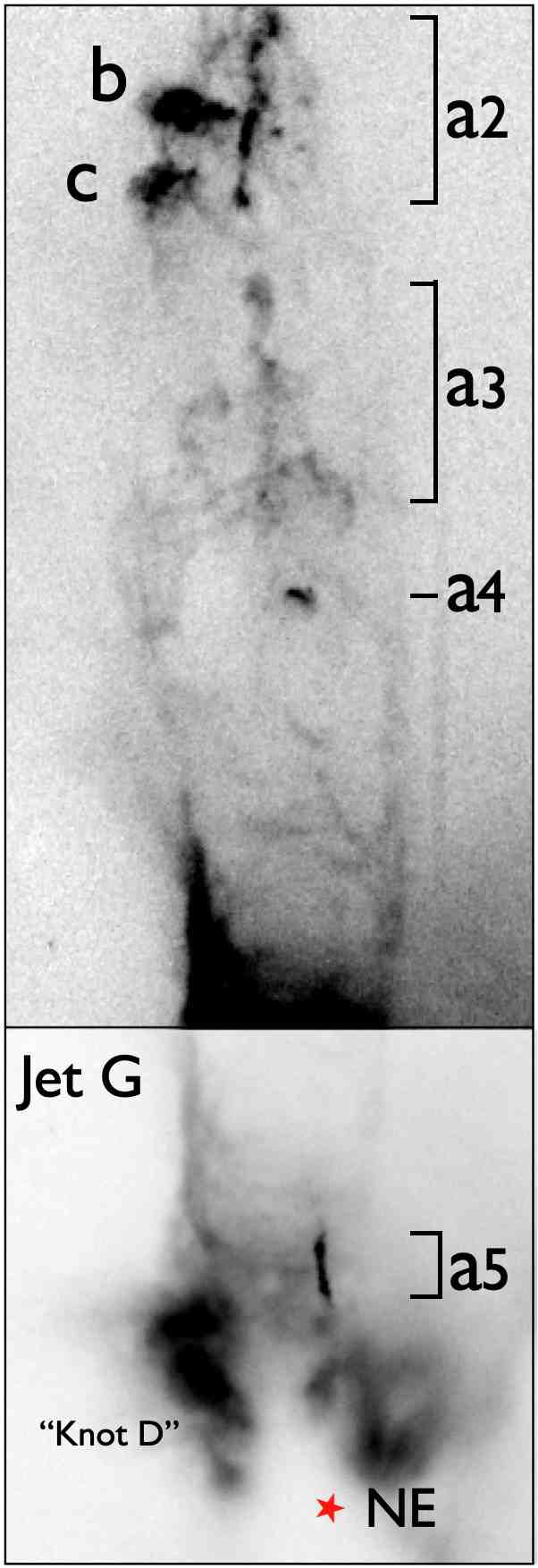}}
\caption{HST near-infrared [\Feii] image of the HH~24 jet~G. The tip of the 
jet (a1) falls outside the WFC3 field, but is seen as the top of the
jet in Figure~\ref{G-subaru}. This complex outflow consists of a
central collimated jet (a1-a5) and two bright bow shocks (b and c)
all wrapped within a wide outflow cavity whose sides are outlined in
[\Feii] emission.  Knot~D was originally so designated by
Herbig (1974), but has turned out to be an H$\alpha$-bright reflection
nebula illuminated by the embedded VLA source NE. The bright lower
part of the figure is shown with a different cut.  The height of the
figure is about 70\arcsec.
\label{G-hst}}
\end{figure}

Jet~G has an unusual morphology. Figure~\ref{G-subaru} shows two cuts
of a deep image from the Subaru 8m telescope, which reveals four main
features of the jet, (1) a central axis with fragments of a long
collimated flow which we denote Ga1-a5 (see
Figure~\ref{G-hst}), (2) an envelope surrounding the entire flow, (3)
several knots that are off-center from the main axis, in particular
the pair of knots labeled Gb and Gc in Figure~\ref{G-hst}, which shows
a 1.644~$\mu$m [\Feii] HST image, and (4) a large bright and diffuse
S-shaped region at the base of the jet, which corresponds to Herbig's
knot D.

Knot~D was observed spectroscopically by Jones et al. (1987) who found
that it is mainly a reflected continuum with H$\alpha$ and H$\beta$ in
emission. Polarimetry by Strom et al. (1974a,b), Schmidt \& Miller
(1979), and Scarrott (1987) suggested that SSV~63 is a likely source
of the reflected light (see Section~\ref{subsec:reflection}), but
Jones et al. (1987) argued that another embedded source should exist
on the axis of the G flow. Their proposed position is only 3 arcsec
from the NE radio continuum source found later by Reipurth et
al. (2002), and lying on the axis of the G outflow (see
Figure~\ref{G-hst}).

The linear chain of knots denoted a1-a5 in Figure~\ref{G-hst} includes
a fragment (a5) of a jet near the source NE. While this appearance is
similar to many other ill-defined jets, an unusual feature is the
envelope that surrounds the jet, seen well in
Figure~\ref{G-subaru}. The distance from the tip of the jet to source
NE is 75~arcsec, corresponding to 30,000~AU in projection. The width
of the envelope at its widest is about 14~arcsec, corresponding to
5600~AU. Near its base, much of this envelope near its base is
illuminated by light from source NE, and there appears to be several
rings or corrugations in its lower part.  Presumably this is an
outflow cavity originating from source NE.  The two brightest knots in
Jet~G are located off the axis of the Ga knot chain.

Unique among the HH~24 jets, jet~G has a bright component
of H$_2$ emission, see Figure~\ref{opt-ir-mosaic}. The apex of jet~G,
labeled a1, is dominant in H$_2$, and closer to the source, around a3,
prominent wings of H$_2$ indicate the presence of low-velocity
shocks.

Jones et al. (1987) obtained long-slit spectroscopy of the central a1 - a5
knots, and found very high blueshifted heliocentric velocities of -130
to -140 \kms. We have obtained spectra of the off-axis Gb-knot, and
find a low velocity of $\sim$0 \kms. Our proper
motion measurements of the a-knots indicate motions of 100-200 \kms,
but the b and c off-axis-knots are stationary within the errors. They
are both very bright in [\Sii], indicating that they are
low-excitation shocks. They seem to be oriented towards the north-east,
and when tracing a line backwards one finds the near-IR YSO IRS~1 (see
Section~\ref{subsec:Ha1-4}). However, the lack of measurable proper motions
makes it impossible to establish a possible association with this source.



\subsection{Other Jets}\label{subsec:otherjets}

In addition to the above major shocked outflows, there are three additional
rather inconspicuous flows, J, X, and L. The two first are discussed
in Section~\ref{subsec:association} and the third in
Section~\ref{sec:hh24mms}.


\subsection{Parsec-scale Outflows}\label{subsec:giant-flows}



Many well-collimated HH jets are associated with distant bow shocks that
can be more than one parsec from their driving sources. Such giant jets
provide fossil records of the mass loss and accretion histories of their
driving sources (Reipurth, Bally, and Devine 1997). The formation of
these giant terminal working surfaces is discussed in
Section~\ref{sec:discussion}.

The HH 24 jet complex is not an exception to this. Several distant
shocks, found by Herbig (1974, HH~19,20,21), Strom et al. (1986,
HH~37), and Reipurth \& Graham (1988, HH~70), are located to the north
of the HH 24 complex (Figure~\ref{overview}). In their study of the HH
objects in this region, Jones et al. (1987) recognized the probable
relation of these objects to the HH~24 jets. Basic properties of the
various distant components of these giant flows, known as well as new,
are given in Table~\ref{table:giant-bowshocks}, and are discussed in
more detail below. 

\subsubsection{HH 19}

HH~19 is a bright and highly structured object, with the appearance of
a large fractured bow shock (Figure~\ref{overview}).  Between HH~19
and SSV~63 is a faint group of knots, labeled HH~19-O by Eisloeffel \&
Mundt (1997). The faint but well collimated jet~J points within a few
degrees towards HH~19. This jet is launched by source Wb, which is
therefore also the likely driving source of HH~19. This identification was
supported by the proper motion measurements of Jones et al. (1987),
who found tangential velocity vectors of the HH~19 complex of 60-90
\kms\ directed away, to within a few degrees, from the SSV~63 multiple
system.  The distance of HH~19 from source Wb is $\sim$400~arcsec,
corresponding to a projected distance of 0.77~pc.

\begin{figure}
\centerline{\includegraphics[angle=0,width=8.3cm]{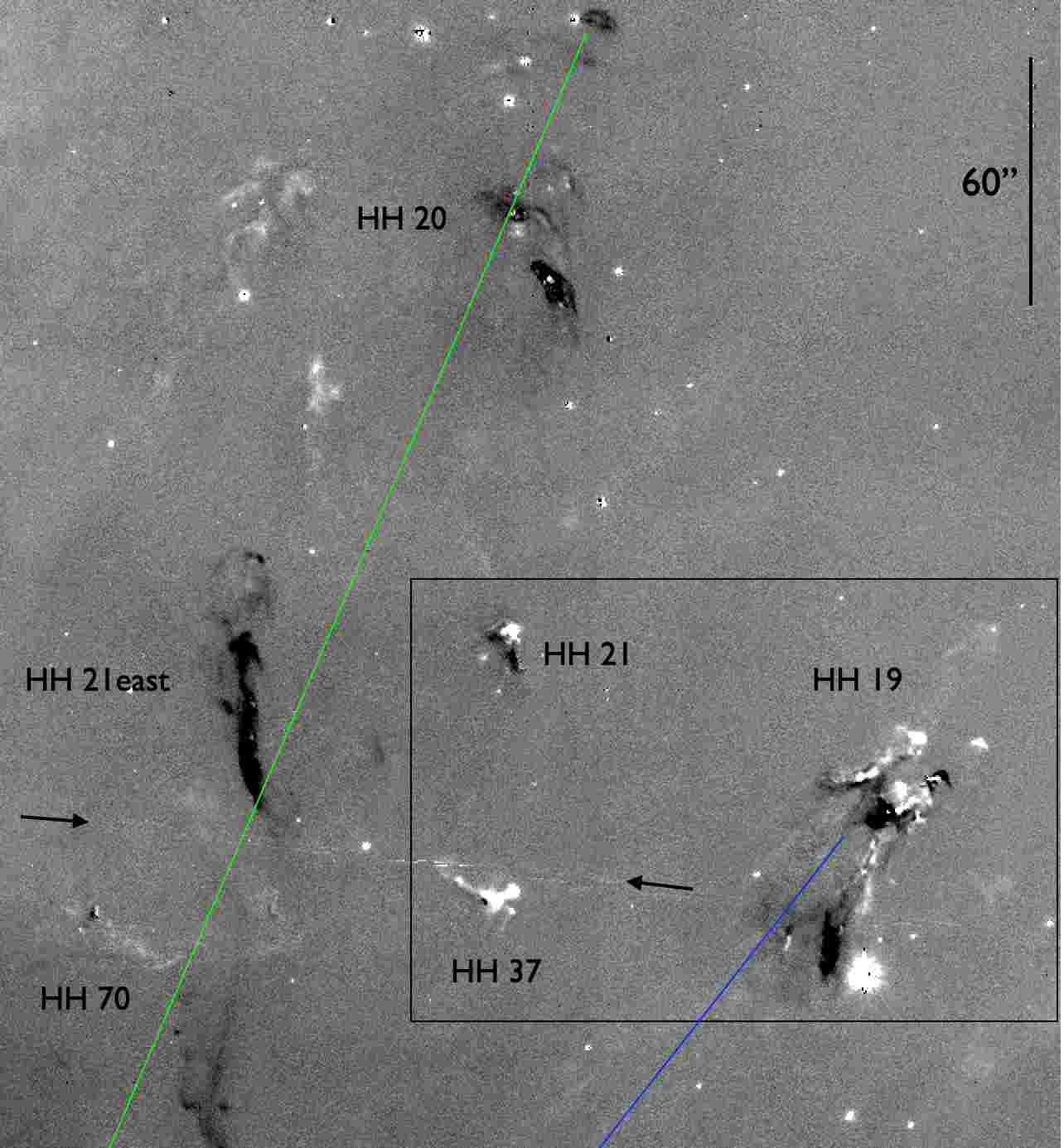}}
\caption{The two fractured giant bow shocks driven by the two jets C
(axis shown in green) and J (axis blue) shown in an H$\alpha$--[\Sii]
image obtained with the Subaru telescope. Black is H$\alpha$-strong
and white is [\Sii]-strong. The two arrows mark defects in the
CCD. The box indicates the area shown in Figure~\ref{hh19-21-37}. 
North is up and east is left.
\label{distant-bowshocks}}
\end{figure}

\begin{figure}
\centerline{\includegraphics[angle=0,width=8.3cm]{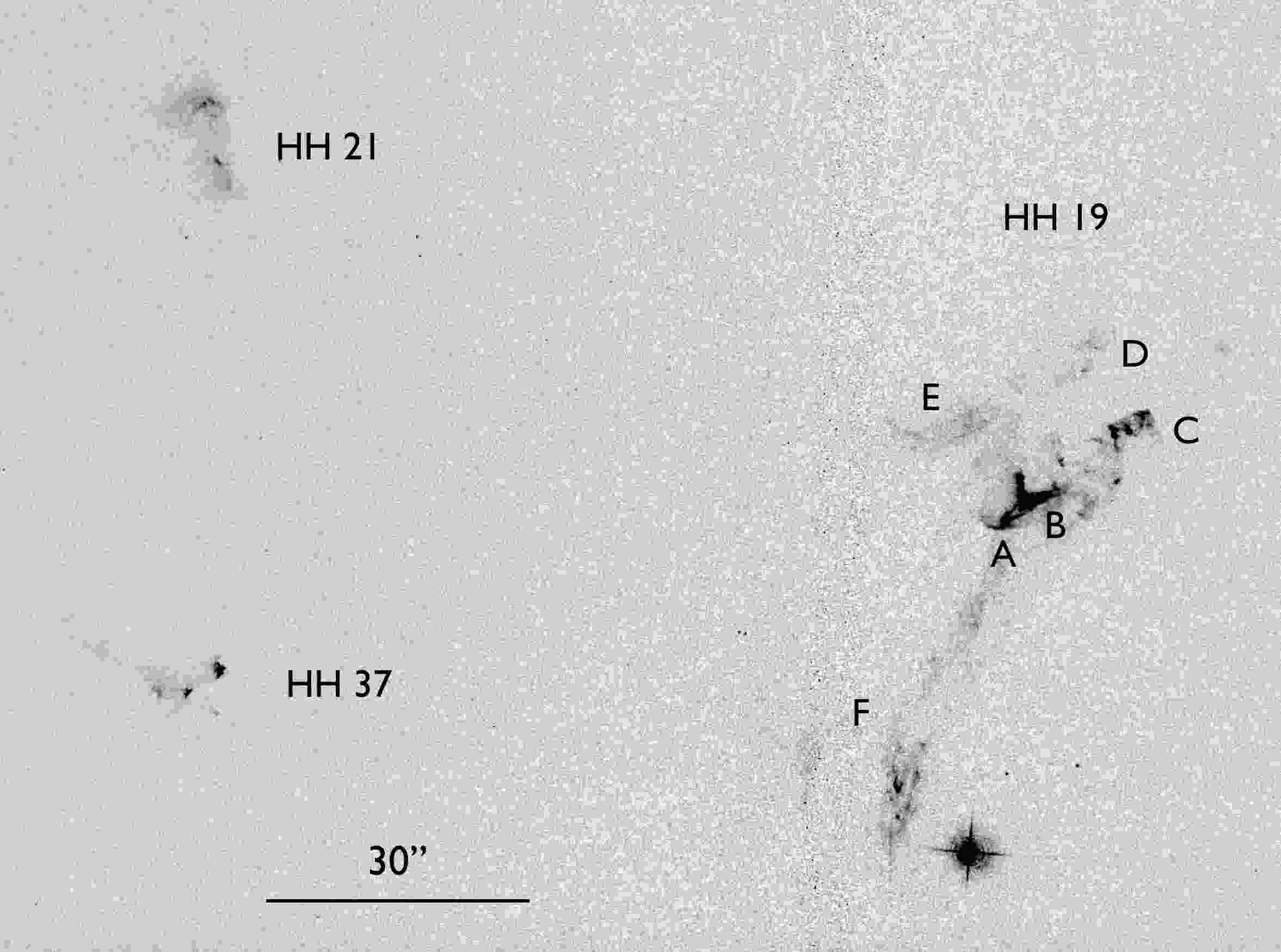}}
\caption{Distant bow shocks HH 19, HH 21, and HH 37 from an H$\alpha$ image obtained with the HST as a parallel ACS observation. The annotation of HH~19 is from Mundt et al. (1984). North is up and east is left.  
\label{hh19-21-37}}
\end{figure}

Figure~\ref{distant-bowshocks} shows an H$\alpha$--[\Sii] difference
image including HH~19. While some HH working surfaces have clean
morphologies, with H$\alpha$-strong bow shocks and  weaker [\Sii]
jet shocks (e.g., HH~34, Reipurth \& Heathcote 1992), HH~19's highly
fractured structure does not show such simple patterns. The complexity
of the individual shocks in HH~19 is further illustrated in
Figure~\ref{hh19-21-37}, which shows an H$\alpha$ image that was
fortuitously obtained in parallel-mode with ACS while the HH~24 jets
were imaged with WFC3. Some features appear to have forward facing
bow-shapes, while others are backward facing. These latter structures
tend to show little or no proper motion while the forward facing
shocks exhibit the fastest motions. It seems that some ejecta associated
with jet~J are overrunning either stationary, or slowly moving, 
dense globules of material.

Our measurements indicate a mean tangential velocity of HH~19 around
100~\kms\, but with large internal variations, and directed straight
away from the SSV~63 core along the axis of jet~J. 
Assuming that this velocity is representative of the
motion of HH~19 since it was launched, it indicates an age of
$\sim$8,000~yr. 

Our spectra show that HH~19 is blueshifted, as already noted by Jones
et al. (1987), with velocities ranging from -100 to +29 \kms\ and
with a peak around -15 to -20 \kms\ in the Orion reference frame.  
This suggests that the flow is moving close to the plane of the sky,
at an angle of roughly 10$^\circ$.

\begin{figure*}[h!]
\centerline{\includegraphics[angle=0,width=18cm]{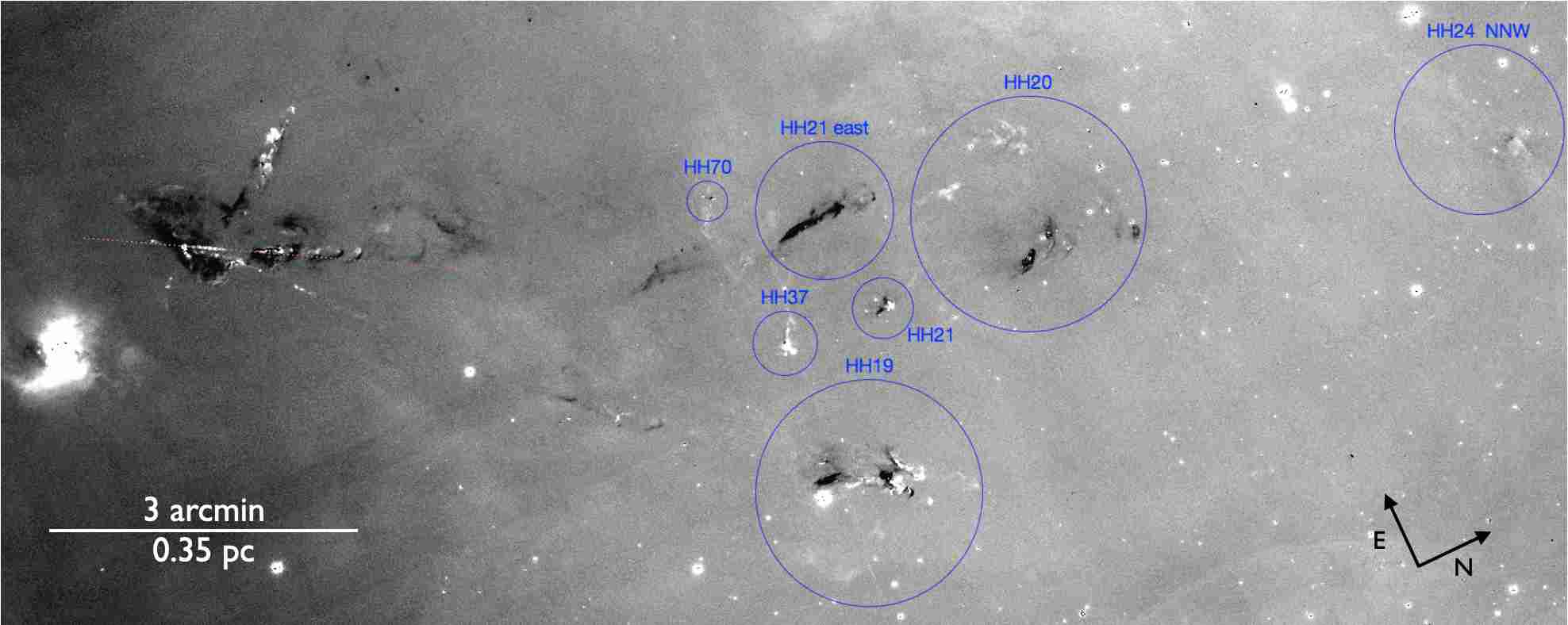}}
\caption{A large complex of shocks is found north of HH~24. The group HH~20, 21, 37, 70 forms a giant fractured bow shock. Further north, a distant faint shock is detected, here labeled NNW. The projected distance from source EA to the most distant shock HH~24~NNW is 1.45~pc.          
These shocks are associated with the HH~24C jet that is
pointing towards them. HH~19 is a giant bow shock associated with the
HH~24J jet.  Figure based on H$\alpha$ (black) and [\Sii] (white)
Subaru images.
\label{giantshocks-N}}
\end{figure*}

\begin{figure*} 
\centerline{\includegraphics[angle=0,width=18cm]{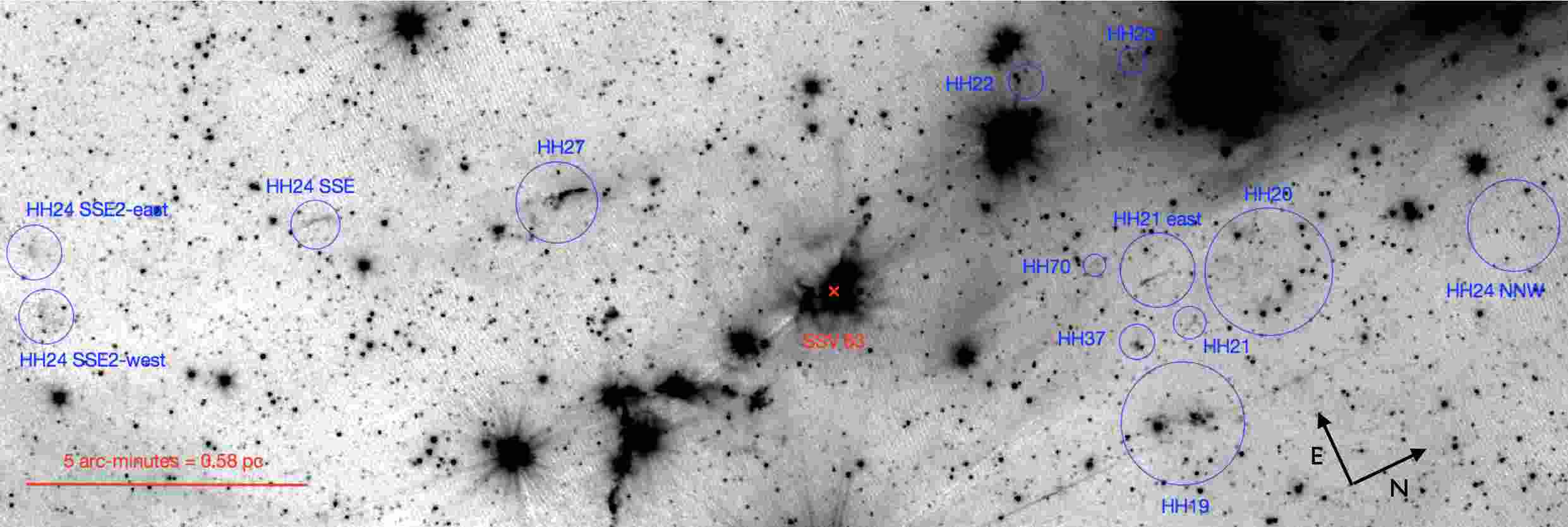}}
\caption{The HH~27 shock is a counterpart to the HH~19 terminal bow
shock for the HH~24J jet. Further south, a filamentary shock, here
labeled HH~24~SSE, is located. Even further south, two faint
nebulosities, labeled HH~24 SSE2e and SSE2w, are found. The figure is a
composite from Spitzer IRAC1 and IRAC2 images.  
\label{giantshocks-S}} 
\end{figure*}

\subsubsection{HH 27}

On the opposite side of source~Wb, along the axis of the J-jet and at
a distance of ~$\sim$320~arcsec (0.62~pc), is the bright compact HH
object HH~27 (Figure~\ref{overview}). Based on this location, it
appears highly likely that HH~19 and HH~27 form opposite working
surfaces in a giant outflow with a combined projected extent of
$\sim$1.4~pc. Whereas HH~19 is blueshifted, HH~27 is redshifted,
showing a broad H$\alpha$ line profile with a peak velocity in the
Orion Nebula rest frame of about +32~\kms. The 0.15~pc difference in
extent of the blue- and red-shifted lobes may be related to HH~19
moving out of the L1630 cloud, whereas HH~27 may still be closely
associated with the cloud. This is supported by Jones et al. (1987)
who found HH~27 to be the highest extinction object in the 
region, with an A$_v$$\sim$3, based on measurements of Balmer decrements.

Despite the presence of bright, compact knots in HH~27, the absence of
nearby reference stars means that only an upper bound on its
tangential velocity  of V$_{PM} <$60 \kms\ could be measured, a limit
consistent with the object moving into a cloud.

\subsubsection{Extensions of HH~24C}

The shocks in the HH~24C jet grow fainter and wider as they move to
the NNW of source Ea. Several working surfaces with H$\alpha$-bright
bow shocks sitting as shoulders on [\Sii]-rich jet shocks are evident
in Figure~\ref{definitions}. Beyond those, the flow appears as a very
faint and diffuse filigreed bubble of shocks reaching as far as
140~arcsec (55,000 AU = 0.27~pc in projection) from source Ea
(Figure~\ref{overview}). Such a structure may result from a wider
outflow interacting with the surface of the L1630 cloud.

\subsubsection{HH 20, 21, 37, 70, NNW}

Further downstream there is what appears to be a giant fractured bow
shock encompassing HH~20, 21, 37, and 70, see
Figure~\ref{overview}. Our spectra show that HH~20 is blueshifted,
with line profiles peaking at a velocity of about -120~\kms,
confirming the early work of Jones et al. (1987).  The most distant
shock in the HH~20 complex is $\sim$530~arcsec (1.02~pc in projection)
from source~Ea. We concur with Jones et al. (1987) and Eisloeffel \&
Mundt (1997) that these distant shocks are likely driven by
SSV~63. The tangential velocities of the components of the HH~20
complex 
are on average
around 130~\kms, indicating a dynamical age of 6800 yr, again assuming
a constant velocity over time. However, there is a large dispersion in
motion among the various features. For HH~21, 37 and 70 the motions
are so slow that no measurable proper motions could be determined. For
HH~20, tangential velocities are in the range $\sim$50-100~\kms . The
north-south oriented filament, HH~21~east shows coherent motion towards
the north with a speed larger than 100 \kms.  However, the
northern-most knot exhibits apparent motion towards
PA$\sim$-24~$\deg$.  This may be due to fading of one part of the
shock and brightening of another part towards the west.

We have obtained widefield images to the NNW and
SSE of HH~24 in search of further shocks, and have identified several
along the E/C jet axis. Figure~\ref{giantshocks-N} shows the sum of
our deep H$\alpha$ and [\Sii] images with SuprimeCam where we identify
yet another faint shock, dubbed HH24-NNW, along the E/C jet axis, at a
distance of 750~arcsec, or 1.46~pc in projection. The object is too diffuse
for proper motion to be measured.


While within 1\arcmin\ of source Ea, knots in the jets C and E exhibit
tangential motions of about 250 to 300 \kms , the various HH objects
located farther away from the SSV~63 core show a systematic decline of the
proper motions with increasing distance from the SSV~63 core.  This
behavior is similar to what is observed in other parsec-scale
protostellar outflows and indicates deceleration of the ejecta as they
interact with slower moving or stationary media.

\begin{figure}[t] 
\centerline{\includegraphics[angle=0,width=6cm]{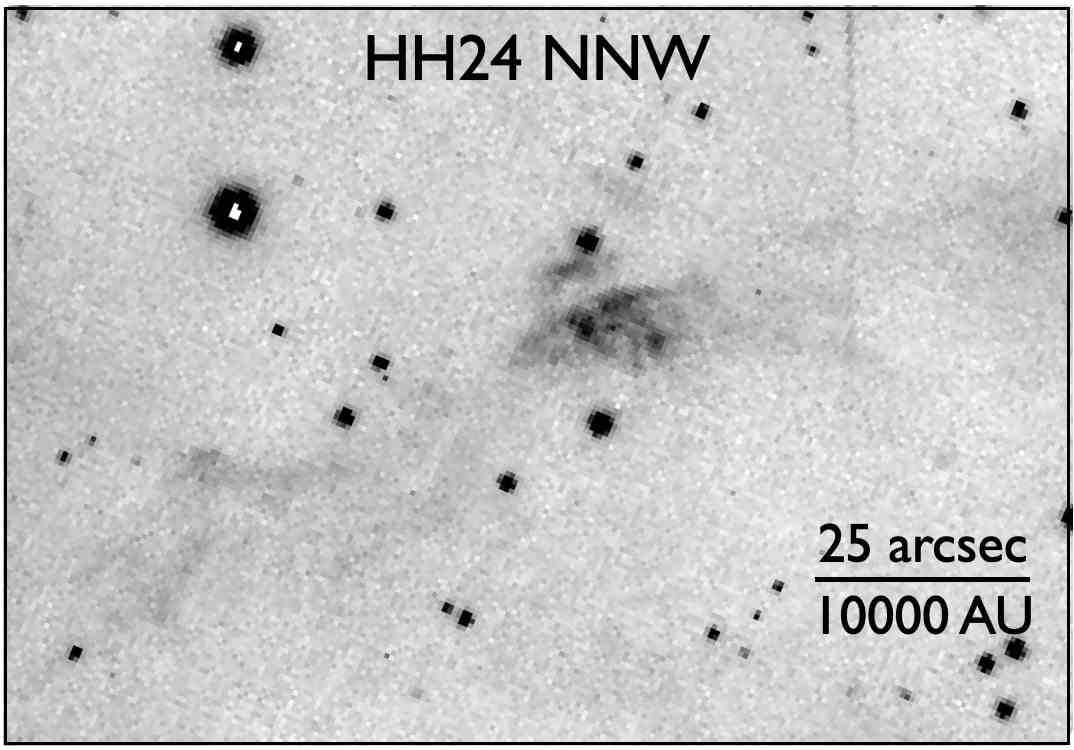}}
\centerline{\includegraphics[angle=0,width=6cm]{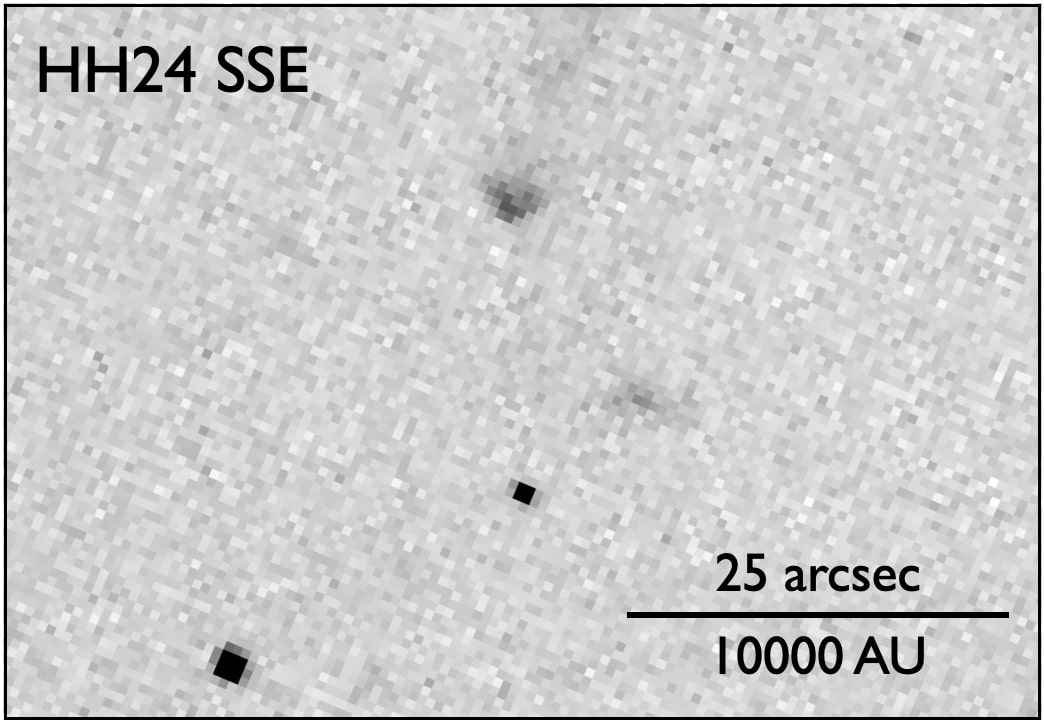}}
\caption{Optical images of distant shocks in the HH 24 complex.  
Top: HH~24~NNW, which is a low-excitation object, in a
[\Sii] image. Bottom: HH~24~SSE, which is a
high-excitation object, in H$\alpha$. North is up and east is
left.  
\label{NNW-SSE}} 
\end{figure}



\begin{deluxetable*}{lccccccc}
  

\tablecaption{Giant Bow Shocks$^a$\label{table:giant-bowshocks}}
\tablewidth{0pt}
\tablehead{
    \colhead{Shock}            &
    \colhead{$\alpha$$_{2000}$} & 
    \colhead{$\delta$$_{2000}$} &			   
    \colhead{Assoc. Jet} &
    \colhead{Source} &
    \colhead{Pos.Angle} &
    \colhead{Sep.[$''$]} &
    \colhead{Length [pc]$^b$} 
}
      
\startdata
HH 19       & 5:45:49.6 & -00:05:11  & Jet J &  Wb   & 317 & 398  & 0.77    \\
HH 20       & 5:45:55.6 & -00:02:47  & Jet C &  Ea   & 336 & 477  & 0.92     \\
HH 21       & 5:45:55.7 & -00:04:27  & Jet C &  Ea   & 330 & 387  & 0.75     \\
HH 21east   & 5:45:59.8 & -00:04:46  & Jet C &  Ea   & 338 & 343  & 0.67     \\
HH 27       & 5:46:22.9 & -00:13:44  & Jet J &  Wb   & 135 & 319  & 0.62     \\
HH 37       & 5:45:56.0 & -00:05:32  & Jet C &  Ea   & 325 & 330  & 0.64     \\
HH 70       & 5:46:02.3 & -00:05:36  & Jet C &  Ea   & 341 & 283  & 0.55     \\
HH 24 NNW   & 5:45:51.3 & +00:01:41  & Jet C &  Ea   & 340 & 750  & 1.45     \\
HH 24 SSE   & 5:46:28.6 & -00:17:53  & Jet E &  Ea   & 147 & 503  & 0.98    \\
HH 24 SSE2e & 5:46:35.3 & -00:22:47  & Jet E & Ea  & 152 & 863  & 1.67   \\
HH 24 SSE2w & 5:46:31.0 & -00:23:04  & Jet E & Ea  & 157 & 851  & 1.65   \\
\enddata
\tablecomments{a: All objects are very extended; coordinates
refer to  bright features or the geometric center of an object. All objects 
were measured on optical images except HH~24 SSE2e and HH~24 SSE2w, which were measured on Spitzer IRAC2 images. b: Projected length.}
\end{deluxetable*}



\subsubsection{HH 24 SSE, SSE2e, SSE2w}

In the southern part of the HH~24 complex we have discovered three
more distant knots, labeled SSE, SSE2e, and SSE2w. They are shown on
Figure~\ref{giantshocks-S}, which is a composite from Spitzer IRAC1
(3.6~$\mu$m) and IRAC2 (4.5~$\mu$m) images, where these distant shocks
are more pronounced. The projected distance of SSE from source Ea is
0.98~pc, and from our optical images we determine a tangential motion
of roughly 150 \kms . Assuming a constant velocity the age of this knot
is $\sim$8200~yr.  The projected distance of the SSE2 pair from source
Ea is 1.66~pc. Thus, the total extent of the HH~24 E/C flow is 3.1~pc,
making it among the largest HH flows known. Figure~\ref{NNW-SSE} shows
optical close-ups of the individual NNW and SSE shocks. The NNW shock
has a very large extent of $>$40,000~AU, and is likely the northern
terminal bow shock for the HH~24~E/C jet pair.  In contrast, the SSE
shock just consists of two knots, located well behind the two most
distant shocks, SSE1 and 2, which likely together form the southern
terminus of the E/C jet pair. We discuss how these multiple working
surfaces have been formed in Section~\ref{sec:discussion}.



\vspace{5mm}

\begin{deluxetable}{lll}
\tablecaption{Images used for Proper Motions of Giant Bow Shocks\label{table:images-pm-giant-bowshocks}}
\tablecolumns{3}
\tablewidth{0pt}
\tablehead{
   \colhead{Date} &
   \colhead{MJD }  &
   \colhead{Instrument \& Filter} 
  }
\startdata
   18 December 2001   & 52261  &  CTIO 4m Mosaic \Ha , [\Sii ]                 \\
   06 January 2006	& 53741  &  Subaru Suprimecam \Ha  , [\Sii ]            \\ 
   18 February 2014  & 56706  &  HST WFC3/ACS  [\Feii ], \Ha\ , [\Sii ]    \\
   03 February 2016  & 57421  &  HST WFC3/ACS  [\Feii ], \Ha\              \\
   01 December 2021 & 59549 &  APO ARCTIC [\Sii ]      \\
  \enddata
\end{deluxetable}



\begin{deluxetable*}{lcccl}
\tablecaption{Parsec-Scale Components \& Proper Motions\label{table:pm-giant-bowshocks}}
\tablecolumns{5}

\tablewidth{0pt}
\tablehead{
  \colhead{R.A.  \& Dec.} &
  \colhead{PM$^a$ }  &
  \colhead{V$^a$}    & 
  \colhead{PA}     &
  \colhead{Comments} 
  \\
  \colhead{ (J2000) } &
  \colhead{ (mas $\rm yr^{-1}$)  }  &
  \colhead{ (\kms ) }    & 
  \colhead{ (deg.) }    &
  \colhead{  }
}
\startdata
    5:46:35.2     -0:22:43  &   -         &    -       &    -     &    HH 24 SSE2-east.  South terminus \\
    5:46:30.6     -0:23:06  &   -         &    -       &    -     &    HH 24 SSE2-west.  South terminus \\
    5:46:28.8     -0:18:05  &  61       & 156      &  115  &    HH 24 SSE1   \\
    5:46:22.7     -0:13:43  & $<$30  &  $<$60 &  -      &    HH 27  \\
    5:45:56.2     -0:07:18  & 83        &  157     &  -43   &  jet J; faint bow  \\ 
    5:45:49.6     -0:05:11  & 46        &    87     &  -23   &  jet J;  HH 19 S \\
    5:45:49.1     -0:04:53  & 54        &  102     &   -25  &  jet J;  HH 19 N.  Northwest terminus \\
    5:45:69.0     -0:04:32  & 57        &   108    &  -24   &  HH 21east \\
    5:45:59.8     -0:05:02  & 49        &     93    &  -3     &  HH 21east  E1 (\Ha ) \\
    5:45:59.8     -0:04:55  &   74      &    140   &  -7      &  HH 21east E2 (\Ha ) \\
    5:45:59.8     -0:04:31  &  131     &     248  &  -5      &  HH 21east E3 (\Ha ) \\
    5:45:69.0     -0:04:28  &  132     &     250   & -11    &  HH 21east N-tip (\Ha ) \\
    5:45:55.7     -0:04:26  & $<$30  &  $<$60 &  -       &  HH 21  \\
    5:45:58.5     -0:03:22  & 55        &  104     &     9    & HH 20 S  \\
    5:45:55.0     -0:03:02  & 92        &   175    &  -16    & HH 20 NW1 \\
    5:45:55.6     -0:02:47  & 59        &   112     &  -22   & HH 20 NW2 \\
    5:45:54.2     -0:02:01  & 73        &   139    &     0    & HH 20 N \\
    5:45:51.1     +0:01:42 & $<$30  &  $<$60 &  -        & HH24 NNW.   North terminus 
   \enddata
\tablecomments{a: no motion detected is marked as --}
\end{deluxetable*}


\subsubsection{Proper Motions of Distant Bow Shocks}\label{subsubsec:parsec-PM} 

We have three epochs of groundbased optical images spanning from 2001
to 2021 that cover parts of these parsec-scale shocks surrounding the HH~24
complex (see Table~\ref{table:images-pm-giant-bowshocks}).  Images
obtained with the Blanco 4-meter telescope at CTIO in 2001, the Subaru
8-meter telescope in 2006, and the Apache Point Observatory (APO)
3.5-meter in 2021 were used for proper motion measurements of these
distant HH objects.  The time interval between the 2001 and 2021
images was 19.95 years

Table~\ref{table:pm-giant-bowshocks} lists the positions and proper
motions of features measured on the 2001 Blanco 4m, the 2006 Subaru, and
2021 APO images.  At a distance of 400~pc, a displacement of 1\arcsec\
in a time interval of 19.95 years corresponds to a speed of 93.5 \kms
.  The uncertainties of the tangential velocities vary from about 20
to as much as 60 \kms\ owing to the diffuse structure of some of the
features, residual distortions in the images, and the lack of close-by
field stars to use for image registration.

 \begin{figure}
\centerline{\includegraphics[angle=0,width=8.3cm]{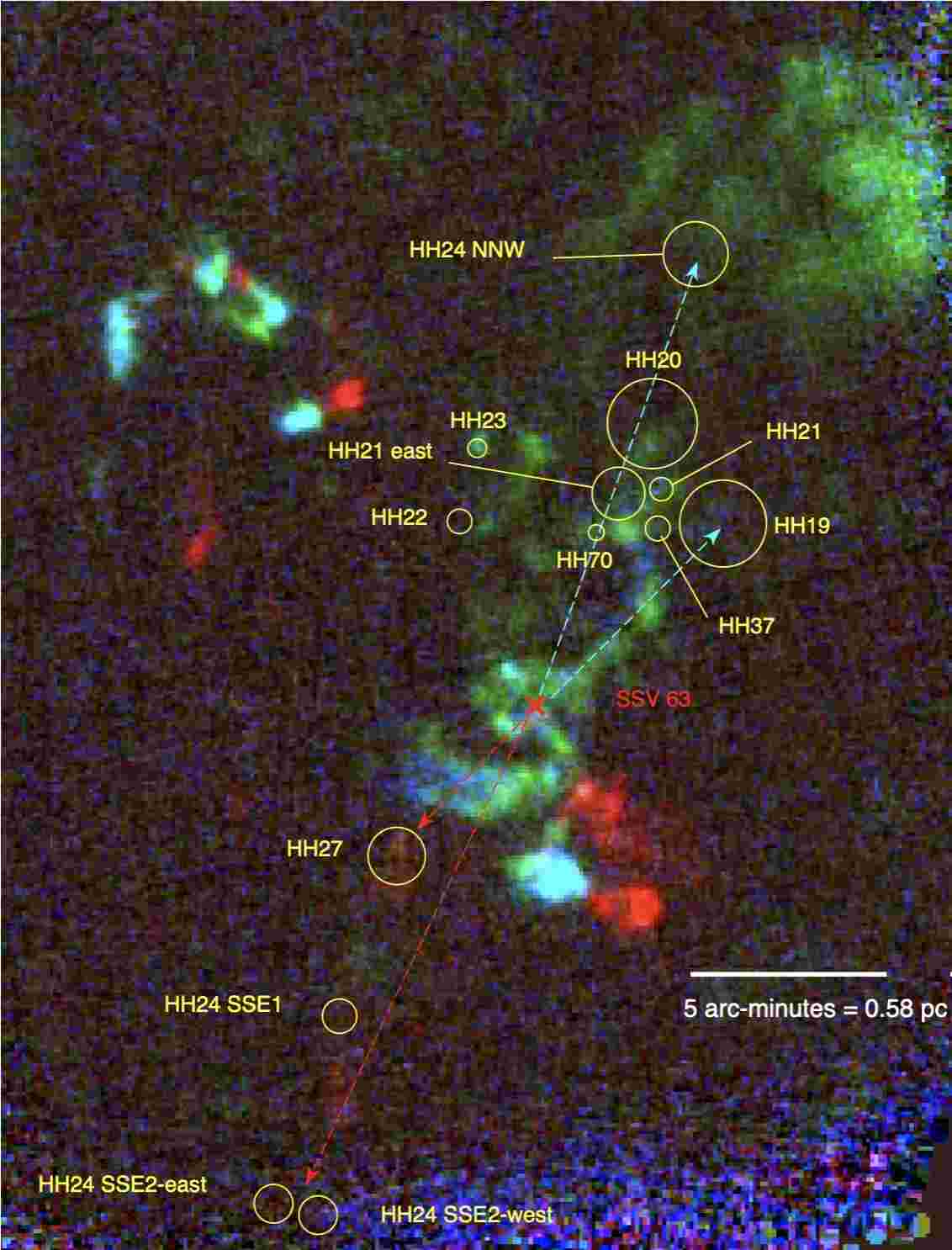}}
\caption{Three-color image showing `high-velocity' J=3-2 CO emission
associated with the parsec-scale outflows from the SSV~63 cloud core
and the HH~24 jets.  V$_{lsr}$ = 0 to 5 \kms\ is shown in blue; V$_{lsr}$
= 5 to 7.5 \kms\ is shown in green; V$_{lsr}$= 15 to 20 \kms\ is shown
in red.  The various HH objects that may be associated with the
extended outflows from SSV~63 are marked.  Dashed blue lines show the
blueshifted HH components associated with jets C and J.  Dashed red lines
show redshifted components associated with their counterflows. 
North is up and east is left. Data from Stanke et al. (2022).
\label{ALCOHOLS}}
 \end {figure}

\subsubsection{Parsec-scale CO Outflows}\label{subsubsec:parsec-CO}

Stanke et al. (2022) have mapped the entire Orion B molecular cloud in
the J=3-2 CO transition at 346 GHz with the APEX telescope (the
ALCOHOLS survey).  The beam size of this survey is $\sim$19\arcsec .
Figure~\ref{ALCOHOLS} shows `high-velocity' CO emission in the
vicinity of the SSV~63 cloud core.  Towards NNW, there is a low-radial
velocity counterpart to jet~J, also blueshifted as the HH
objects.  We find that a clumpy, low velocity bubble appears to
surround the various distant HH objects likely powered by jet~C.
Faint, redshifted emission is associated with the counterflows.
The impact of the SSV~63 outflows on the Orion B cloud has been very
significant, and not only in the immediate vicinity of the sources,
where cavities have been blown out (Figure~\ref{pressrelease}).  A
detailed analysis of these giant molecular outflows is, however,
beyond the scope of this paper.

A number of smaller, and presumably younger bipolar outflows are also
seen in this part of the Orion B cloud.












\vspace{0.3cm}

\section{THE CLASS 0 SOURCE HH 24 MMS}\label{sec:hh24mms}

Forty~arcsec south of the SSV~63 complex lies a very bright submm source,
HH~24~MMS, discovered at 1300~$\mu$m by Chini et al. (1993).
Bontemps et al. (1995) and Chandler et al. (1995) detected a VLA
source towards HH~24~MMS at 3.6~cm and 7~mm, respectively, both in C/D
configuration. Ward-Thompson et al. (1995) obtained an improved
position at 350~$\mu$m, showing that the VLA source is coincident with
the submm source, and identified it as a deeply embedded Class~0
source. Reipurth et al. (2002b) detected the source at 3.6~cm with the
VLA-A and provided a more accurate position for HH~24~MMS.  Two
additional nearby faint sources were detected with high-resolution VLA-C/D
observations at 6.9~mm by Kang et al. (2008).


\begin{figure}
\centerline{\includegraphics[angle=0,width=5cm]{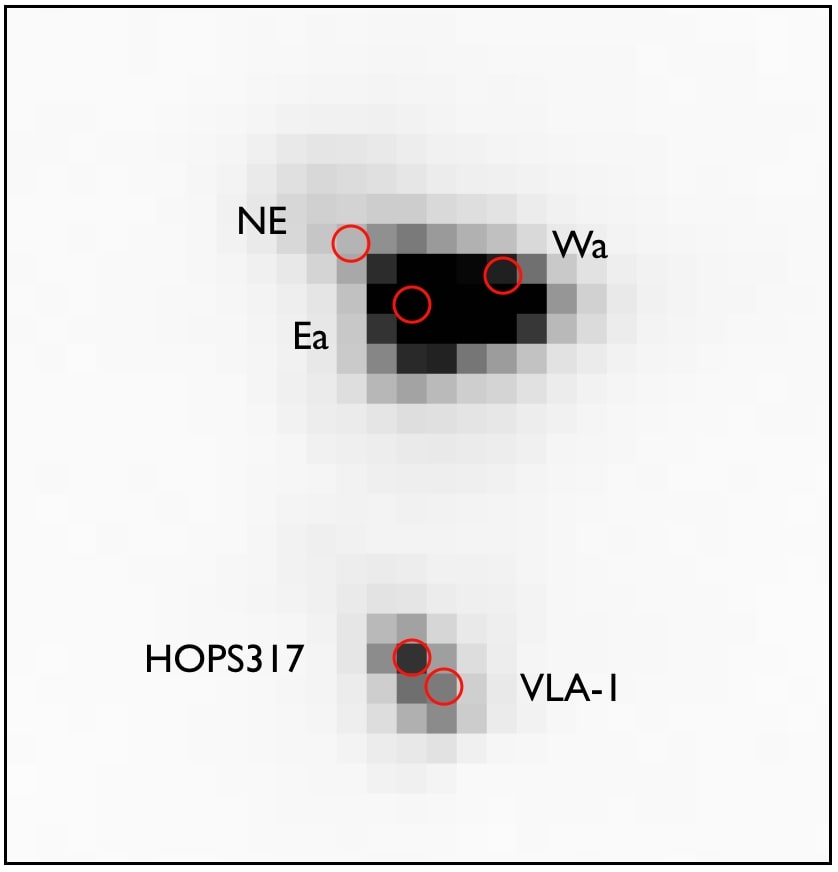}}
\caption{Herschel 70~$\mu$m image of the HH~24 source SSV~63 and
HH~24~MMS. At 70~$\mu$m the two sources HOPS~317 and MMS VLA~1 are just
resolved, but at 160~$\mu$m the two sources are unresolved.  The width of the figure is 85\arcsec.
\label{herschel}} 
\end{figure}

Near HH~24~MMS, Furlan et al. (2016) identified on Herschel images a
cool source, HOPS~317, which was previously discovered with Spitzer
and identified as the near-infrared source 2MASS-J05460852--0010390.
They concluded that it is a Class~0 source with a total luminosity of
10.6~L$_\odot$, a bolometric temperature of T$_{bol}$=47.5~K, and an
extinction A$_V$=41.5 mag.  However, examination of the Herschel
images show that HH~24~MMS and HOPS~317 are two separate sources,
$\sim$5~arcsec apart. While the two sources are just resolved at
70~$\mu$m, with HOPS~317 being the dominant source, at 160~$\mu$m they
are blended, see Figure~\ref{herschel}.  Hsieh et
al. (2021) observed the region with ALMA and in addition to separating
HOPS~317 and HH~24~MMS, they found a third source about 12\arcsec\ to
the northwest, which they dub HH24mms-NW (their Figure~3b). It could
be that HH~24~MMS forms a small multiple system, possibly
non-hierarchical, and if so unstable.


\begin{figure}
\centerline{\includegraphics[angle=0,width=7cm]{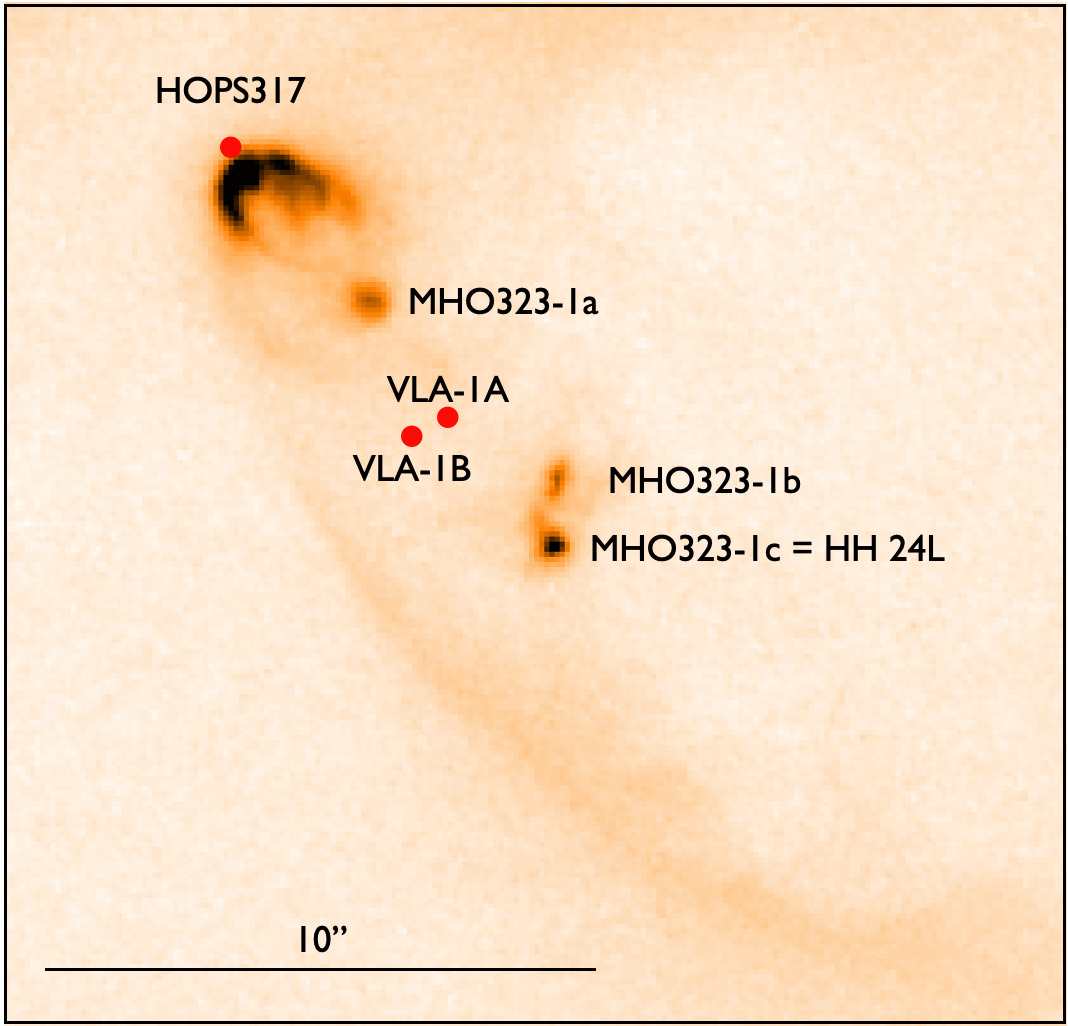}}
\caption{HST near-infrared image of the HH 24 MMS region obtained with WFC3 
in the [\Feii] 1.644 $\mu$m line.  The knot MHO~323-1c is an optically
visible HH object here labeled HH~24L. The protostar HOPS~317 is seen
to illuminate an outflow cavity which contains several objects in the
HH~24L flow (see Figure~\ref{definitions}). It is evident that the VLA
source(s), associated with HH~24~MMS, and HOPS~317 are separate
sources. The labels VLA-1A and VLA-1B refer to the positions marked in
Figure~\ref{vla-hh24mms}.
\label{hh24mms-hst}}
\end{figure}

\begin{figure}
\centerline{\includegraphics[angle=0,width=8.3cm]{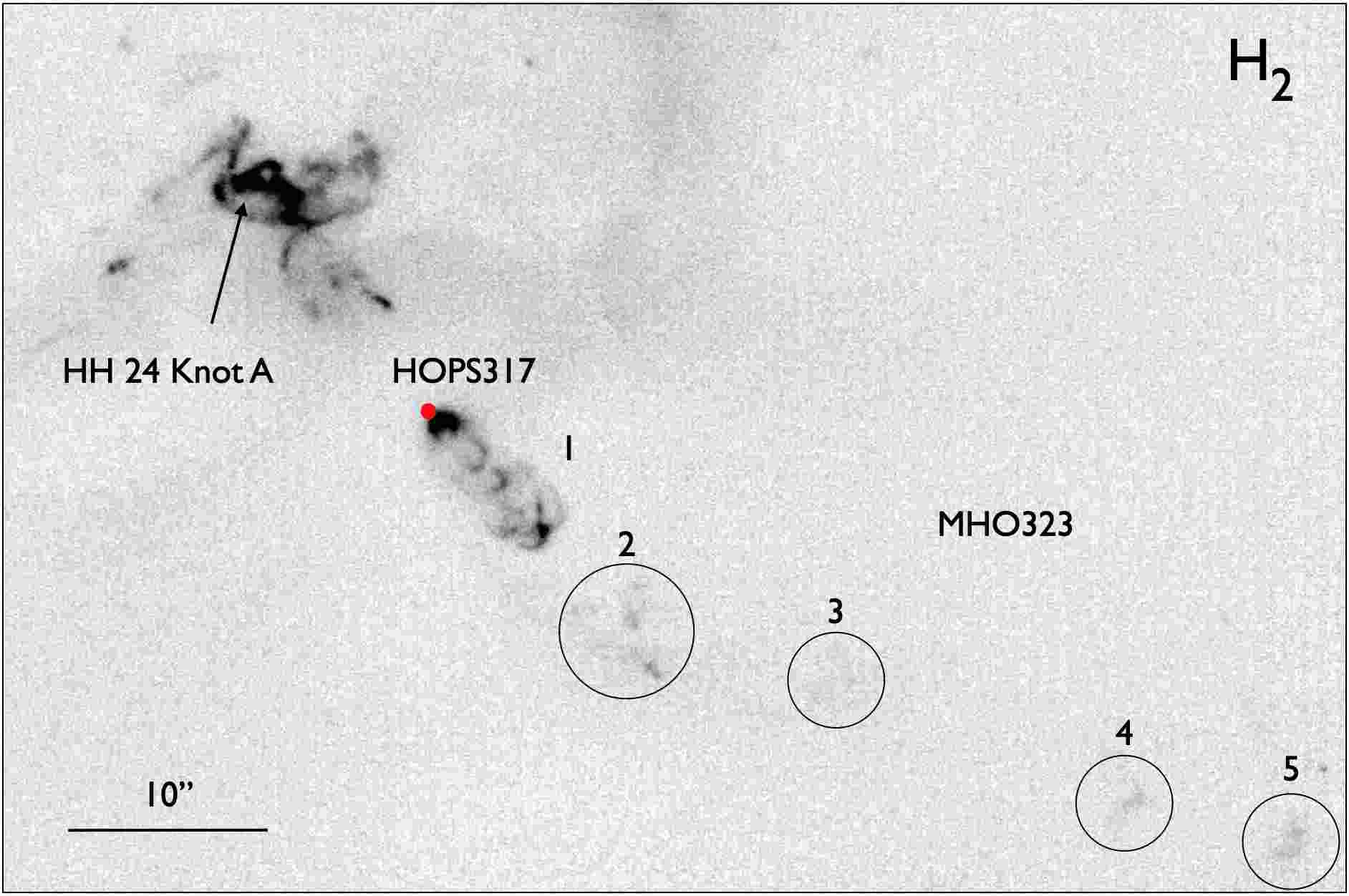}}
\caption{An H$_2$ image of the HH 24 MMS region obtained with NIRI on
the Gemini-N telescope. The HH~24L flow is very extended at near-IR
wavelengths, and further H$_2$ knots beyond knot~5 can be seen in
Figure~\ref{opt-ir-mosaic}.
\label{hh24mms-gemini}}
\end{figure}

Figure~\ref{hh24mms-hst} shows an infrared image obtained with 
WFC3 on HST through a [\Feii] 1.64~$\mu$m filter. The image shows an
illuminated outflow cavity with a bright apex opening out from HOPS~317
and several emission-line knots, the brightest of which is an
optically visible Herbig-Haro knot here designated HH~24L. 
The HH object is located 9\arcsec\ from HOPS~317,
which at a distance of 400~pc corresponds to a projected separation of
3600~AU. If the flow moves with a tangential velocity of about 100~\kms,
typical of HH flows, then it was ejected from this source $\sim$170~years ago.



Figure~\ref{hh24mms-gemini} is an image in the H$_2$ 1-0 S(1) line at
2.12~$\mu$m of the same region, which shows that the HH~24L flow from
HOPS~317 is much more pronounced in H$_2$ near the source, showing a
chain of small nested bow shocks and a series of more distant knots,
with additional knots apparent in Figure~\ref{opt-ir-mosaic}. The
molecular hydrogen flow emanating from HOPS~317 is known as MHO~323
and we here extend the notation to the four fainter outflow
components.\footnote{The MHO catalog is maintained by Dirk Froebrich
and is available at http://astro.kent.ac.uk/$\sim$df/MHCat and is
described in Davis et al. (2010)} It is noteworthy that the position
angles of the four outermost knots steadily increase with distance
from HOPS~317, suggesting precession of the source and indicative of a
close binary companion. Alternatively the flow may be deflected near
knot~2.

In the opposite direction, several H$_2$ knots are seen along the
principal flow axis, including a bow shaped H$_2$ structure that is
intertwined with the bright HH~24A knot located on the HH~24E flow
axis. As discussed in Section~\ref{subsec:HH24A}, it appears that
HH~24A represents, at least partially, the collision of a flow from
HOPS~317 with a stationary cloud.

\begin{figure}
\centerline{\includegraphics[angle=0,width=5cm]{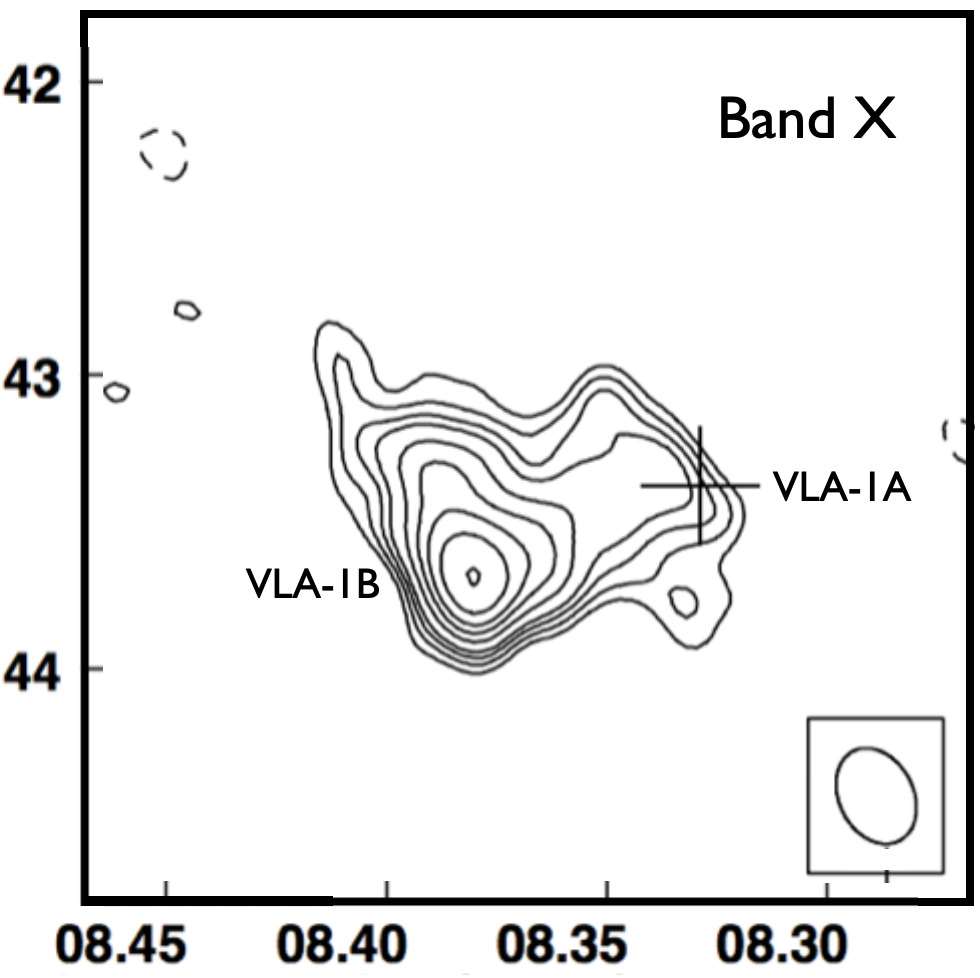}}
\caption{A VLA X-band image of the HH 24 MMS region from 2019. The position of the earlier epoch observation from 2000 of Reipurth et al. (2002b) is shown as a cross and labeled VLA-1A, while the current position is labeled VLA-1B. Right ascension is in seconds at 5h 46m, declination is in arcseconds at -00$^\circ$10'.  
\label{vla-hh24mms}}
\end{figure}

We have carried out the hitherto deepest and highest resolution
observations of the HH~24~MMS region with the Karl G. Jansky Very
Large Array at 10.0 GHz (X band) and 44.0 GHz (Q band), see Section~2
for details. Figure~\ref{vla-hh24mms} shows the Band-X map revealing
an extended highly structured nebula. The VLA position obtained by
Reipurth et al. (2002b), marked with a cross and labeled VLA-1A, is
0.8~arcsec from the peak of the new observations, labeled VLA-1B.


These observations can be understood in several ways:

{\em a:} The morphology seen in Figure~\ref{vla-hh24mms} is
reminiscent of a bow shock pointing back towards the SSV~63W
sources about 40 arcsec to the NNW. If the radio continuum emission
is due to shocks it is most likely free-free emission (e.g.,
Rodr\'{i}guez et al. 1999), in which case the shift of the peak
emission from 2000 to 2019 could be flickering of the shocks, as seen
in many HH objects (e.g., Raga et al. 2016a). However, if the shock
originates in SSV~63W, it would be a remarkable coincidence that it
happens to coincide with a bright embedded submm source.

{\em b:} Alternatively, the shocks may be local, driven by outflow
from the submm source. However, the extended emission has a spectral
index between 9.0 and 11.0~GHz of 2.9$\pm$1.2, which seems too steep
for free-free emission, in particular because for diffuse emission one
expects an optically-thin flat spectrum.  The index between 10 and
44~GHz has a value of 2.9$\pm$0.1, confirming the steepness
(Table~\ref{table:vla-parameters}).

{\em c:} The shift in position may be due to motion of the source. The
two positions are measured 18.75 years apart, indicating a projected
velocity of 15~\kms. Such a high velocity would require the source
to have been ejected from a small multiple system, but no other
sources are found near HH~24~MMS from the presumed direction of motion.

{\em d:} It is conceivable that HH~24~MMS is a binary with a
separation of 0.8 arcsec, corresponding to a projected separation of
320~AU. Such binaries are common among young stars. If so, the
components could be variable, as is sometimes seen in young radio
continuum sources (Anglada et al. 2018). In 2000 the western source
would have been the brighter of the two, while in 2019 the eastern
source was brighter.

{\em e:} Finally, HH~24~MMS may
be irradiating its near environment, and the extended 3.6~cm emission
could be dust heated by radiation from the submm source. Circumstellar
material close to the source could obscure the light and create a
lighthouse effect, and if the dust grains are small the heating and
cooling would be rapid and thus variable. The diffuse low-level
emission seen in Figure~\ref{vla-hh24mms} from the deep 2019
observations would seem to favor such an interpretation.

None of the above scenarios can be firmly rejected, although some are
more unlikely than others. We conclude that variable heating of dust
is the most likely explanation of the observations.

HOPS~317 and HH~24~MMS are separated by 5$\arcsec$,
corresponding to 2000~AU in projection. They are currently bound to
their host core, but as they accrete mass and the core shrinks it is
likely that they eventually become bound as a binary with a shrinking
orbit due to dynamical friction (e.g., Stahler 2010, Sadavoy \&
Stahler 2017). It is conceivable that, in the future when the cloud
disperses, HOPS~317 and HH~24~MMS will become bound to the SSV~63
multiple, thus forming a wide multiple system, not unlike the well
known wide high-order multiple system of Mizar and Alcor (Mamajek et
al. 2010).

\vspace{-0.5cm}


%
%



\vspace{0.3cm}

\section{KINEMATICS OF NEARBY LOW-MASS STARS AND BROWN DWARFS}\label{sec:halo}

As we will discuss in Section~\ref{sec:ALMA-I}, the stars in the
SSV~63 system have significant masses, between 0.9 and
2.1~M$_\odot$. With such massive members one would expect to find a
large number of low-mass objects if the initial mass function is close
to normal. However, the only potential low-mass objects are the
components S (Section~\ref{subsec:imaging}) and N
({Section~\ref{sec:ALMA-I}). In view of this disparity we have carried
out a deep slitless grism survey using GMOS on the Gemini-N telescope
to search for faint H$\alpha$ emission stars in the area of SSV~63,
for details see Section~\ref{sec:observations}. H$\alpha$ emission was
detected in only 5 stars, marked as H$\alpha$~1-5 in
Figure~\ref{kilder}, and with coordinates and near- and mid-infrared
photometry in Table~\ref{table:coordinates}.  Fang et al. (2009)
obtained low-resolution spectra of 4 of these sources, and our results
concur with theirs. We find that H$\alpha$~1 is a CTTS with spectral
type M3.5, H$\alpha$~2 is a strong-lined brown dwarf with spectral
type M7, H$\alpha$~3 is a CTTS with spectral type M4.5, H$\alpha$~4 is
a WTTS with spectral type M4.5, and H$\alpha$~5 is a WTTS borderline
brown dwarf with spectral type M5.5. Spectra of the two objects with
the latest spectral types are shown in
Figure~\ref{halpha5-2spectra}. All these five objects are optically faint
red objects (Table~\ref{table:coordinates}).\footnote{Most of the
sources discussed in this section, are listed as YSOs in Table~4 of
Megeath et al. (2012) with the following IDs: Wa=$\#$3168,
Ea=$\#$3167, NE=$\#$3169, H$\alpha$~1=$\#$3177, H$\alpha$~2=$\#$3176,
H$\alpha$~3=$\#$3175, IRS~1=$\#$3170, IRS~2=$\#$3171.  The sources Wb
and Eb are not listed, probably because they could not be resolved
from Wa and Ea, respectively.  H$\alpha$~4 and H$\alpha$~5 are also
not listed, probably because they are too faint for reliable
photometry with Spitzer (they are, however, detected by WISE, see
Table~\ref{table:coordinates}.)}

\begin{figure}
\centerline{\includegraphics[angle=0,width=8.3cm]{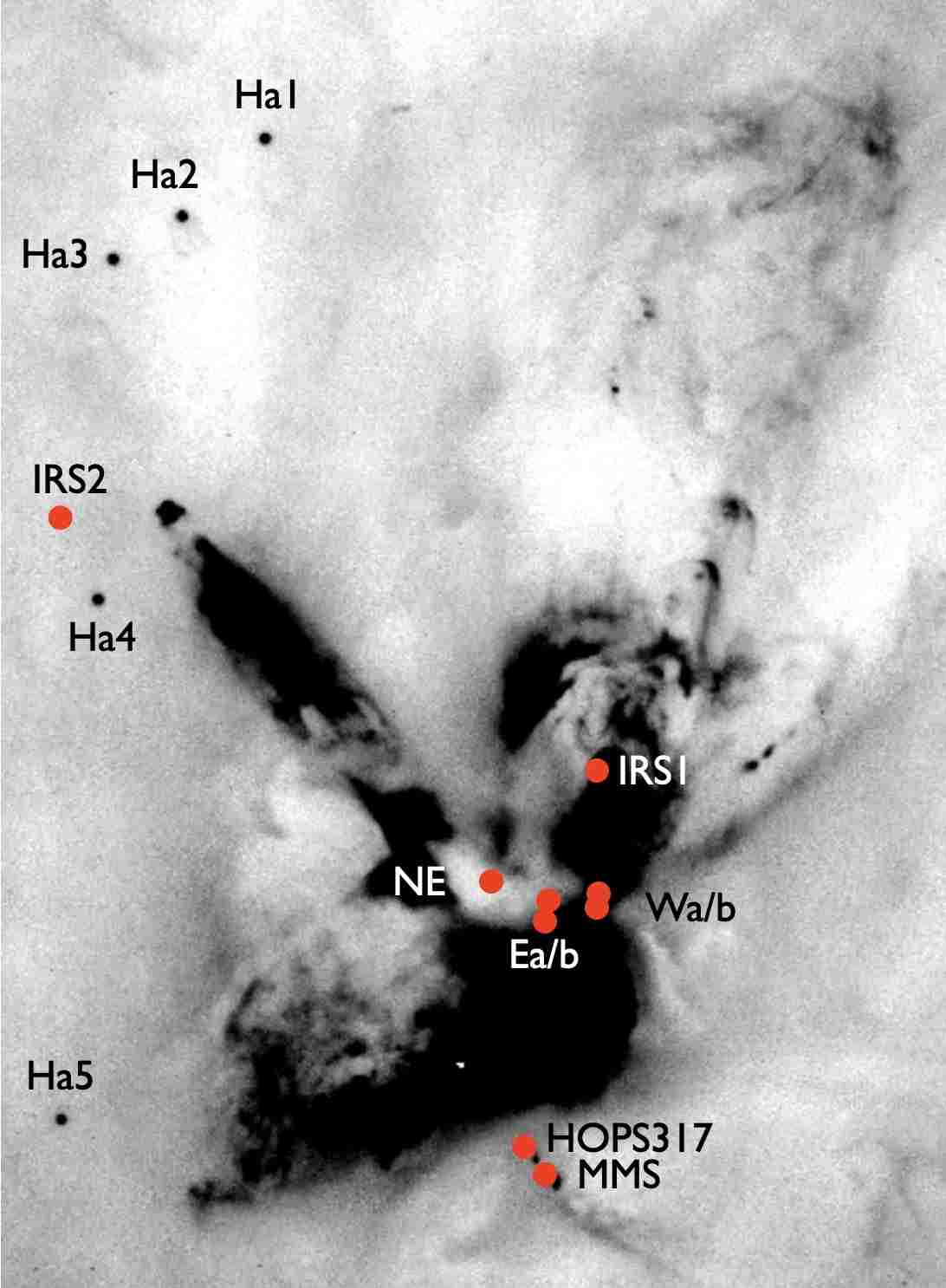}}
\caption{Identification of the optical H$\alpha$~1-5 sources and additional infrared sources in the HH 24 region, marked on an H$\alpha$+[\Sii] image from the Subaru telescope. North is up and east is left. 
\label{kilder}}
\end{figure}

\begin{figure}
\epsscale{2.0}
\centerline{\includegraphics[angle=0,width=8cm]{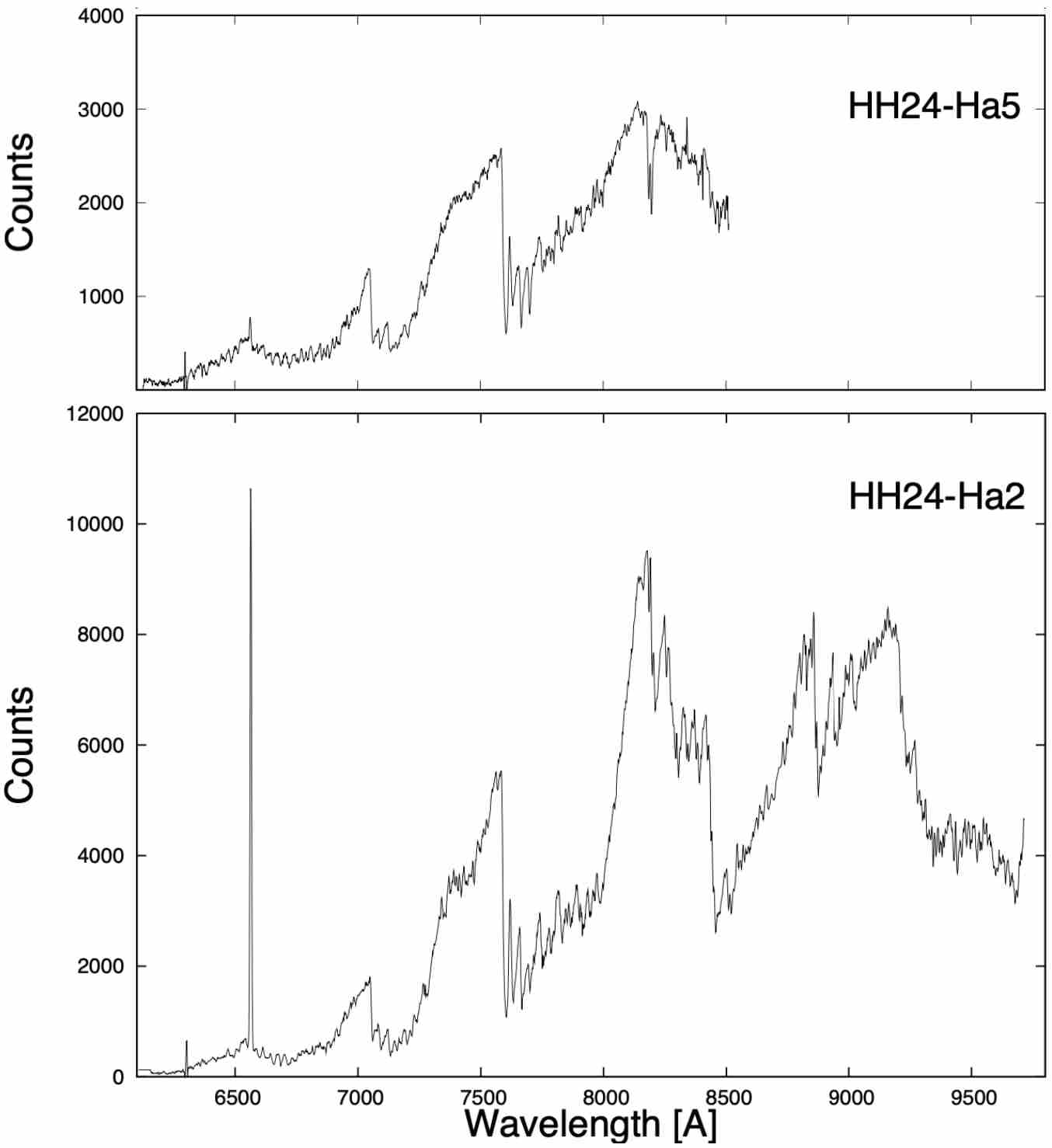}}
\caption{Optical spectra of the M5.5 borderline brown dwarf H$\alpha$~5 and the M7 brown dwarf H$\alpha$~2 obtained with GMOS on Gemini-N. 
\label{halpha5-2spectra}}
\end{figure}

H$\alpha$~1 - 5 are located far from any of the dense cloud cores in
the region (Figure~\ref{fig850micron}), suggesting that they have
traveled to their current locations from elsewhere. We have examined
the Gaia EDR3 catalog, and find that Gaia has detected all of the five
H$\alpha$ emitters.

\begin{figure}
\centerline{\includegraphics[angle=0,width=8.3cm]{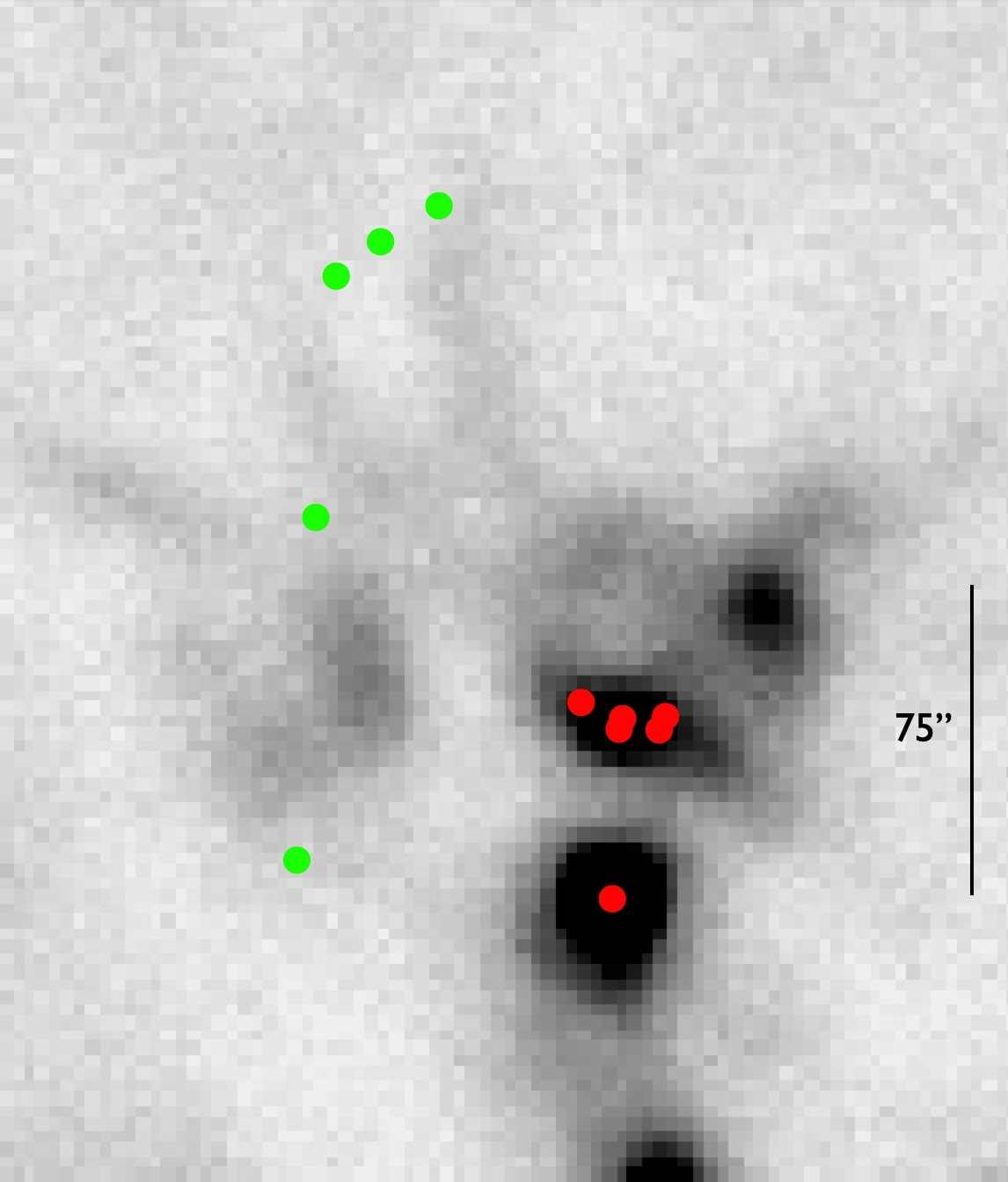}}
\caption{The cloud core in which the HH 24 multiple system is embedded
and its surroundings are seen here in a 850~$\mu$m dust continuum
image from SCUBA2, courtesy Helen Kirk (see Kirk et al. 2016a,b). The
components Wa/b, Ea/b, and NE are marked in red, as is HH 24 MMS
to the south, while the five optically visible H$\alpha$ emission
stars are marked in green. Note how the multiple system is associated
with a very dense core, while the H$\alpha$ emission stars are located far from any
dense cloud cores. The dimensions of the figure are 0.46 $\times$
0.51~pc. North is up and east is left. 
\label{fig850micron}}
\end{figure}

\subsection{The Runaway Borderline Brown Dwarf HH24-H$\alpha$5}\label{subsec:Ha5}

One object, H$\alpha$~5, immediately stands out because it has a very
large, well determined proper motion determined in Gaia DR3 as
0.7839$\pm$0.0420 mas/yr in a reference frame determined by the motion
of 129 YSOs in L1630 from Fang et al. (2009). This corresponds to a
tangential velocity of $v_{tan}$ = 26.1$\pm$1.4~\kms\ at the assumed
distance of 400~pc. Recently a number of such low-mass runaway and
walkaway stars have been found near the ONC (McBride \& Kounkel 2019,
Schoettler et al. 2020), who estimate that 1-2\% of the cluster
members they studied are runaway stars. What is particularly
interesting about H$\alpha$~5 is that its proper motion vector, with a
position angle of 121$^\circ$, points directly away from the HH~24
cloud core (Figure~\ref{halpha5-origin}). One member of the SSV~63
multiple system, source~NE, is located within a 2$\sigma$ uncertainty
cone around the H$\alpha$~5 trajectory. It therefore appears very
likely that H$\alpha$~5 was ejected from source~NE about 5800~yr
ago. If so, it implies that either NE or H$\alpha$~5 is a close
binary.\footnote{It should be noted that there is another, more
distant, star marginally within the uncertainty cone, namely the
source labeled IRS~1 in Figure~\ref{halpha5-origin}, also known as
WISE J054607.76-000937.7. It is a highly extincted YSO showing a
mid-infrared excess.}

\begin{figure*}
\epsscale{1.5}
\centerline{\includegraphics[angle=0,width=12cm]{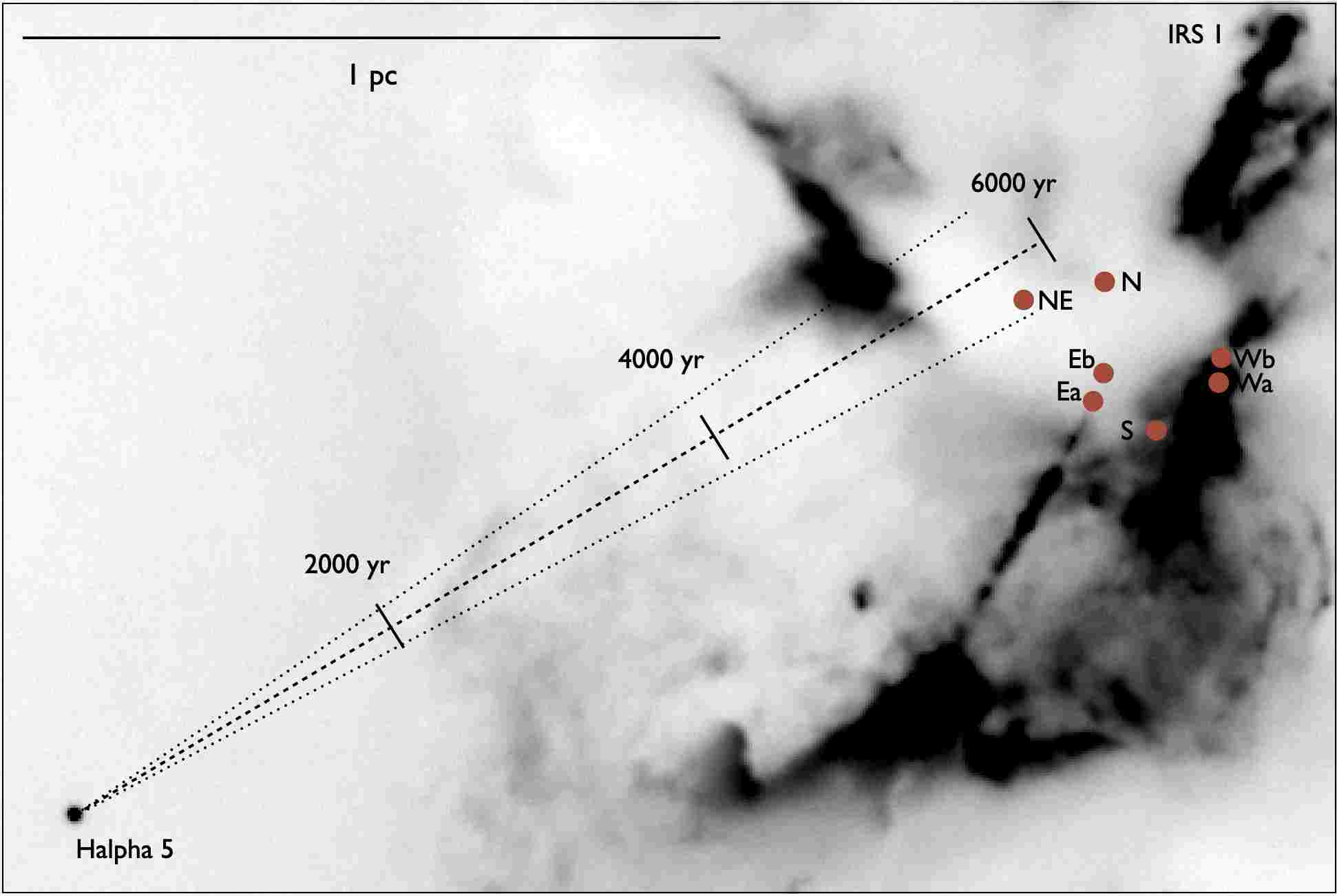}}
\caption{ The borderline brown dwarf H$\alpha$~5 moves away from the SSV~63 multiple system with a tangential velocity of about 26~\kms. At this
speed it was $\sim$5800~yr ago close to the NE source, from which it
was likely ejected in a triple interaction. The dotted lines represent
a 2$\sigma$ error on the Gaia measurement. The image is a sum of an
H$\alpha$ and a [\Sii] exposure with the Subaru telescope.
\label{halpha5-origin}}
\end{figure*}

\begin{figure}
\centerline{\includegraphics[angle=0,width=6cm]{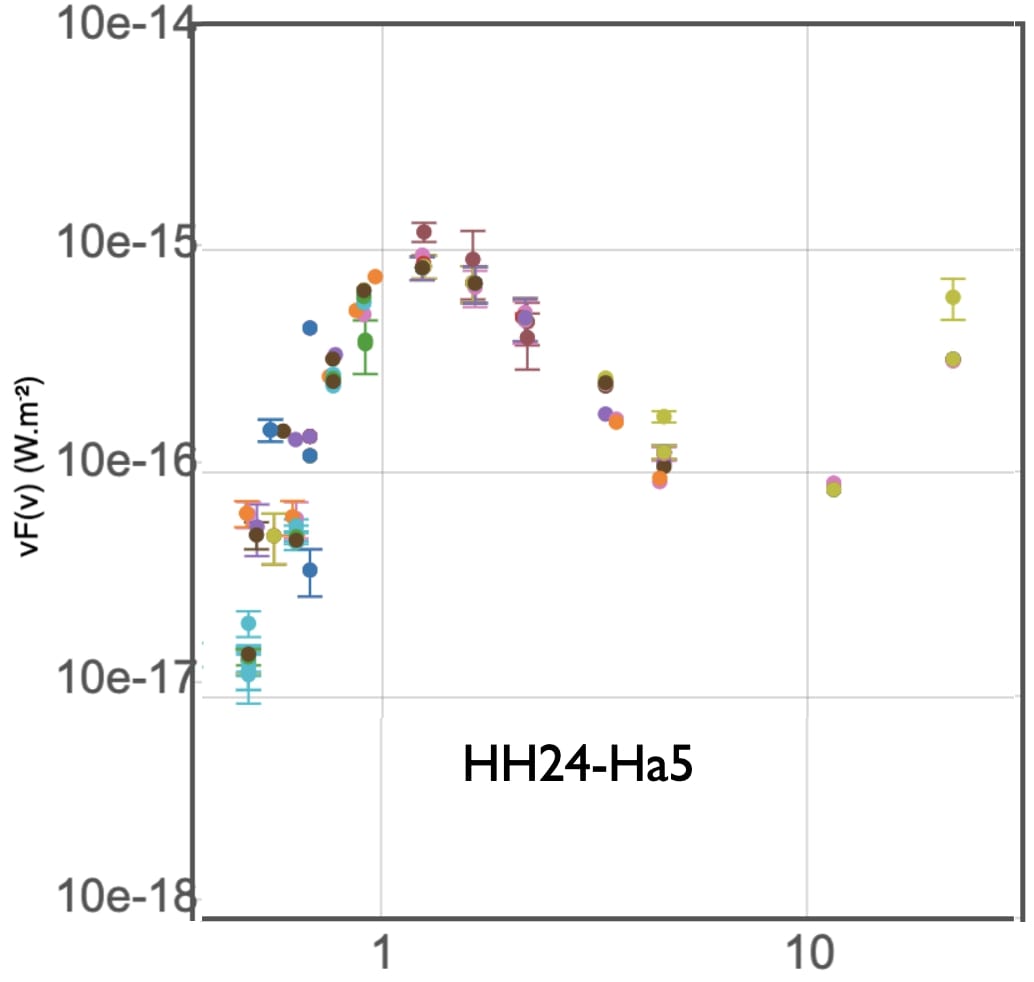}}
\caption{  The energy distribution of H$\alpha$~5 obtained with the Vizier Photometry Viewer. The majority of data points are from SDSS, PanSTARRS, 2MASS, WISE, and Spitzer. The distribution is a clean Planck curve out to 5~$\mu$m, but the WISE 12 and 22~$\mu$m data points show a steeply rising infrared excess from circumstellar material. The abscissa is wavelength in microns. 
\label{halpha5-SED}}
\end{figure}

\begin{figure*}
\centerline{\includegraphics[angle=0,width=16cm]{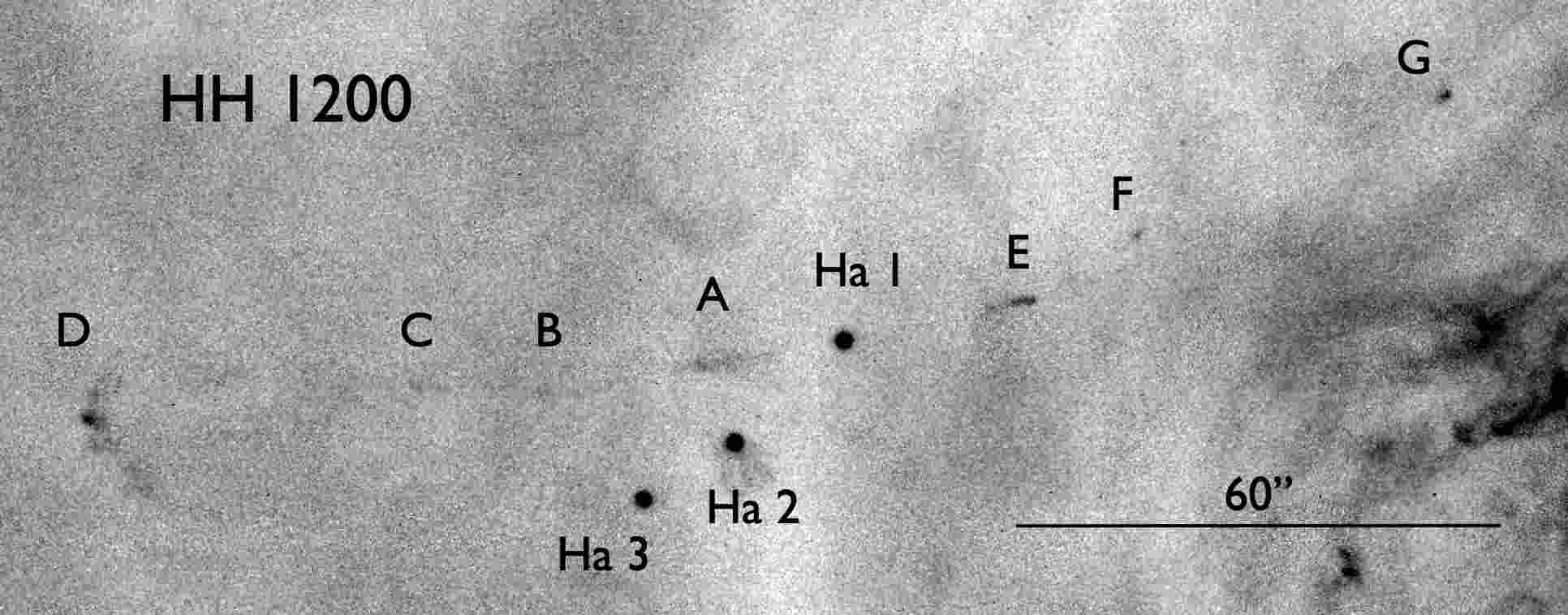}}
\caption{  The faint curved HH 1200 jet emanating from source
H$\alpha$~1 as seen on a deep H$\alpha$ image obtained at the Subaru
telescope. The distance between knots D and G is 172 arcsec,
corresponding to a projected separation of 0.33~pc. 
North is up and east is left.  
\label{hh1200}}
\end{figure*}

Source~NE is a protostellar object, so if H$\alpha$~5 was once part of a triple
system including NE, it follows that it is itself also a protostellar object.
Reipurth et al. (2010) posited that dynamical breakups during the
embedded phase could produce optically visible low-mass orphaned
protostars drifting away from their birthsites. H$\alpha$~5 appears to
be a fine case of such an orphaned protostar. The escape velocity from
a $\sim$10~M$_\odot$ core of gas and stars (see
Section~\ref{sec:efficiency}) is about 1.5~\kms, and it follows that
H$\alpha$~5 is escaping from the system.

The spectral class to effective temperature conversion established by
Herczeg \& Hillenbrand (2014) indicates that a spectral type of M5.5
corresponds to an effective temperature of about 2900~K. The
evolutionary models of Baraffe et al. (2015) show that this is very
close to the temperature for a 1 million year old object at the
hydrogen-burning limit. So is H$\alpha$~5 a brown dwarf? Unfortunately
the uncertainties involved are too large to allow a firm
answer. First, even though both our spectrum and that of Fang et
al. (2009) agree on the spectral classification, a much higher
spectral resolution would be needed for a more accurate
classification. Second, for models at 1 Myr or younger, the
sensitivity to initial conditions is significant, and the accretion
history of an object adds further uncertainty. Third, the temperature
of about 2900~K determined for a 1 Myr old object at the hydrogen
burning limit is model dependent. Taken together, the best that can be
said is that H$\alpha$~5 hovers right around the stellar/substellar
boundary.

Figure~\ref{halpha5-SED} shows the spectral energy distribution of
H$\alpha$~5. At wavelengths out to 5~$\mu$m it follows a Planck curve,
but the WISE 12 and 22~$\mu$m data points show a strong mid-infrared
excess. The indication is that H$\alpha$~5 is having circumstellar
material, but is missing an inner disk, thus resembling a transitional
disk. Reipurth \& Clarke (2001) suggested that brown dwarfs ejected in
a triple interaction would lose some of their disks in the process,
ending up with truncated disks, which was confirmed in a detailed
numerical study by Umbreit et al. (2011), see also Steinhausen et
al. (2012). It is conceivable that the disk around H$\alpha$~5 is in
the process of re-assembling after being perturbed during the
ejection.



\begin{deluxetable}{lrrrcr}
\tablecaption{Gaia EDR3 Proper Motions for H$\alpha$ 1-5\label{table:gaia}}
\tablecolumns{6}
\tablewidth{0pt}
\tablehead{
   \colhead{Star} &
   \colhead{PM($\alpha$)}  &
   \colhead{PM($\delta$)} &
   \colhead{V$_{tan}$$^a$} &
   \colhead{PA$^a$}
\\
   \colhead{} &
   \colhead{mas/yr } &
   \colhead{mas/yr} &
   \colhead{\kms } &
   \colhead{deg}
  }
\startdata
H$\alpha$~1 &   -0.711 $\pm$1.255 & -0.556 $\pm$0.901 &  0.5 $\pm$2.9 & 313.9 \\
H$\alpha$~2 &    1.175 $\pm$0.302 & -0.048 $\pm$0.249 &  3.5 $\pm$0.7 &  67.8 \\
H$\alpha$~3 &    0.544 $\pm$0.535 & -0.392 $\pm$0.438 &  2.1 $\pm$1.3 &  71.8 \\
H$\alpha$~4 &   -3.438 $\pm$0.437 & -3.094 $\pm$0.368 &  7.1 $\pm$1.1 & 231.1 \\
H$\alpha$~5 &   11.176 $\pm$0.560 & -7.949 $\pm$0.476 & 26.1 $\pm$1.4 & 121.6 
\enddata
\tablecomments{a: For calculation of space motion and position angle, the Gaia EDR3 proper motions listed in this table were corrected for the bulk motion of the L1630 cloud ($\alpha$ -0.519, $\delta$ -0.741) determined from Gaia proper motions of 129 YSOs associated with the cloud.}
\end{deluxetable}


\vspace{-0.5cm}

\subsection{Other H$\alpha$ Emission Stars and Infrared Sources}\label{subsec:Ha1-4}

Gaia EDR3 proper motions for the other 4 H$\alpha$ emitters are given
in Table~\ref{table:gaia}.  As can be seen, none of the
objects have particularly high velocities, and none are pointing
directly away from the SSV~63 multiple system.  Thus none are runaway
or walkaway stars. However, the SSV~63 cloud core is the nearest
high-density region to these young stars, so they could have been born
in the core and drifted away, perhaps nudged along by the more massive
stars. Assuming an approximate projected separation of about
100$\arcsec$ from SSV~63 and a mass of stars and cloud core of about
10~M$_\odot$, the orbital speed of a bound object is around 0.5~\kms,
so at least some of these H$\alpha$ emission stars may be weakly bound
to the SSV~63 system. That the velocity vectors do not point away from
SSV~63 could be due to the highly irregular mass distribution of stars
and gas in the region. Future Gaia releases will improve on the
accuracy of proper motions for these very faint objects. Two are
worthy of some comments.

H$\alpha$~1 is associated with a very faint, but highly collimated HH
flow, here called HH~1200. Figure~\ref{hh1200} is a part of our deep
Subaru H$\alpha$ image and shows that HH~1200 is a bent jet, with two
symmetric lobes, the eastern (containing knots A,B,C, D) with a length
of 81~arcsec (0.16~pc) and the western (knots E, F, G) with a length
of 93~arcsec (0.18~pc). The eastern lobe terminates in knot G, which
has a clear bow shock morphology. HH~1200 is much brighter in
H$\alpha$ than in [\Sii], and is thus a high-excitation flow.


H$\alpha$~2 has a spectral type of M7, and for an assumed age of
$\leq$1~Myr, its spectral type indicates that it is a very young brown
dwarf (see Figure~\ref{halpha5-2spectra}). It is also very bright at
mid-infrared wavelengths, suggesting the presence of circumstellar
material. H$\alpha$~2 has been detected as an X-ray source with Chandra
by Simon et al. (2004, their source \#16), whereas none of the other 4
H$\alpha$ emission stars were detected.

Among the numerous near- and mid-infrared sources detected in 2MASS,
WISE, and Spitzer images in L1630, two sources close to SSV~63 should
be mentioned. IRS~1 is a faint optically visible star, classified as a
disk-bearing star in Megeath et al. (2012),  but bright at 
near-infrared wavelengths (Figures~\ref{kilder} and
\ref{halpha5-origin}). As we speculated in
Section~\ref{subsec:jetG} it is potentially the driving source of two
of the shocks in the G-jet.  IRS~2, marked in Figure~\ref{kilder},
also has a steeply rising energy distribution and is classified as a
young star by Megeath et al. (2012).  We note that it is a binary with
a fainter companion 0.8$\arcsec$ distant at PA = 325$^\circ$.

\vspace{0.3cm}

\section{CORE MASS AND STAR FORMATION EFFICIENCY}\label{sec:efficiency}

The SSV~63 multiple system is located in a cloud core that is part of
a north-south molecular ridge active in star formation in the Orion-B
cloud. The region has been studied in various transitions
including CO, C$^{18}$O, CS, and HCO$^+$ by Gibb \& Heaton (1993),
Gibb et al. (1995), and Gibb \& Little (1998). Sub-mm dust continuum
observations of the region have been reported by Chini et al. (1993),
Lis et al. (1999), and Kirk et al. (2016a,b). The molecular ridge has
been sculpted by the many molecular outflows in the region (see
Figure~\ref{overview}). 

The cloud core in which SSV~63 resides is being torn apart by multiple
jets, as seen at optical and infrared wavelengths in
Figures~\ref{gemini} and \ref{pressrelease}, where the remnant of the
core and associated outflow cavities are seen illuminated by the
embedded sources. The core has also been significantly churned by the
random motions of the stars in the non-hierarchical multiple
system. If they are moving with characteristic velocities around 1 \kms, 
stars like Ea and NE with 2~M$_\odot$ will have a Bondi radius of
$\sim$1800~AU and core crossing times of the order of 40,000~yr.
Hence the stars will have traversed the core maybe a dozen times or
more since their formation.

K\"onyves et al. (2020) used the Herschel Gould Belt Survey of the
Orion~B cloud to study the numerous cores in this complex. By
combining PACS 70 and 160~$\mu$m and SPIRE 250, 350, and 500~$\mu$m
data they were able to derive not only column densities but also dust
temperatures. Their core \#1025 corresponds to the HH~24 core for
which they determine a mean core radius of 0.019~pc (diameter
$\sim$20~arcsec), a dust temperature of 16.3~K, and a core mass of
2.31~M$_\odot$.

The 850~$\mu$m map of the SSV~63 cloud core by Kirk et al. (2016a,b),
see Figure~\ref{fig850micron}, shows clearly that the core is better
described as an ellipse, which we fit with semi-minor and semi-major axes
of 12.4\arcsec\ $\times$ 31.6\arcsec\ at a PA=70$^\circ$. This area
produces a 850~$\mu$m flux of 1.737~Jy. Using the T$_{dust}$ = 16.3~K of
K\"onives et al. and using the mass formula of Lane et al. (2016,
their Eqn~1) then yields a current mass of 3.3~M$_\odot$, which we
adopt here. Given the various uncertainties involved, this is probably
accurate to within a factor of two.






Assuming that the masses of all the components of SSV~63 stars adds up
to roughly 7 M$_\odot$ (see Section~\ref{sec:ALMA-I}), we can in
principle estimate the star formation efficiency of the cloud core. If
we further assume that the original core mass is the current mass plus
the mass of the stars born in the core, that is, of the order of
10~M$_\odot$, we obtain a very high star formation
efficiency. It makes little difference that the mass lost in outflows
from the stars has not been included, as it is relatively small. But,
more importantly, the core is not isolated from the surroundings and,
as will be shown in Section~\ref{subsec:streamer}, it appears that the
core is being continually fed gas from its environment.  One possible
scenario is that the initial small starburst that has taken place in
the HH 24 core may have been triggered by infall of gas onto the core,
and has continued at the rate that gas has become available, with
source Eb being the most recent member of the small cluster. Whether
star formation has proceeded in a static or a dynamic scenario, it
appears that gas has been converted into stars at a high
efficiency.

Eventually, as will be discussed later, the sources Ea and NE will
emerge as young late-type Herbig Ae stars
surrounded by a halo of loosely bound
lower mass stars, as is frequently seen around Herbig Ae stars
(Hillenbrand 1995, Hillenbrand et al. 1995). Testi et al. (1997) found that the
clustering of YSOs around Herbig Ae/Be stars depends on their mass,
with Be stars having significantly richer environments than Ae stars;
in their sample of 6 Herbig Ae stars the mean number of components was
4.

\begin{figure}[b]
\centerline{\includegraphics[angle=0,width=8.3cm]{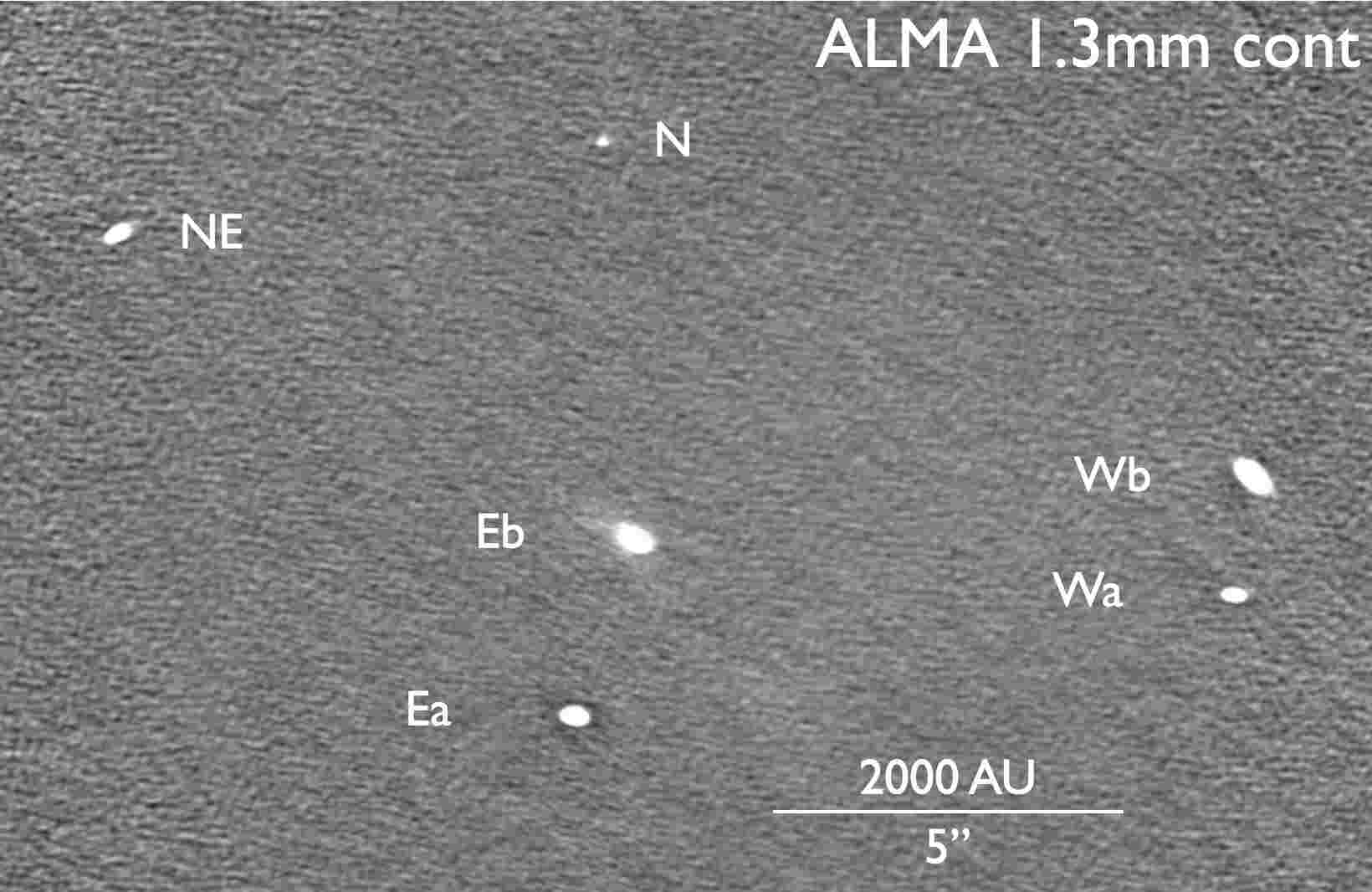}}
\caption{ An ALMA self-calibrated continuum 1.3mm image showing the
principal submm sources of the SSV~63 multiple system. All except
source~S are detected.
\label{ALMA-continuum}} 
\end{figure}

\vspace{0.3cm}

\section{ALMA 1.3 MM OBSERVATIONS OF CIRCUMSTELLAR DISKS}\label{sec:ALMA-I}

\subsection{Continuum emission}\label{subsec:continuum}

Six continuum compact sources were detected with our ALMA observations
at 1.3 mm (Figure~\ref{ALMA-continuum}).  These are the five
sources Ea, Eb, Wa, Wb, and NE, as well as the new source N
(Section~\ref{subsec:sourceN}). Source~S was not detected by ALMA. A
two-dimensional Gaussian function was fitted to each continuum compact
source, and the center, integrated flux, and deconvolved size were
measured (Table~\ref{table:cont-fit}).  Only source N was not
resolved. The total fluxes of the residuals after subtracting the
fitted Gaussian functions from the observed maps are comparable to or
less than the uncertainties of the fitted fluxes, although the
observed continuum intensity distributions in source NE, Ea, Eb, and
Wb cannot be well reproduced with a Gaussian function. It is
noteworthy that the major axis of these resolved continuum sources is
almost precisely perpendicular to the jets associated with
them. In addition, in the sources NE, Ea, Eb, and Wb, the compact
C$^{18}$O emission coincident with the compact continuum emission is
observed and shows a clear velocity gradient along the major axis of
the continuum emission (Section~\ref{subsec:c18o}). Thus, these
compact continuum components likely trace the circumstellar disks
around the protostars.  The inclination angles of the circumstellar
disks were estimated from the ratio of the major and minor axes of the
continuum emission.

\vspace{1cm}

\begin{deluxetable*}{cccccccccc}
\tablecaption{Gaussian fitting of the 1.3 mm continuum emission\label{table:cont-fit}}
\centering
\tablewidth{0pt}
\tablehead{Source & RA & Dec & Flux & PA & Major & Minor & Residual & $i$ & $\alpha$ \\ 
 & (ICRS) & (ICRS) & (mJy) & ($\arcdeg$) & (mas) & (mas) & (mJy)}
\startdata
NE & 05:46:08.921 & $-$00:09:56.11 & 13.0$\pm$1.4 & 129.3$\pm$1.2 & 141$\pm$3 & 48$\pm$4 & $-$0.3 & 70.2$\pm$1.5 & 0.0$\pm$0.8\\
Ea & 05:46:08.485 & $-$00:10:03.04 & 48.9$\pm$0.7 & 58.1$\pm$0.5 & 109$\pm$1 & 66$\pm$1 & $-$1.7 & 52.9$\pm$0.4  & 1.7$\pm$0.3\\
Eb & 05:46:08.427 & $-$00:10:00.50 & 11.7$\pm$0.7 & 238.7$\pm$2.4 & 440$\pm$12 & 279$\pm$14 & 1.9 & 50.7$\pm$1.9 & 3.1$\pm$0.3\\
Wa & 05:46:07.854 & $-$00:10:01.30 & 14.2$\pm$1.0 & 74.9$\pm$1.4 & 96$\pm$2 & 42$\pm$2 & $-$0.2 & 63.9$\pm$1.3 &1.7$\pm$0.5 \\ 
Wb & 05:46:07.836 & $-$00:09:59.59 & 49.2$\pm$0.7 & 223.1$\pm$0.3 & 244$\pm$1 & 98$\pm$1 & 0.4 & 66.4$\pm$0.3  & 1.8$\pm$0.3 \\
N & 05:46:08.457 & $-$00:09:54.80 & 0.9$\pm$0.1 & \nodata & \nodata & \nodata & $-$0.07 & \nodata
\enddata 
\tablecomments{PA is the position angle of the major axis from north to east. Major and minor axes are the deconvolved FWHM widths. The fluxes in the residual maps were computed in an area of approximately twice of the apparent size of the continuum emission. $i$ is the inclination angle to the plane of the sky computed from the ratio of the major and minor axes. $\alpha$ is the spectral index between 1.3 and 0.9 mm. Source N is not resolved and not detected at 0.9 mm. The uncertainty of $\alpha$ includes the uncertainty of the absolute flux calibration of 10\%.}
\end{deluxetable*}\label{cont_fit}


The same region was also observed with ALMA at 0.9 mm in 
\citet{Tobin20}. Source N was not detected at 0.9 mm, and the other
sources were detected and resolved with the ALMA 0.9 mm
observations. The deconvolved orientations and sizes measured at 1.3
mm are consistent with those at 0.9 mm within the uncertainties. The
spectral indices of these continuum sources between 0.9 and 1.3 mm
were computed. Except for source Eb, all the continuum sources have
spectral indices $\lesssim$2, suggesting that the continuum emission
is optically thick. The 1.3 mm continuum emission in source Eb is
likely optically thin, and the total (dust+gas) mass ($M_{\rm 1.3mm}$)
of the circumstellar material around source Eb is estimated as
\begin{equation}
M_{\rm 1.3mm} = \frac{D^2 F_\nu}{\kappa B_{\nu}(T_{\rm d})}, 
\end{equation}
where $D$ is the distance, $F_\nu$ is the continuum flux at 1.3 mm, $\kappa$ is the dust mass opacity, and $B_{\nu}(T_{\rm d})$ is the Planck function at a temperature $T_{\rm d}$.
$\kappa$ at 1.3 mm is adopted to be 0.019 g$^{-1}$ cm$^2$ \citep{Beckwith90}, which includes a gas-to-dust mass ratio of 100.
$T_{\rm d}$ is assumed to be 20--94 K. 
$T_{\rm d}$ of 94 K was estimated from the stellar luminosity, 
which can be considered as an upper limit because the protostellar source was resolved to be a multiple system \citep{Tobin20}.
$M_{\rm 1.3mm}$ in source Eb was estimated to be 3--18 $M_{\rm Jupiter}$.

\begin{figure}
\centerline{\includegraphics[angle=0,width=8cm]{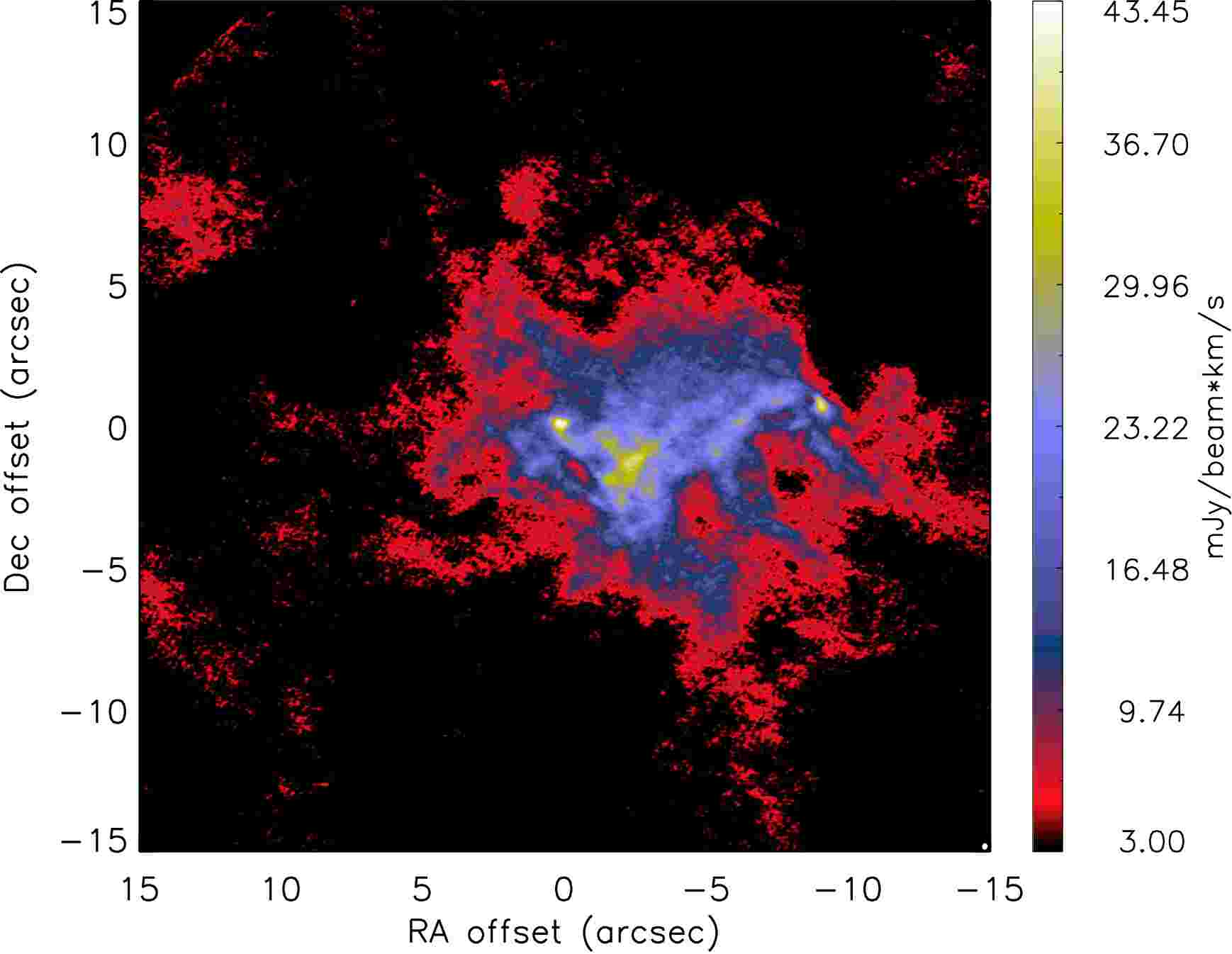}}
\caption{Integrated intensity map of the C$^{18}$O emission in the HH
24 region obtained with the ALMA observations. The integrated velocity
range is 8.5 to 11.5 km s$^{-1}$. The map is centered at source
Eb. 
\label{c18o-clouds}}
\end{figure}

\subsection{C$^{18}$O (2--1) emission}\label{subsec:c18o} 

Extended C$^{18}$O emission associated with the large-scale clouds is
detected at $V_{\rm LSR} \sim 8.5\mbox{--}11.5$ km s$^{-1}$
(Figure~\ref{c18o-clouds}).  At higher velocities relative to the cloud
velocity, compact C$^{18}$O emission is seen around sources NE, Ea, Eb,
and Wb. In these four sources, the high-velocity blue- and redshifted
C$^{18}$O emission is well aligned along the major axis of the
continuum emission (Figure~\ref{c18o-ne}a, \ref{c18o-ea}a,
\ref{c18o-eb}a, and
\ref{c18o-wb}a), which likely traces the disk rotation. We constructed
kinematical models of a geometrically-thin Keplerian disk and
performed fitting to the high-velocity C$^{18}$O emission in sources 
NE, Ea, Eb, and Wb to measure stellar mass ($M_\star$) and systemic
velocity ($V_{\rm sys}$).

Two disk models with different intensity profiles, Gaussian and
power-law functions, were adopted. For each source, the center,
orientation and inclination angle of the model disks were adopted from
the continuum results (Table~\ref{table:cont-fit}) and were fixed in our
disk models. Thus, the free parameters in our disk models are
$M_\star$, $V_{\rm sys}$, and additional parameters to describe the
intensity profiles (three and two parameters for the power-law and
Gaussian profiles, respectively). The fitting was performed with the
velocity channel maps, and only the velocity channels without
significant extended C$^{18}$O emission were included in the
fitting. The velocity range for the fitting of each source is listed
in Table~\ref{table:line-fit}.  We generated velocity channel maps of the
disk models, and convolved the model channel maps to the same beam
sizes as the observed maps.  Then, the residuals were calculated
within a 1$\arcsec$ region centered at the continuum peak after
subtracting the model maps from the observed maps.  We searched for
the best-fit parameters by minimizing the residuals. We did not
simulate ALMA observations and sample the {\em uv} coverage on the
model channel maps because the disk sizes are smaller than the maximum
recoverable angular scale of the observations.

\begin{deluxetable*}{cccccc}
\vspace{-0.8cm}
\tablecaption{Disk properties and stellar masses\label{table:alma-disk}}
\centering
\tablehead{Source & $M_{\rm disk}$ & $R_{\rm dust}$ & $R_{\rm gas}$ & $M_\star$ & $V_{\rm lsr}$ \\
 & ($M_{\rm J}$) & (au) & (au) & ($M_\sun$) & (km\,s$^{-1}$) }
\startdata
NE & \nodata & 51$\pm$1 & 245$^{+12}_{-17}$ & 2.1$^{+0.2}_{-0.1}$ & 9.6$\pm$0.1 \\
Ea & \nodata & 39$\pm$1 & 161$^{+3}_{-14}$ & 2.0$\pm$0.1 & 9.6$\pm$0.1 \\
Eb & 3--18 & 159$\pm$4 & 332$^{+23}_{-3}$ & 1.3$\pm$0.1 & 10.8$\pm$0.1 \\
Wa & \nodata & 35$\pm$1 & \nodata & \nodata & \nodata \\
Wb & \nodata & 81$\pm$1 & 492$^{+4}_{-12}$ & 0.9$\pm$0.1 & 9.3$\pm$0.1 
\enddata 
\tablecomments{Except for Source Eb, the continuum disks are optically thick at 1.3 mm, so the disk mass cannot be estimated. The C$^{18}$O fitting results with the Gaussian intensity profile are adopted here for comparison with the continuum disk size. The disk radius is defined as twice the 1$\sigma$ width of the best-fit Gaussian profile.}
\end{deluxetable*}

\begin{figure*}
\centerline{\includegraphics[angle=0,width=15cm]{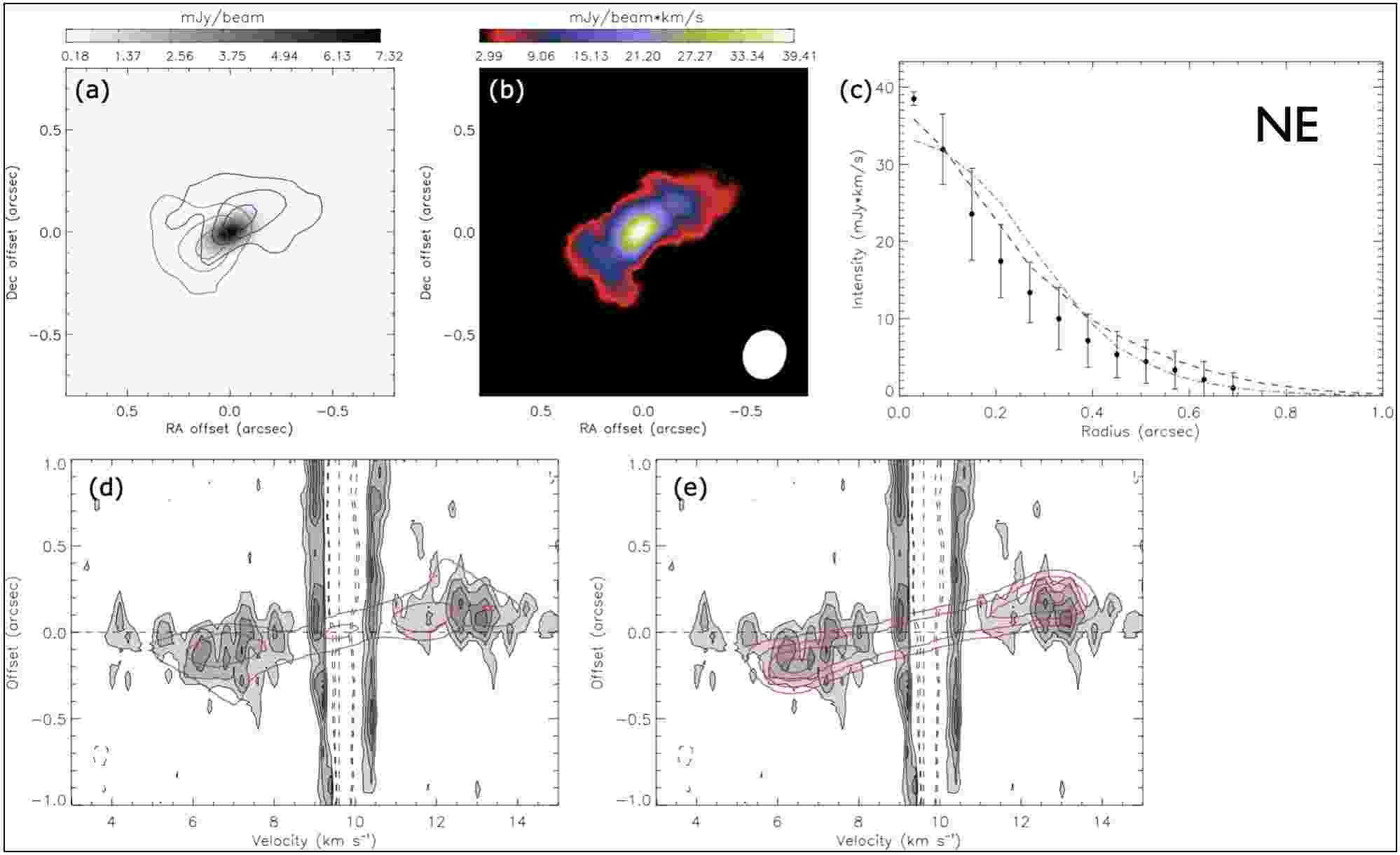}}
\caption{(a)
Integrated intensity maps for source NE of the blue- and redshifted
high-velocity C$^{18}$O emission (blue and red contours) overlaid on
the 1.3 mm continuum map of source NE. The integrated velocity ranges
of the blue- and redshifted high-velocity C$^{18}$O emission are
listed in Table~\ref{table:line-fit}, where the velocity ranges
adopted in the model fitting are listed. The contour levels start from
4$\sigma$ in steps of 4$\sigma$ to 20$\sigma$ and then in steps of
10$\sigma$, where 1$\sigma$ is 1.2 mJy beam$^{-1}$ km s$^{-1}$. (b)
Total integrated intensity map of the C$^{18}$O emission in the
disk. A Keplerian mask generated based on our best-fit disk model was
applied to the C$^{18}$O velocity channel maps to minimize the
contamination from the cloud emission. A white ellipse shows the beam
size. (c) Azimuthally averaged intensity profiles of the C$^{18}$O
emission in the disk (data points) extracted from (b). Blue and red
dashed lines present the intensity profiles extracted from the maps of
the model disks with the power-law and Gaussian intensity profiles,
respectively. (d) and (e) PV diagrams of the C$^{18}$O emission along
the major axis of the disk (gray scale) in comparison with those
extracted from the best-fit disk models (red contours) with the
power-law and Gaussian intensity profiles, respectively. The contour
levels start from 2$\sigma$ in steps of 1$\sigma$, where 1$\sigma$ is
1.7 mJy beam$^{-1}$. 
\label{c18o-ne}} 
\end{figure*}

\begin{figure*}
\centerline{\includegraphics[angle=0,width=15cm]{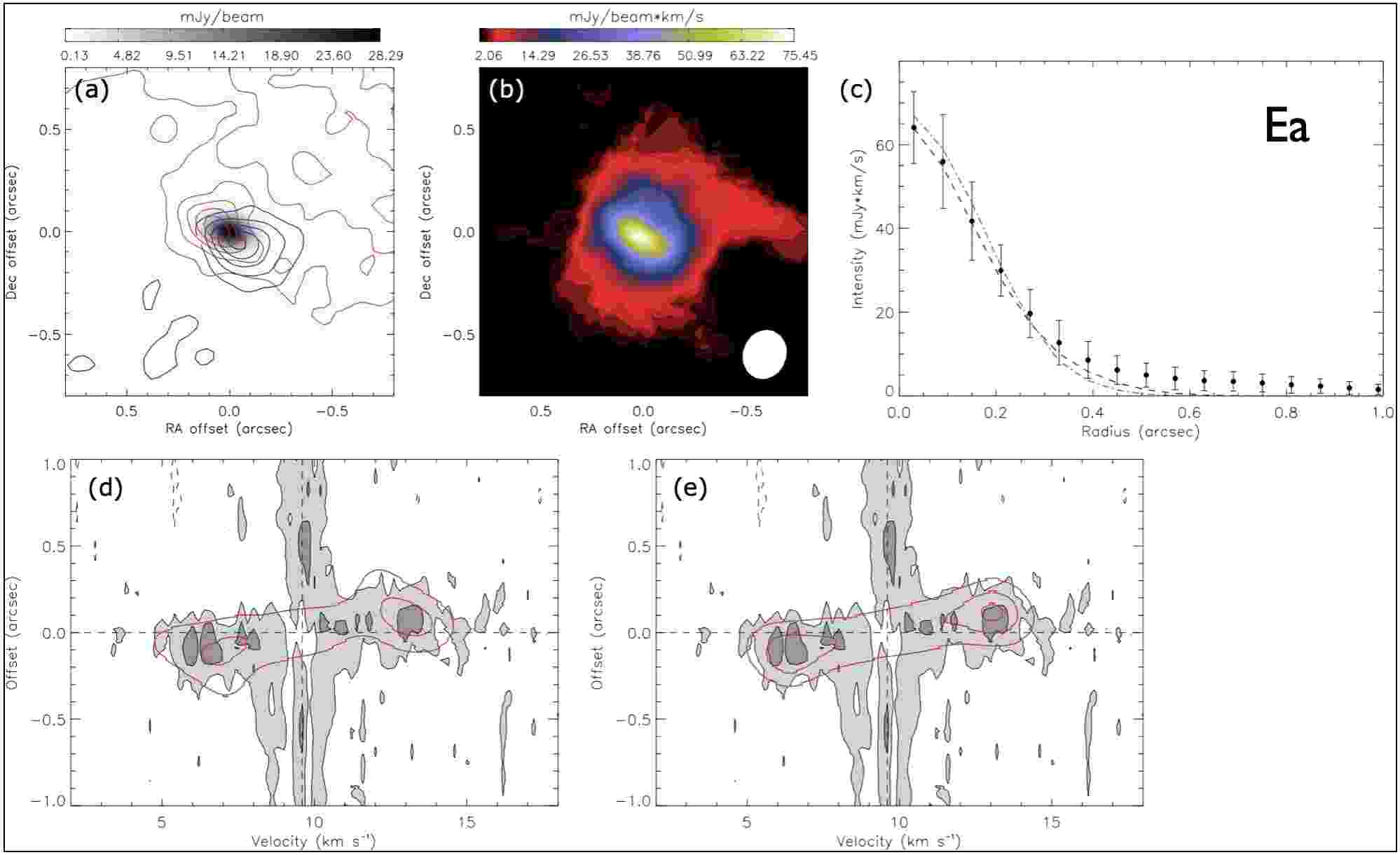}}
\caption{Same as Figure~\ref{c18o-ne} but for Source Ea. In (a), 1$\sigma$ is 1.3 mJy beam$^{-1}$ km s$^{-1}$. In (d) and (e), the contour levels start from 2$\sigma$ in steps of 3$\sigma$. 
\label{c18o-ea}}
\end{figure*}

\begin{figure*}
\centerline{\includegraphics[angle=0,width=15cm]{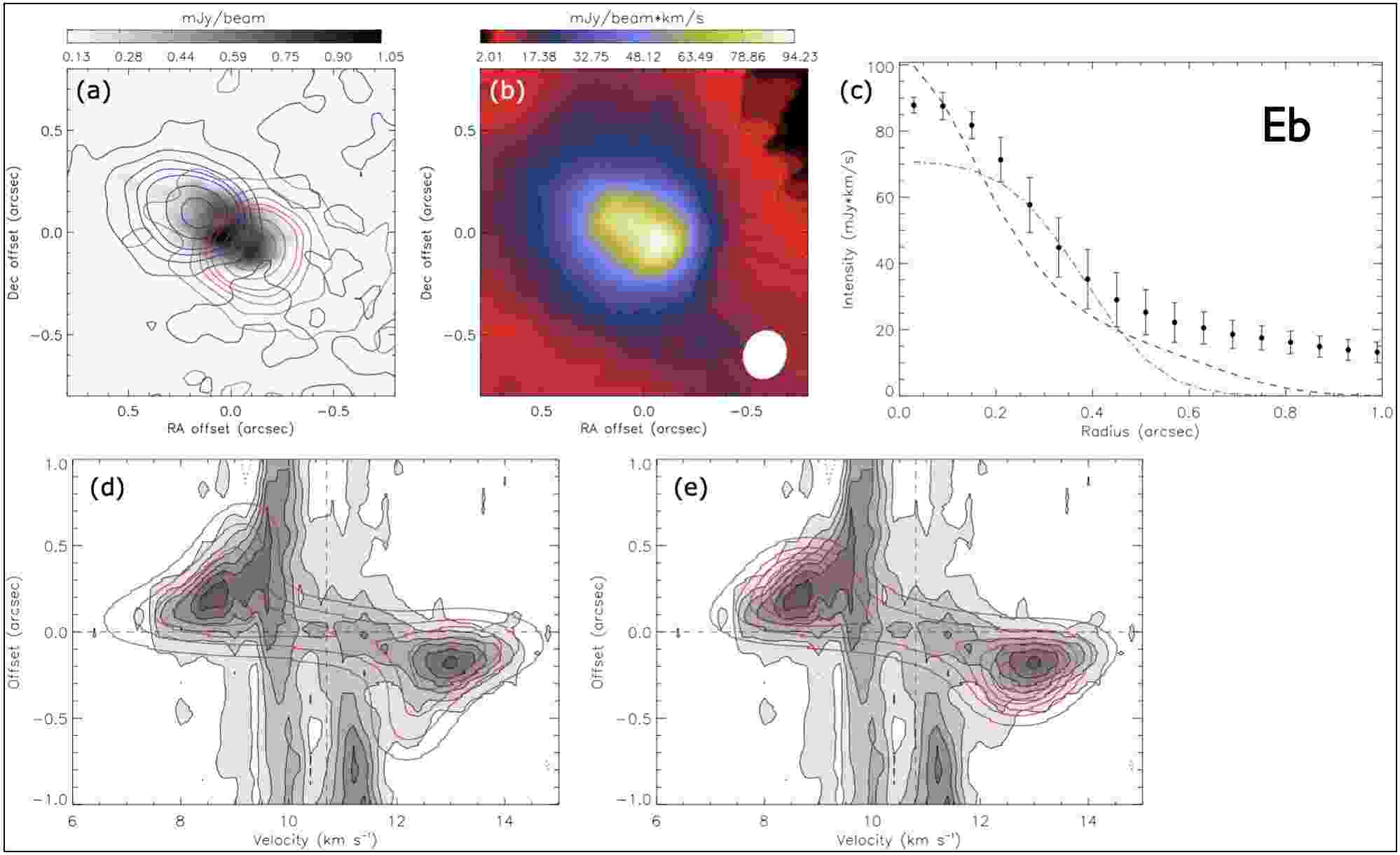}}
\caption{Same as Figure~\ref{c18o-ne} but for Source Eb. In (a), 1$\sigma$ is 1 mJy beam$^{-1}$ km s$^{-1}$. In (d) and (e), the contour levels start from 2$\sigma$ in steps of 3$\sigma$. 
\label{c18o-eb}}
\end{figure*}

The best-fit parameters are listed in Table \ref{table:line-fit}.  We
found that for source Ea, the outer radius of the disk model with a
power-law intensity profile ($R_{\rm out}$) could not be constrained
with our fitting, so it was fixed to be 0\farcs4. We confirmed that
the best-fit $M_\star$ and $V_{\rm sys}$ of source Ea are not
sensitive to the choice of the fixed outer radius, and that the
results remain unchanged when the outer radius is adopted to be
0\farcs3 or 0\farcs6. The uncertainties in the disk orientation and
inclination are included in the error propagation in our fitting,
although they are not free parameters in our disk models. We note that
the uncertainty of $M_\star$ in Table \ref{table:line-fit} does not
include the uncertainty due to the geometrically thin approximation in
our disk models. There could be an additional uncertainty in $M_\star$
of 10\%--20\% if the C$^{18}$O emission traces a flared disk,
especially when the disk is highly inclined
\citep[e.g.,][]{Braun21}. Nonetheless, the mass estimates
clearly show that both source NE and Ea, with masses of about
2~M$_\odot$, are much more massive than a T~Tauri star, and in fact
will later emerge from the cloud core as young Herbig~Ae
stars. Sources Eb and Wb will become observable as massive T~Tauri
stars. Given that the sources may still experience significant
accretion, these could be conservative estimates.  

To reveal the distributions of the C$^{18}$O emission in the disks
with least contamination from an ambient envelope or cloud emission, we
constructed Keplerian masks based on our best-fit disk models and
applied them to the observed velocity channel maps. The total
integrated intensity maps of the C$^{18}$O emission after applying the
Keplerian masks are shown in Figures~\ref{c18o-ne}b, \ref{c18o-ea}b,
\ref{c18o-eb}b, and \ref{c18o-wb}b, but diffuse emission can still be
seen in source Ea, Eb, and Wb. We extracted azimuthally averaged
intensity profiles of the C$^{18}$O emission from the Keplerian masked
maps. The observed intensity profiles in source NE and source Ea could
be fitted with our simple disk models (Figures~\ref{c18o-ne}c and
\ref{c18o-ea}c), while those in source Eb and Wb could not be fully
explained with simple Gaussian or power-law functions
(Figures~\ref{c18o-eb}c and \ref{c18o-wb}c). In source Eb and Wb, the
power-law disk models fit the central and outer intensity profiles
better, and the Gaussian disk models describe the intensity profiles
at intermediate radii better. Nevertheless, the best-fit

\begin{figure*}[t]
\centerline{\includegraphics[angle=0,width=15cm]{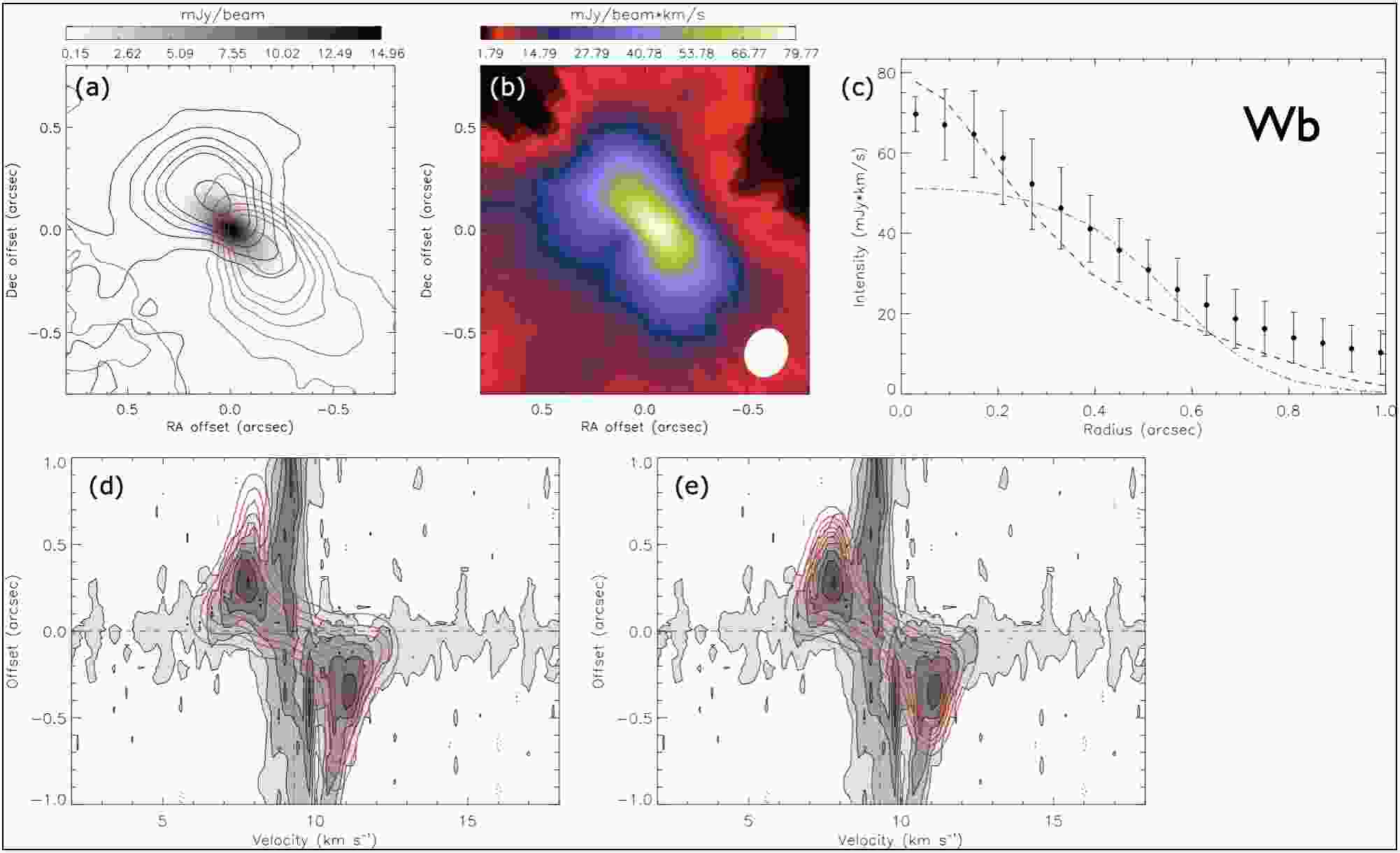}}
\caption{Same as Figure~\ref{c18o-ne} but for Source Wb. In (a), 1$\sigma$ is 1.3 mJy beam$^{-1}$ km s$^{-1}$. In (d) and (e), the contour levels start from 2$\sigma$ in steps of 3$\sigma$. 
\label{c18o-wb}}
\end{figure*}

$M_\star$ and
$V_{\rm sys}$ of all the sources are not sensitive to the intensity
profiles assumed in the disk models. The best-fit $M_\star$ and
$V_{\rm sys}$ from the fitting with the Gaussian and power-law
intensity profiles are consistent within the uncertainty. In panel (d)
and (e) in Figures~\ref{c18o-ne}--\ref{c18o-wb}, the observed
position--velocity (PV) diagrams along the disk major axes are
compared with those extracted from the best-fit disk models.  The
observed velocity structures of the compact C$^{18}$O emission around
source NE, Ea, Eb, and Wb indeed can be explained with Keplerian
rotation, and significant extended emission associated with the
ambient envelopes or clouds is also seen in the PV diagrams.


\begin{deluxetable*}{ccccccc}
\vspace{-0.8cm}
\tablecaption{Fitting of the C$^{18}$O emission with kinematical disk models\label{table:line-fit}}
\centering
\tablewidth{0pt}
\tablehead{\multicolumn{7}{c}{Power-law intensity profile}\\ 
\hline
Source & $R_{\rm out}$ & $M_\star$ & $V_{\rm lsr}$ & $\log I_0$ & $p$ & Velocity ranges \\ 
 & (mas) & ($M_\odot$) & (km s$^{-1}$) & (Jy) &}
\startdata
NE & 701$^{+63}_{-59}$ & 2.1$^{+0.2}_{-0.1}$ & 9.6$\pm$0.1 & $-$3.44$\pm$0.03 & $-$2.0$\pm$0.1 & 4.9--8.3 \& 10.9--14.5 \\ 
Ea & 400$^a$ & 1.9$\pm$0.1 & 9.6$\pm$0.1 & $-$3.56$\pm$0.03 & $-$2.0$\pm$0.1 & 4.5--8.1 \& 10.5--14.5 \\
Eb & 701$^{+24}_{-17}$ & 1.3$\pm$0.1 & 10.7$\pm$0.1 & $-$3.14$\pm$0.01 & $-$1.3$\pm$0.1 & 7.3--9.5 \& 11.5--14.3 \\
Wb & 853$^{+38}_{-17}$ & 0.9$\pm$0.1 & 9.3$\pm$0.1 & $-$3.05$\pm$0.01 & $-$1.3$\pm$0.1 & 4.9--8.5 \& 10.1--14.1 \\
\hline
\multicolumn{7}{c}{Gaussian intensity profile}\\ 
\hline
Source & $\sigma_{\rm R}$ & $M_\star$ & $V_{\rm lsr}$ & $\log I_0$ & & Velocity ranges \\ 
 & (mas) & ($M_\odot$) & (km s$^{-1}$) & (Jy) & \\
\hline
NE & 287$^{+14}_{-20}$ & 2.0$^{+0.2}_{-0.1}$ & 9.6$\pm$0.1 & $-$3.1$\pm$0.05 && 4.9--8.3 \& 10.9--14.5 \\
Ea & 188$^{+3}_{-16}$ & 2.0$\pm$0.1 & 9.7$\pm$0.1 & $-$2.79$^{+0.04}_{-0.01}$ &&4.5--8.1 \& 10.5--14.5 \\
Eb & 388$^{+27}_{-4}$ & 1.3$\pm$0.1 & 10.8$\pm$0.1 & $-$2.94$^{+0.01}_{-0.04}$ && 7.3--9.5 \& 11.5--14.3 \\
Wb & 576$^{+5}_{-14}$ & 0.9$\pm$0.1 & 9.3$\pm$0.1 & $-$3.11$\pm$0.01 && 4.9--8.5 \& 10.1--14.1
\enddata 
\tablenotetext{a}{R$_{out}$ for source Ea could not be constrained by model fitting and was fixed at 400~mas.}
\tablecomments{The intensity profile of the model disks is adopted to be power-law or Gaussian functions. The power-law function is described with a power-law index $p$, an outer radius $R_{\rm out}$, and the intensity at a radius of 100 au in a logarithmic scale $\log I_0$. The Gaussian function is described with the 1$\sigma$ width $\sigma_{\rm R}$ and the peak intensity in a logarithmic scale $\log I_0$. $M_\star$ and $V_{\rm sys}$ are stellar mass and systemic velocity, respectively. $V$ ranges are the velocity ranges included in the fitting. The velocity channel maps at velocities close to $V_{\rm sys}$ were excluded in the fitting to avoid cloud contamination. The uncertainty does not include the systematic uncertainty due to the geometrically thin assumption. If the C$^{18}$O emitting surface is flared with a scale height ($h/r$) larger than 0.1, there is an additional uncertainty in $M_\star$ of 10\%--20\%, especially when the disk is more inclined.}
\end{deluxetable*}


\subsection{Summary of source properties}\label{subsec:sourcesummary}

Table~\ref{table:alma-disk} compares the disk sizes, $M_\star$ and $V_{\rm
sys}$ in our targets in the HH~24 region. The best-fit parameters of
the C$^{18}$O disk models with Gaussian intensity profiles are adopted
here for comparison with the continuum results which were also fitted
with the Gaussian functions.  The disk radius is defined as twice the
1$\sigma$ width of the fitted Gaussian function, the same as that in
\citet{Tobin20}.  The radii of the gaseous disks traced by the
C$^{18}$O emission are two to six times larger than those of the dusty
disks traced by the continuum emission. This is similar to the
observations of several T Tauri disks
\citep[e.g.,][]{Sanchis21}. Nevertheless, the significant cloud and/or
envelope contamination is seen in the C$^{18}$O emission in our data,
so the disk components cannot be fully separated from the ambient gas
(e.g., panel (b) in Figures~\ref{c18o-ne}--\ref{c18o-wb}), which
introduces an uncertainty in our estimated radii of the gaseous
disks. Observations at higher resolutions and sensitivity and more
detailed models are needed to fully separate the disk and envelope
components.

\vspace{0.3cm}

\section{CORE KINEMATICS AND MOLECULAR OUTFLOWS}\label{sec:ALMA-II}


Our ALMA data also includes observations of the J=2-1
transitions of $^{12}$CO, $^{13}$CO, and C$^{18}$O, the J=5-4
transition of SiO, and the 3(0,3)-2(0,2) transition of H$_2$CO
(Table~\ref{table:almamaps}). To study the core and outflows, we used
our ALMA data from the compact figuration observations of these
molecular lines (with a synthesized beam of about 0.5\arcsec\
$\times$0.8\arcsec) as these data are more sensitive to extended
structures; the higher resolution data over-resolved some of the
outflow features.  Figure~\ref{alma-fov} shows an outline of the
primary beam of the ALMA observations superimposed on the HST [\Feii]
image. The locations of the five brightest 1.3 mm continuum sources
detected in the ALMA pipeline products are marked.

\begin{figure*}
\centerline{\includegraphics[angle=0,width=8cm]{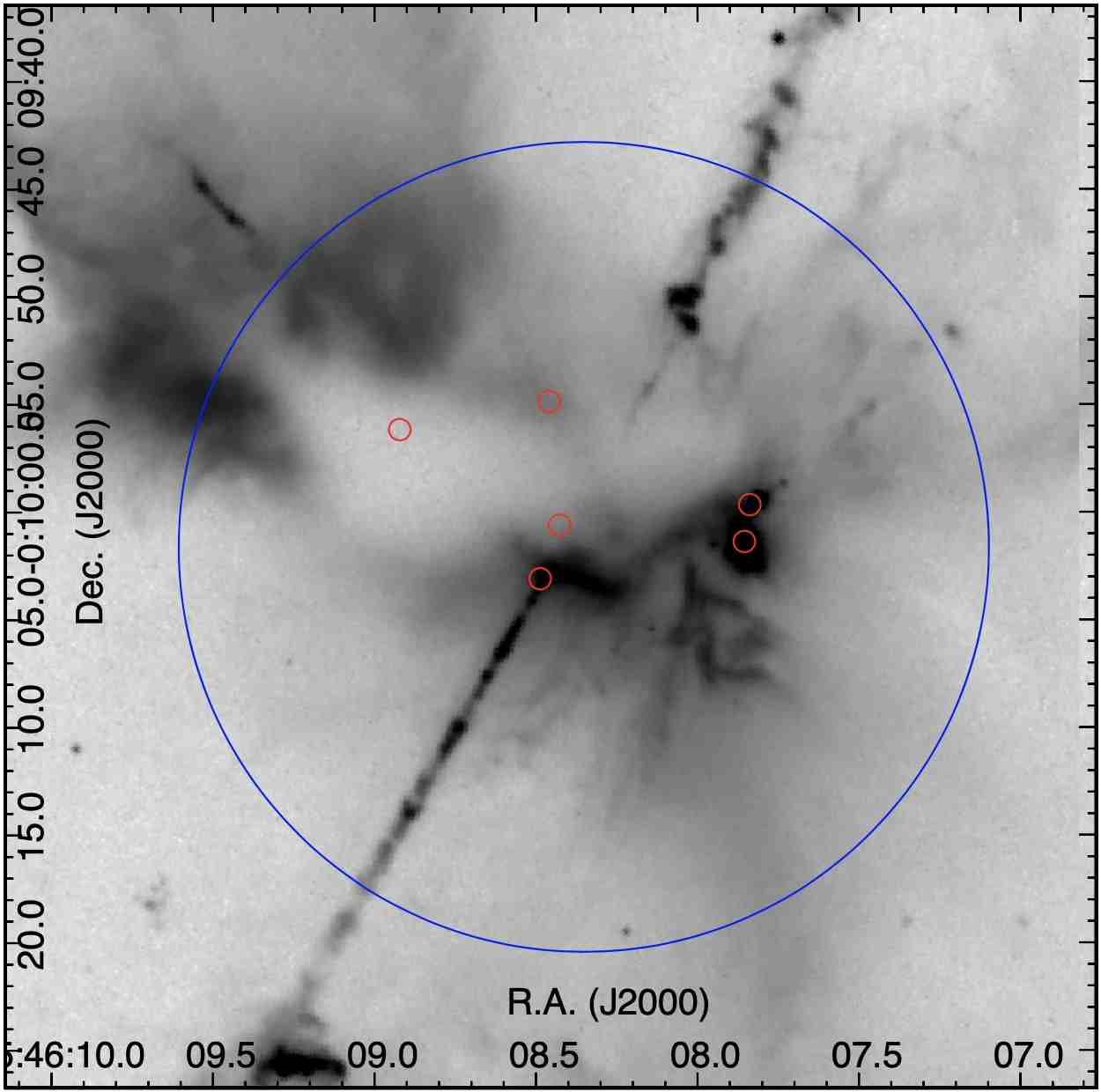}}
\caption{The ALMA primary-beam field of view overlaid on the HST
[\Feii] image.  The center of the ALMA observations is at 5:46:08.35
-00:10:01.7 (2000) and the radius of the field is 20~arcsec,
corresponding to where the sensitivity decreases to 20\% of that of
the phase center. The six sources detected by ALMA at 1.3mm continuum
are marked with red circles.
\label{alma-fov}}
\end{figure*}

\begin{figure*}
\centerline{\includegraphics[angle=0,width=17cm]{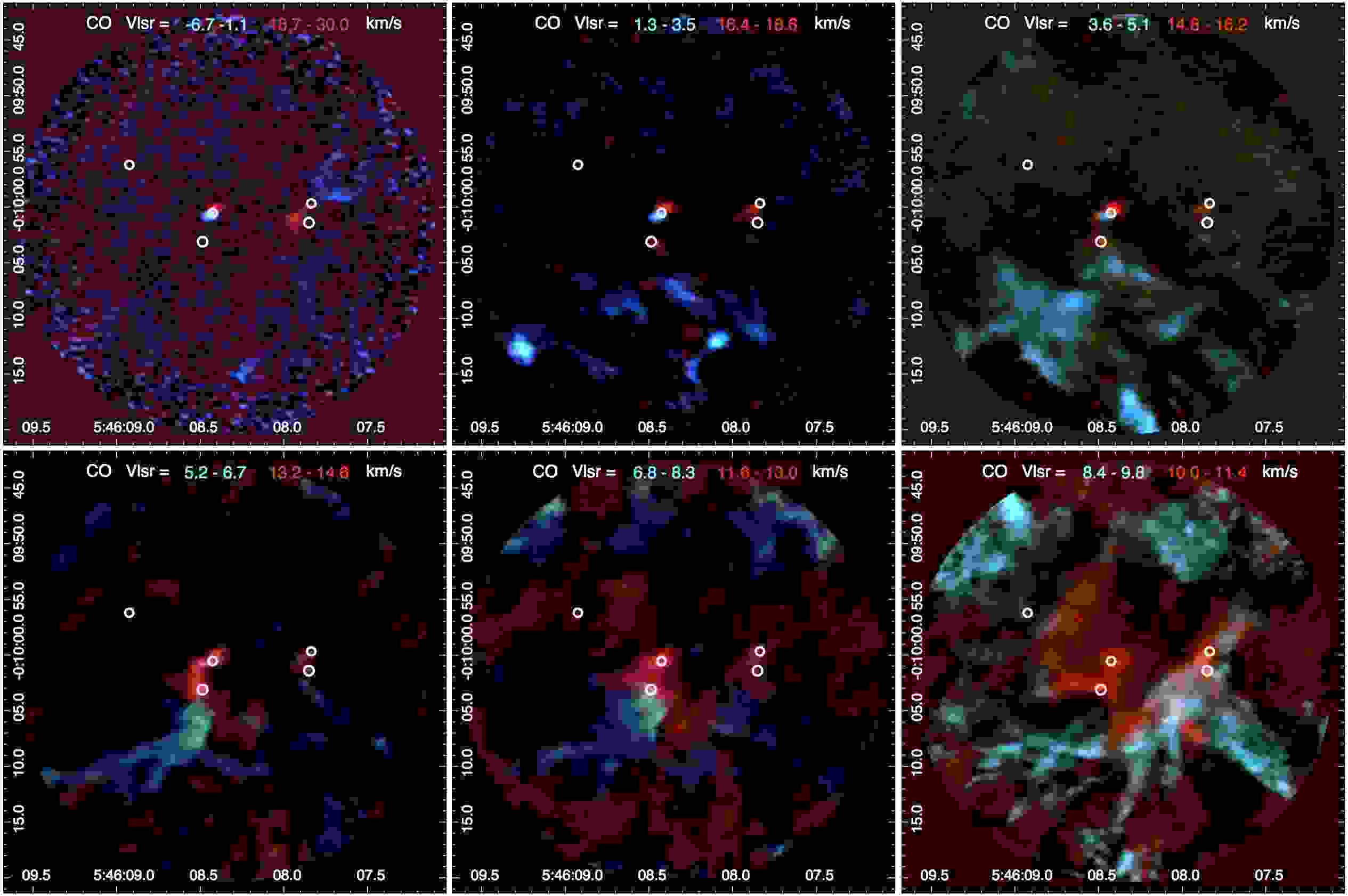}}
\caption{ $^{12}$CO mosaic of outflows in the HH 24 core region as observed with
ALMA in the range -6.7 $<$ V$_{lsr}$ $<$ 30~\kms, with the most extreme blue- and red-shifted velocities in the upper left panel, and the velocities closest to the core emission in the lower right panel. The observations were done with the 12m array and have a spatial resolution of about 0.5~arcsec. 
\label{12co-outflows-mosaic}}
\end{figure*}

\begin{figure*}
\centerline{\includegraphics[angle=0,width=17cm]{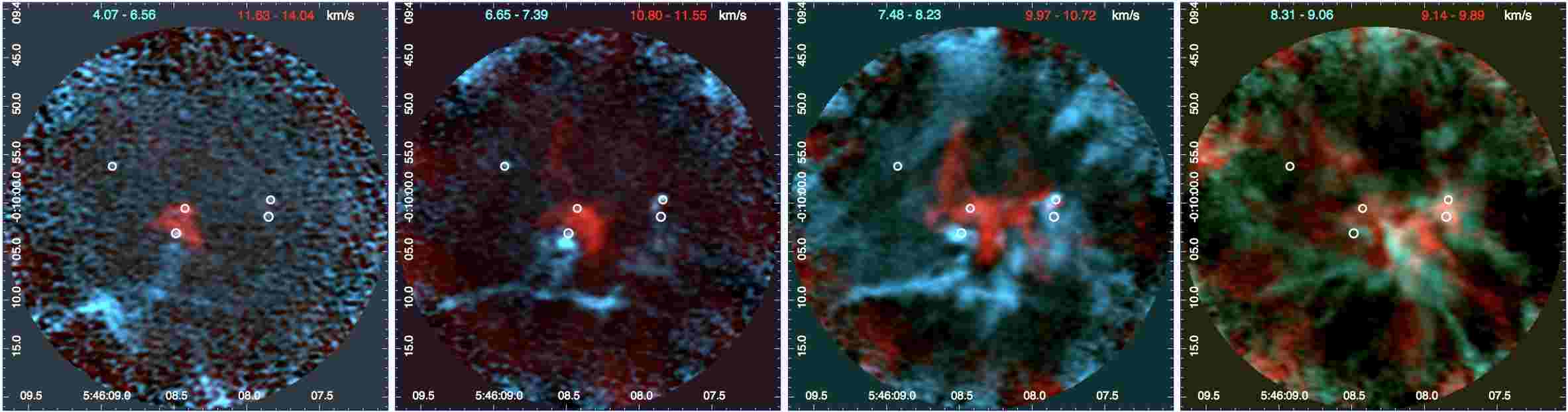}}
\caption{$^{13}$CO mosaic of outflows in the HH 24 core region as observed with
ALMA in the range 4.07 $<$ V$_{lsr}$ $<$ 14.04~\kms, with the highest
blue- and red-shifted velocities in the left panel, and the
velocities closest to the core emission in the right panel. The
observations were done with the 12m array and have a spatial
resolution of about 0.5~arcsec. 
\label{13co-outflows-mosaic}}
\end{figure*}


\begin{figure*}
\centerline{\includegraphics[angle=0,width=17cm]{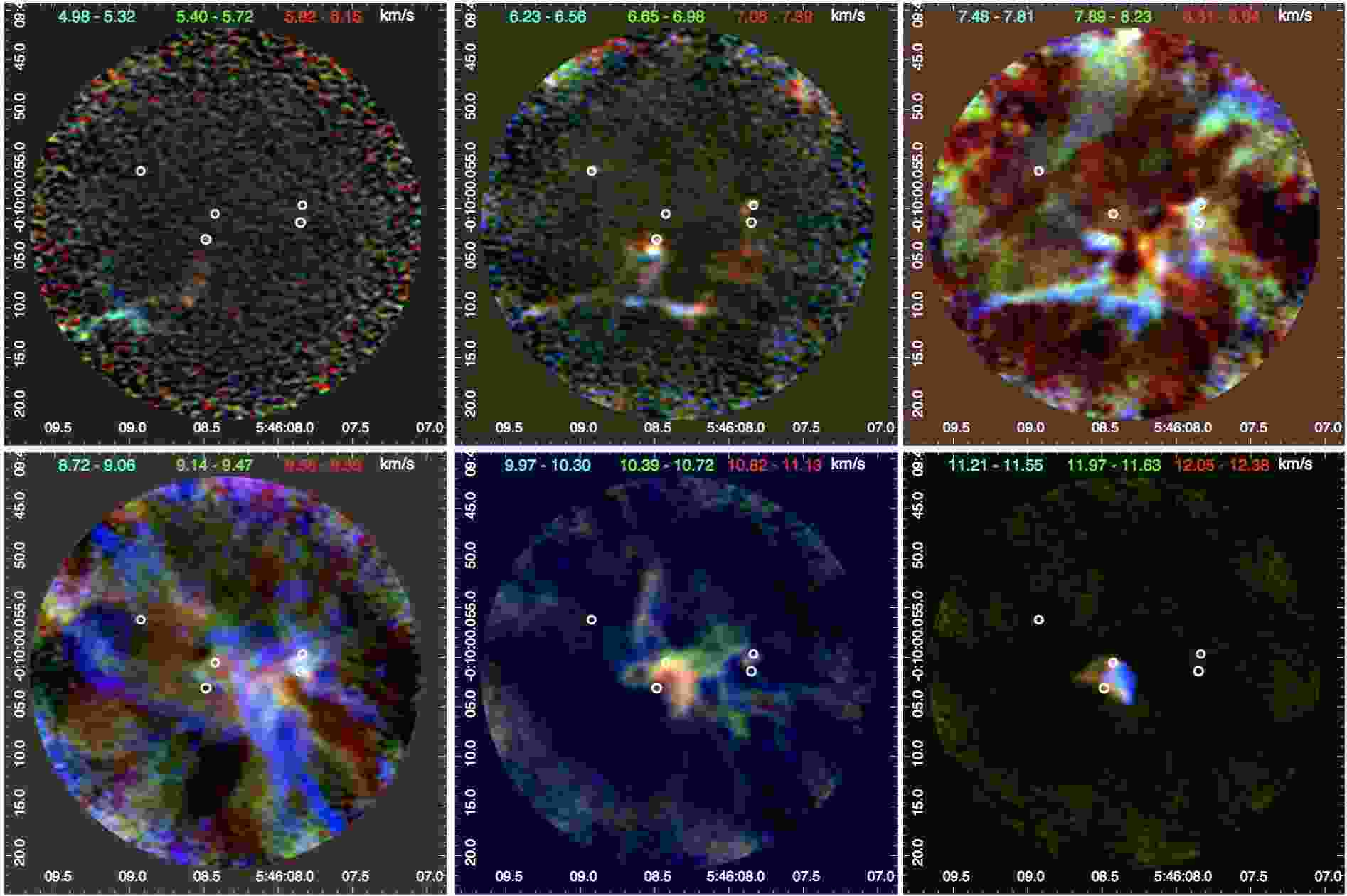}}
\caption{ ALMA $^{13}$CO channel maps of the HH 24 core region 
from V$_{lsr}$ = 4.98 to V$_{lsr}$ = 12.38~\kms, with a velocity
spacing between the panels of $\sim$1.25~\kms. Each panel shows three
velocities, in blue, green, and red as listed in each frame. The
observations were done with the 12m array and have a spatial
resolution of about 0.5~arcsec. 
\label{13co-mosaic}}
\end{figure*}


\begin{figure*}
\centerline{\includegraphics[angle=0,width=17cm]{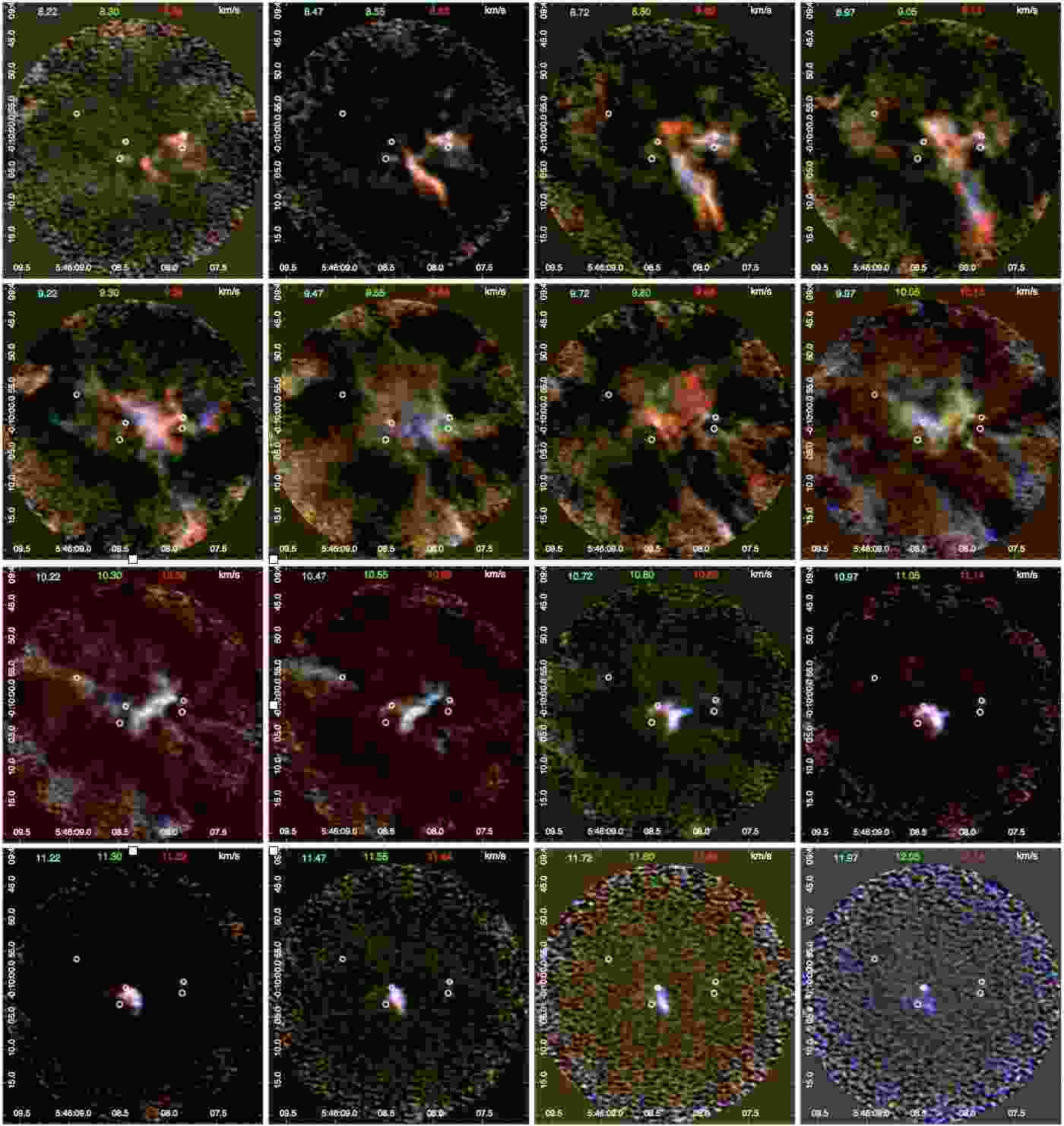}}
\caption{ ALMA C$^{18}$O channel maps of the HH 24 core region 
from V$_{lsr}$ = 8.22 to V$_{lsr}$ = 12.14 \kms. The velocity spacing
between the panels is $\sim$0.25 \kms. Each panel shows three
velocities, in blue, green, and red as listed in each frame.  The
observations were done with the 12m array and have a spatial
resolution of about 0.5~arcsec. 
\label{c18o-mosaic}}
\end{figure*}


\begin{figure*}
\centerline{\includegraphics[angle=0,width=17cm]{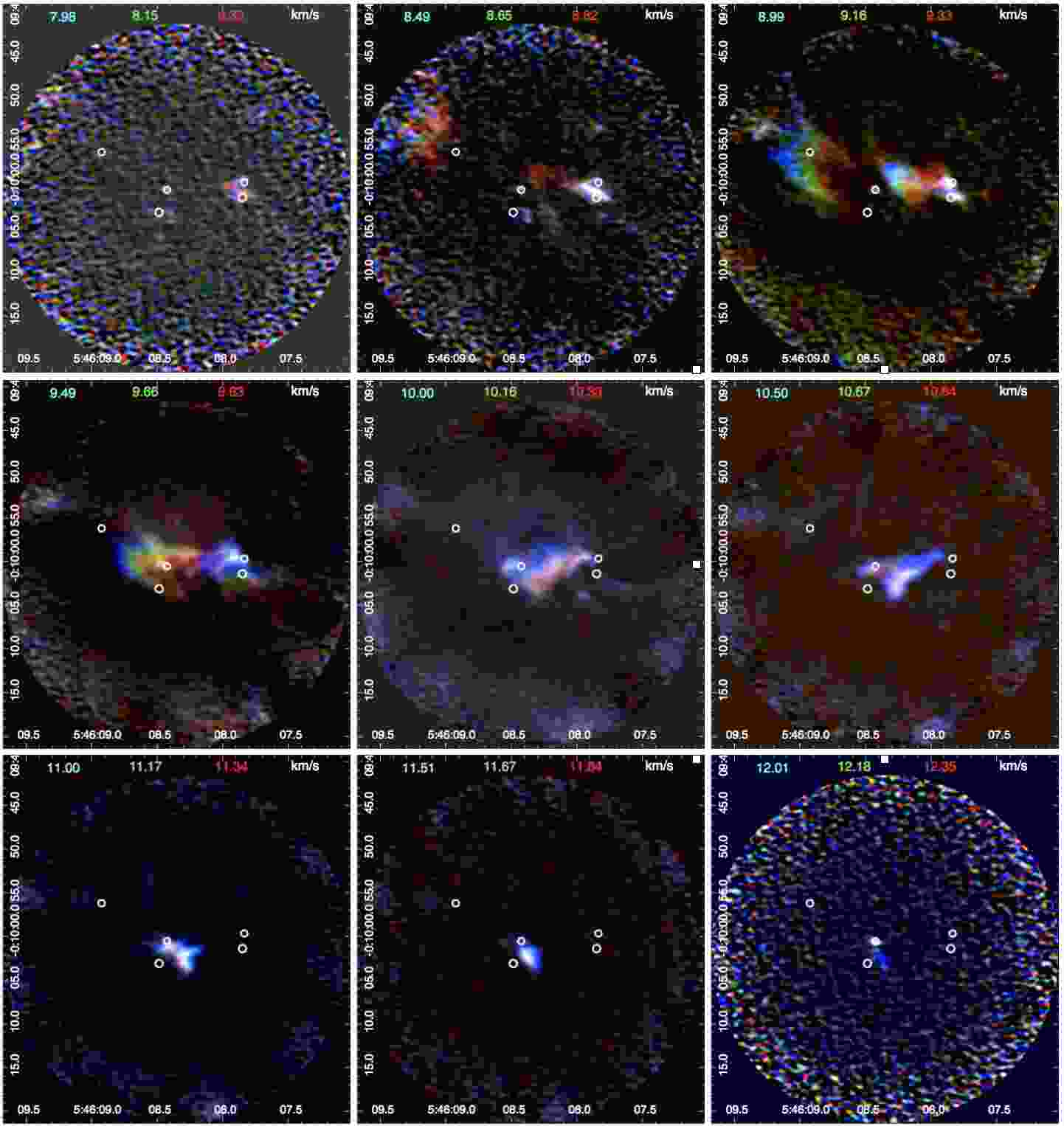}}
\caption{ALMA H$_2$CO channel maps of the HH 24 core region 
from V$_{lsr}$ = 7.98 to V$_{lsr}$ = 12.35~\kms. The velocity spacing
between the panels is $\sim$0.5~\kms. Each panel shows three
velocities, in blue, green, and red as listed in each frame.  The
observations were done with the 12m array and have a spatial
resolution of about 0.5~arcsec. 
\label{h2co-mosaic}}
\end{figure*}


\subsection{Overview of ALMA line data}\label{subsec:overview}

In nearby clouds, the \CO\ lines provide one of the best and most
commonly used tracers of molecular outflows.  \CO\ emission is
sensitive to molecular gas with a density n(H$_2$) $> 10^2$ \cmq .
Figure~\ref{12co-outflows-mosaic} contains six panels showing the velocity
structure of \CO\ emission optimized to show emission produced by
outflows.  Each panel shows both a redshifted and a blueshifted
velocity range indicated by the cyan and red labels at the top.  At
the largest red and blueshifted velocities (top-left), compact flows
are seen to be associated with the sources Eb and Wb.  These show
radial velocities of more than 15 km~s$^{-1}$ with respect to the
velocity of the SSV~63 cloud core in both lobes.
For velocities closer to the 9 to 10 \kms\ cloud velocity, the \CO\
emission becomes impacted by the high optical depth of the \CO\ line
and the loss of large-scale structure resolved out by the ALMA
interferometer.  Gas associated with outflows close to the core radial
velocity tends to be hidden behind the \CO\ photosphere.

Figure \ref{13co-outflows-mosaic} is similar to
Figure~\ref{12co-outflows-mosaic}, but for the $^{13}$CO line, with
each panel showing both a redshifted and a blueshifted velocity range.
with the red- and blueshifted emission from the highest speeds (left
panel) to the lowest speeds (right panel) with respect to the \tco\
line center.  This figure shows that also in the lower opacity \tco\ line,
low velocity flows and cavity walls associated with the outflows
powered by the HH~24 YSOs become apparent.

\tco\ emission is a tracer of molecular gas with a density n(H$_2$)$>
10^3$ \cmq .  Figure~\ref{13co-mosaic} shows the \tco\ emission in the
SSV~63 core covering the radial velocity range from 4.98 to 12.38 \kms.
Each panel shows three adjacent velocity ranges in blue, green, and
red indicated by the corresponding colored labels at the top of each
panel.  

\co\ emission is expected to be optically thin, thus
displaying the kinematic structure of those small-scale features with
a density n(H$_2$)$> 10^3$ \cmq\ that are not resolved out by the
interferometer. Figure~\ref{c18o-mosaic} shows the \co\ data cube as
three adjacent velocity channels in blue, green, and red from
V$_{lsr}$ = 8.22 to 12.14 \kms.

Formaldehyde (H$_2$CO) emission traces gas one to two orders of
magnitude denser than that traced by \co, \tco, and \CO .
Figure~\ref{h2co-mosaic} shows the H$_2$CO data cube as a mosaic where
each panel shows three adjacent velocity channels in blue, green, and
red from V$_{lsr}$ = 7.98 to 12.35 \kms.

\subsection{The MO1 Outflow from Source Eb}\label{subsec:MO1}

The diffuse continuum source Eb powers a compact, arc-second-scale
bipolar \CO\ outflow we label as Molecular Outflow 1 (MO1). MO1 can be
traced $\sim$2\arcsec\ (800 AU) from its source (Figure~\ref{mo1fig}).
The red lobe  is located north-northwest of Eb, while the blue
lobe  is to the south-southeast of the source.  The molecular
outflow axis is perpendicular to the diffuse disk surrounding Eb shown
in Figure \ref{ALMA-continuum}. The position-velocity diagram
(Figure~\ref{mo1-pv}) shows large velocity spikes at red and blue
velocities displaced by less than 1\arcsec \ from the position of
Eb. Given the compact nature of the outflow lobes, we do not see any
other clear velocity structure (i.e., dependence on distance from
source) in the p-v diagram.

The ALMA SiO data cube shows only one feature in the primary beam, a
compact knot of SiO emission associated with the northwest end of the
redshifted lobe of the MO1 flow.  The SiO is confined to a 1.6\arcsec\
by 2.8\arcsec\ region extending from the source to 5:46:08.317,
-0:09:59.68.  The SiO emission peaks at V$_{lsr}$ = 10.7 \kms\ at this
location (thick red circle in Figure \ref{mo1fig}).  A secondary peak
at this velocity nearly coincides with the source (thinner red circle
in Figure \ref{mo1fig}).  Between these two low-velocity peaks, the
SiO spectrum shows a fainter tail of emission extending to 19.3
\kms.

\begin{figure}
\centerline{\includegraphics[angle=0,width=6cm]{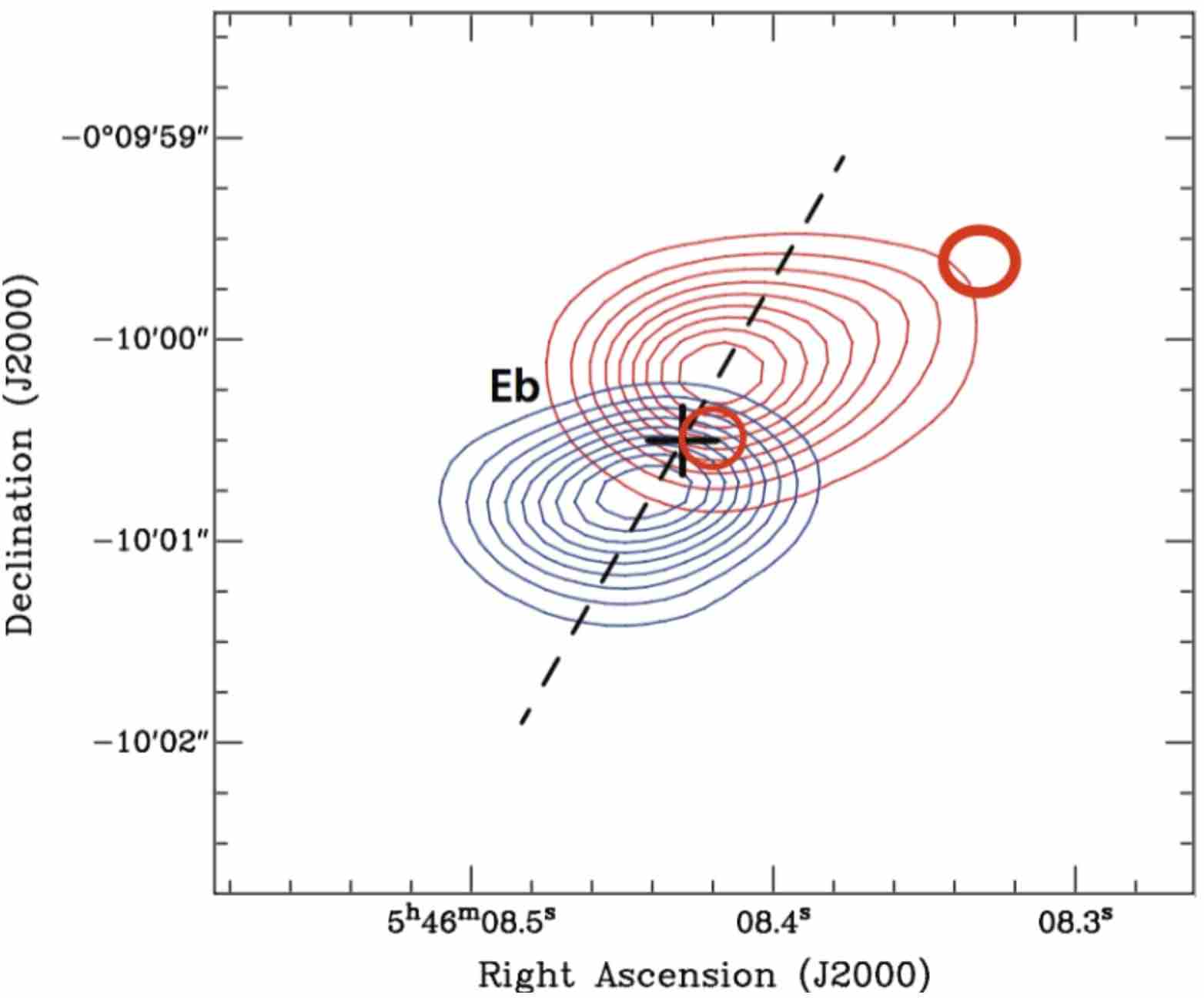}}
\caption{The compact $^{12}$CO molecular outflow MO1 from source Eb. 
 Blue contours show the integrated intensity emission over $-3.0 \leq
 V_{LSR} \leq 4.9$ \kms \/ (with first contour and contour steps of
 0.06 Jy beam$^{-1}$ km~s$^{-1}$), and red contours show the
 integrated intensity emission over $16.3 \leq V_{LSR} \leq 27.1$ \kms
 \/ (with first contour and contour steps of 0.1 and 0.09 Jy
 beam$^{-1}$ km~s$^{-1}$, respectively). The
dashed black line shows the direction along which the
position-velocity diagram shown in Figure~\ref{mo1-pv} is taken.  The
flow is perpendicular to the axis of the disk around source Eb. The
large red ellipse marks the primary peak of SiO emission, and the
smaller circle marks the secondary SiO peak. 
\label{mo1fig}}
\end{figure}

\begin{figure}
\centerline{\includegraphics[angle=0,width=6cm]{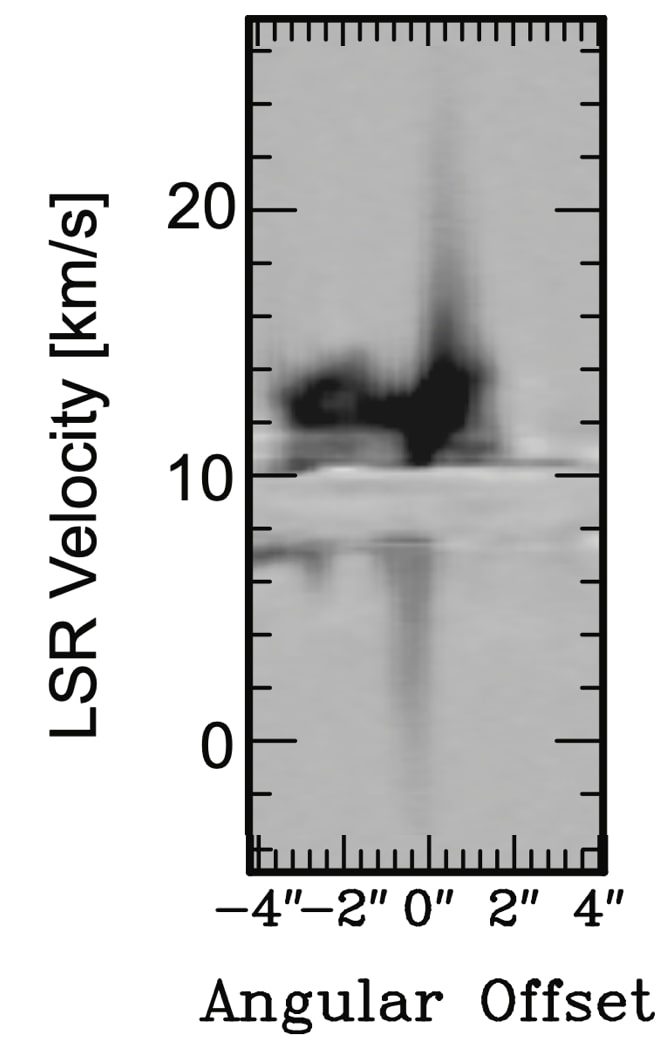}}
\caption{Position-velocity diagram of $^{12}$CO emission along the 
axis of the molecular outflow MO1 from source~Eb. 
\label{mo1-pv}} 
\end{figure}

The minor axis of the source Eb disk and the compact \CO\ outflow is
misaligned with respect to the prominent C and E jets.  Furthermore,
the CO emission has the opposite parity in Doppler shifts: while the
C~jet north-northwest of the SSV~63 core is blueshifted and the E jet
south-southeast of the core is red-shifted, the compact \CO\ flow from
source Eb has the opposite Doppler shifts.  Thus, there is no obvious
connection between molecular outflow MO1 with the E/C jet pair or any
other Herbig-Haro object or near-infrared emission line feature in the
SSV~63 core.

This \CO\ outflow exhibits the lowest and highest radial
velocities with respect to the SSV~63 cloud core in the entire ALMA
field and is the only source powering SiO emission.   The SiO
emission suggests that very recent outflow activity may be impacting
dense gas in the immediate surroundings of this YSO.  The lack of
obvious jets, HH objects or MHOs suggests that accretion and outflow
activity may have been very weak or absent in recent past, say within
the last few hundred or few thousand years.

\begin{figure}
\centerline{\includegraphics[angle=0,width=7cm]{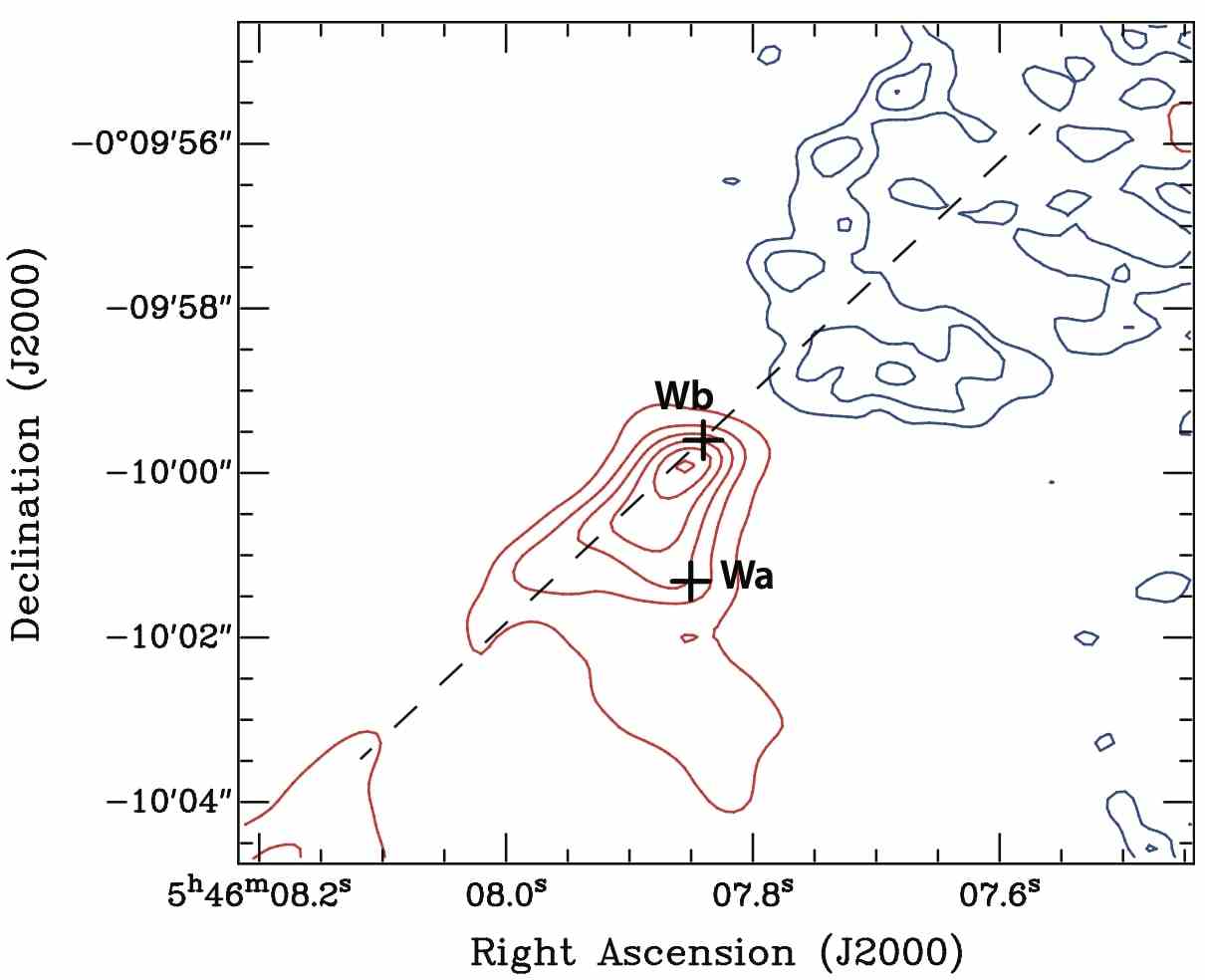}}
\caption{The $^{12}$CO molecular outflow MO2 from Wb. Blue contours
show the integrated intensity emission over $-12.6 \leq V_{LSR} \leq
-0.3$ \kms \/ (with first contour and contour steps of 0.045 and 0.02
Jy beam$^{-1}$ km~s$^{-1}$, respectively), and red contours show the
integrated intensity emission over $11.3 \leq V_{LSR} \leq 27.1$ \kms
\/ (with first contour and contour steps of 0.2 and 0.3 Jy beam$^{-1}$
km~s$^{-1}$, respectively).  The dashed black line shows the direction
along which the position-velocity diagram in Figure~\ref{mo2-pv} is
extracted.  
\label{mo2fig}}
\end{figure}

\begin{figure}
\centerline{\includegraphics[angle=0,width=5cm]{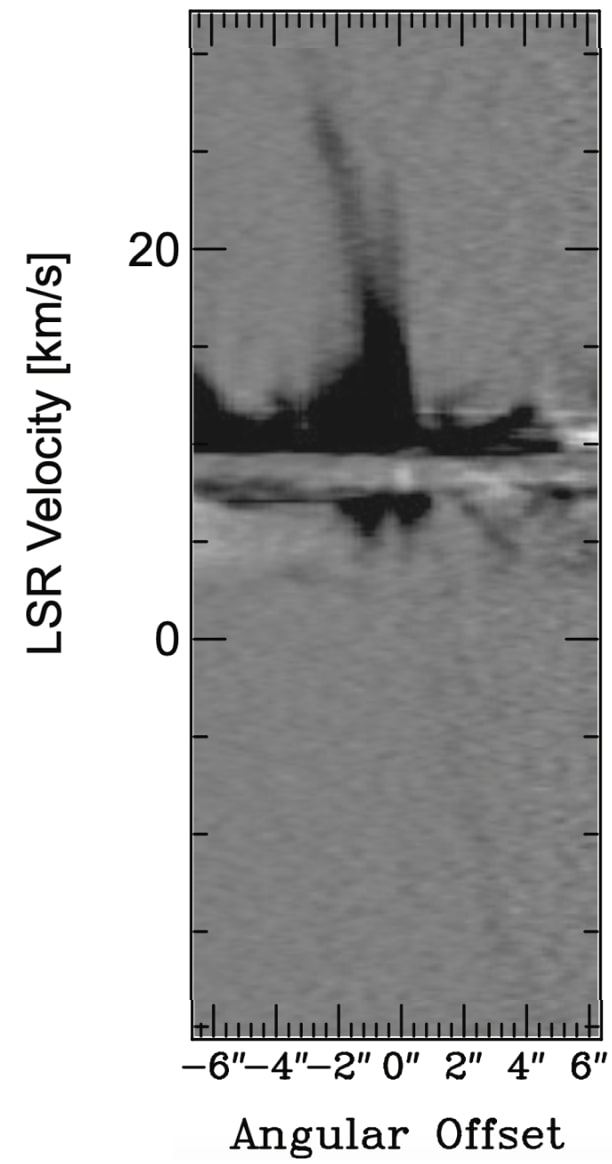}}
\caption{Position-velocity diagram of $^{12}$CO emission along the molecular outflow axis of the molecular outflow MO2 from Wb. 
\label{mo2-pv}}
\end{figure}

\subsection{The MO2 Outflow from Source Wb}\label{subsec:MO2}

The molecular outflow associated with source Wb, labeled MO2, is the
second most prominent molecular outflow from the SSV~63 sources.  MO2
has a significantly more extended morphology than that of MO1 and
exhibits red- and blue-shifted velocities to the southeast and
northwest of Wb, respectively (see Figure~\ref{mo2fig}). The axis of MO2 is approximately perpendicular to the Wb
circumstellar disk major axis. The morphology and kinematics of MO2
are similar to those expected from a molecular outflow formed by
entrainment by a wide-angle wind (as described in Lee et al.~2000).
At the highest velocities relative to the cloud rest-velocity
(upper-left panel in Figure \ref{12co-outflows-mosaic}) there is a compact cone
of redshifted \CO\ emission extending to the southeast, and a more
open cone of blueshifted emission extending towards the northwest
(Figure~\ref{mo2fig}).  The axes of symmetry of these small-scale \CO\
lobes is closely aligned with the orientation and parity of the
optical jet J.  However, the
\CO\ Doppler shifts are more than an order-of-magnitude lower than the
tangential velocities of the jet J knots, and the spatial extent of \CO\
emission that can be related to an outflow from source Wb is at least
two orders-of-magnitude smaller. The channel maps show discrete
emission (i.e., blobs) with higher velocity at larger distances from
the source, and the p-v diagram along the axis of MO2 shows
parabola-like structures (see Figure~\ref{mo2-pv}).

\subsection{The MO3 Outflow from Source N}\label{subsec:MO3}

 A third, clearly defined, compact molecular outflow is powered by
 source N. This compact molecular outflow (denoted MO3) is
 relatively collimated but asymmetric. There is clear redshifted
 emission associated with molecular outflow  from about
 1.5\arcsec \/ out to about 5\arcsec\ (2000 AU) from the source,
 whereas the blue lobe extends from the source out to only about
 1.5\arcsec\ (see Figure~\ref{mo3fig}). The position-velocity diagram
 (Figure~\ref{mo3-pv}) shows a velocity structure in the redshifted
 lobe commonly known as a ``Hubble-wedge'' and usually seen in
 molecular outflows formed through jet bow shock entrainment of
 ambient gas (see, e.g., Lee et al. 2000; Arce \& Goodman 2001). On
 the other hand, the velocity structure of the blue lobe is
 not as clearly defined as that of the red lobe.  There are no obvious
 connections of MO3 to any Herbig-Haro objects or near-IR emission
 line features.

\begin{figure}
\centerline{\includegraphics[angle=0,width=8.3cm]{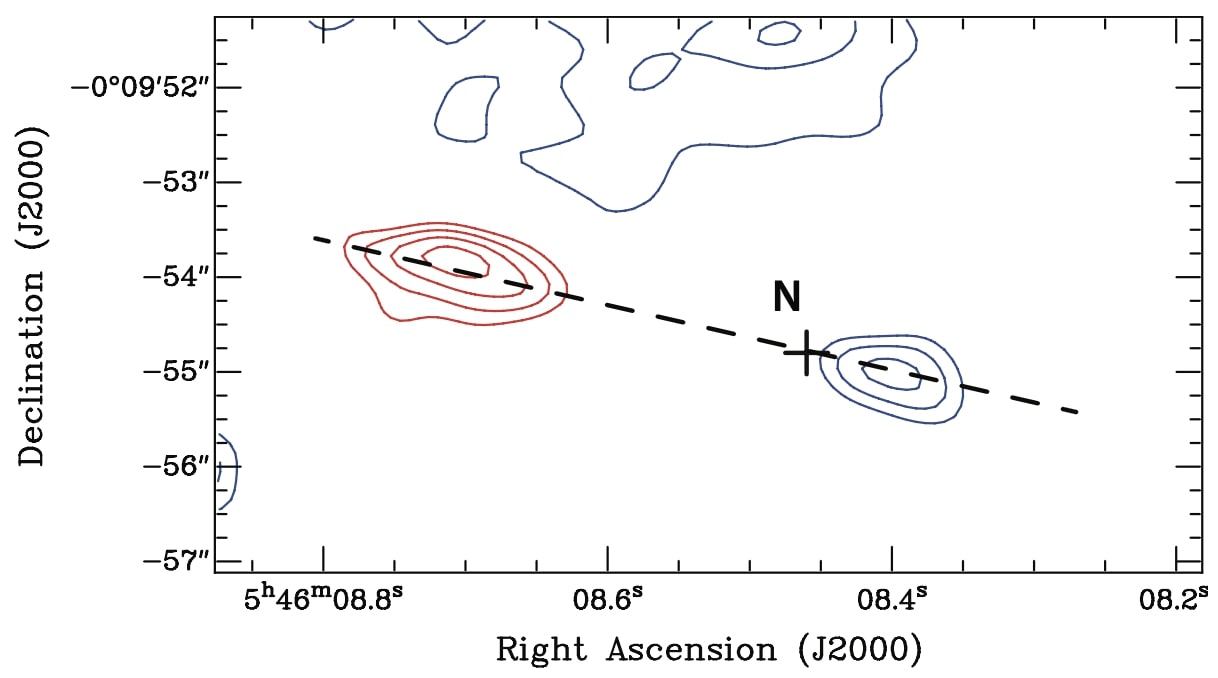}}
\caption{The $^{12}$CO molecular outflow from source N.
 Blue contours show the integrated intensity emission over $4.9 \leq
 V_{LSR} \leq 7.9$ \kms \/ (with first contour and contour steps of
 0.05 Jy beam$^{-1}$ km~s$^{-1}$), and red contours show the
 integrated intensity emission over $12.9\leq V_{LSR} \leq 17.3$ \kms
 \/ (with first contour and contour steps of 0.17 and 0.07 Jy
 beam$^{-1}$ km~s$^{-1}$, respectively). The dashed black line shows
 the direction along which the position-velocity diagram shown in
 Figure~\ref{mo3-pv} is taken.  
\label{mo3fig}}
\end{figure}

\begin{figure}
\centerline{\includegraphics[angle=0,width=8.3cm]{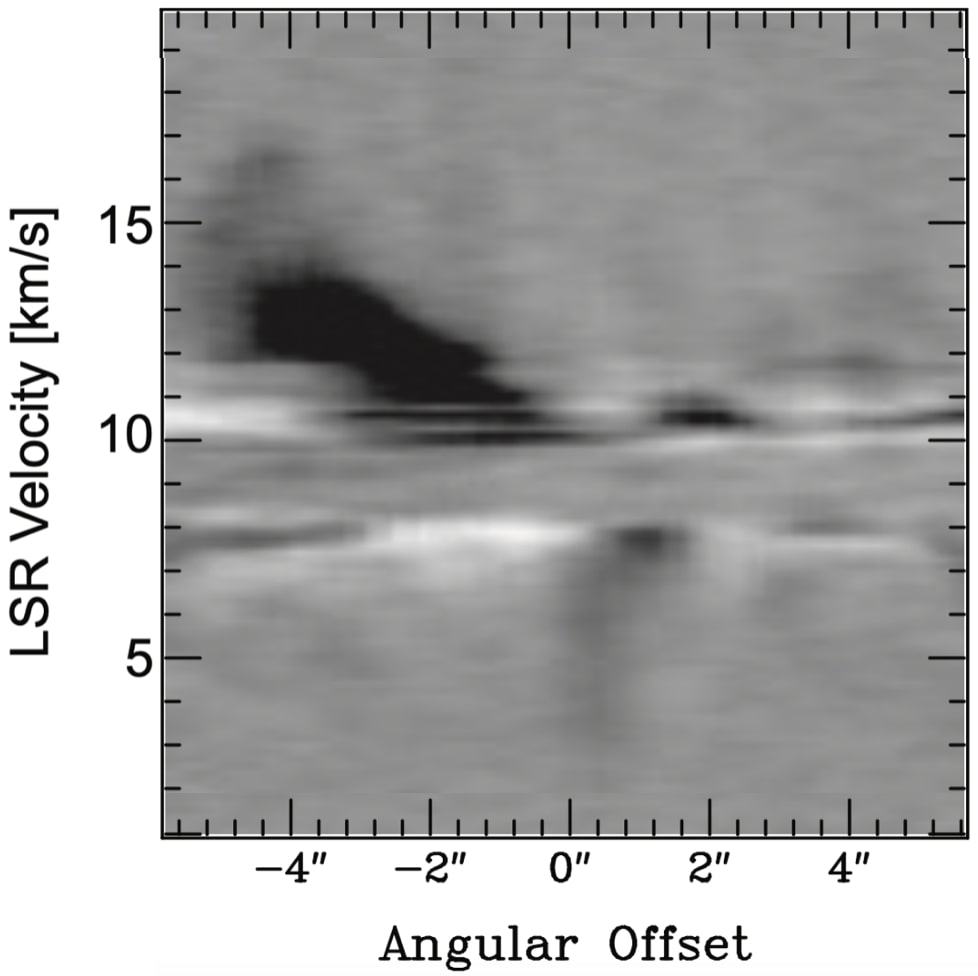}}
\caption{Position-velocity diagram of $^{12}$CO emission along the axis of the molecular outflow MO3 from source~N.   
\label{mo3-pv}}
\end{figure}

The low millimeter flux of source N, the non-detection of this YSO at
visual, IR, or radio wavelengths, combined with the presence of a
compact, low-velocity molecular outflow, suggests that it may be a
sub-stellar object. It could be the youngest of the active accretors
in the SSV~63 cloud core.

\subsection{Outflow from Source Ea?}\label{subsec:Ea-flow} 
 
Most of the low-velocity blueshifted \CO\ emission in the ALMA field
is concentrated in the southern half of the field and extends from
$V_{LSR} \sim $1 to $\sim$8~\kms, thus blueshifted relative to the
cloud velocity and, remarkably, opposite to the redshifted radial
velocity of the optical jet~E emerging from source Ea.
The most intense emission in this radial velocity range is
concentrated south of source Ea, where within about 5\arcsec\ of this
source the emission resembles a clumpy, low-velocity flow, see
Figure~\ref{12co-outflows-mosaic} and Figure~\ref{cavity-walls}.

\begin{figure*}
\centerline{\includegraphics[angle=0,width=18cm]{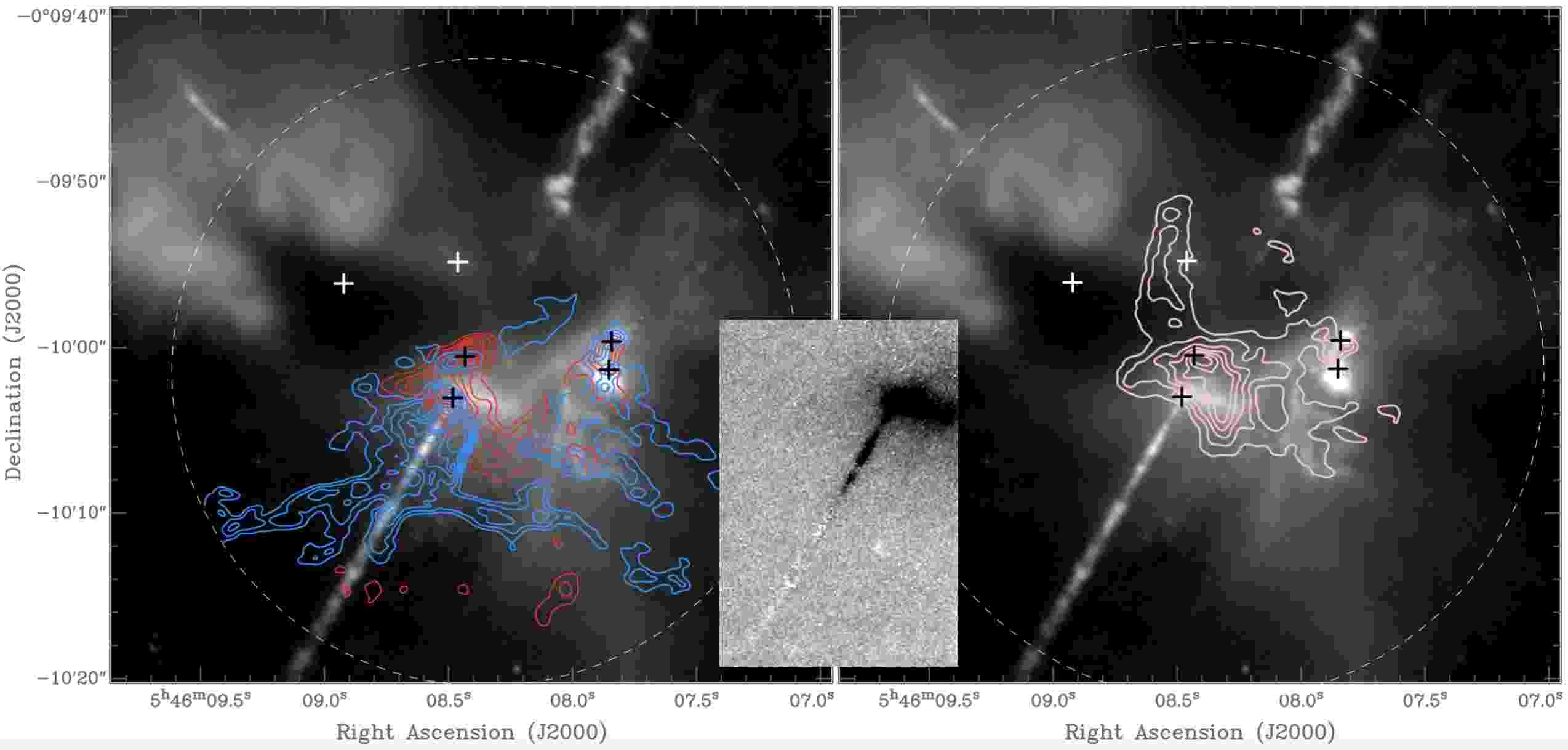}}
\caption{{\em (left)} 
Low-velocity \CO\ possibly associated with the walls of a cavity
surrounding the C and E jets from source Ea plotted on top of the
HST WFC3 (F164N filter) image of the region.  Blue contours show the
integrated intensity emission over $6.5 \leq V_{LSR} \leq 7.1$ \kms \/
(with first contour and contour steps of 0.04 and 0.05 Jy beam$^{-1}$
km~s$^{-1}$, respectively), and red contours show the integrated
intensity emission over $10.6 \leq V_{LSR} \leq 14.4$ \kms \/ (with
first contour and contour steps of 0.3 Jy beam$^{-1}$ km~s$^{-1}$).
Only emission south of declination -00:09:57 (J2000) is
shown. Emission that does not extend beyond 3\arcsec \/ of the map
edge is not shown as it is most likely noise from the low-sensitivity
edge of the map. Crosses show the position of the continuum
sources. 
{\em (right)} 
The \tco\ outflow cavity walls plotted on top
of the HST WFC3 (F164N filter) image of the region. Contours show the
integrated intensity emission over $10.0 \leq V_{LSR} \leq 11.8$ \kms
\/ (with first contour and contour steps of 0.052 and 0.07 Jy beam$^{-1}$
 km~s$^{-1}$, respectively). The dashed circle shows the field-of-view
 of the ALMA map, given by the distance from the center where the
 sensitivity decreases to 20\% of that of the phase center.
Emission within 6\arcsec \/ of the map edge is not shown as it is most
likely noise from the low-sensitivity edge of the map. Crosses show
the position of the continuum sources. 
{\em (center insert)} The HH~24~E jet has a dramatic change in the ratio
of [\Feii] and [\Sii] emission. As discussed in the text, this
primarily reflects changes in extinction. The insert shows the ratio
[\Feii]/[\Sii] of the HST images, such that black is [\Feii] strong
and white is [\Sii] strong. It is clear that the blueshifted $^{12}$CO
emission is associated with high extinction. 
\label{cavity-walls}}
\end{figure*}


The blueshift of the \CO\ emission south of source Ea suggests, in
light of the much faster redshifted velocities of the optical jet~E,
that the CO emission here represents gas that has been deflected
towards us by either a wide-angle wind surrounding the jet, or 
material was ejected at right angles from the axis.  As faster ejecta
in a velocity-variable jet overtakes slower material in the jet beam,
material can be ejected to the side.  Over time, the pressure of such
sideways moving ejecta or a wide-angle wind can create a wide-angle
cavity whose near-side walls would be expanding towards the observer.
As discussed in Section~\ref{subsubsec:structure-E}, the diminishing
ratio of [\Feii]/[\Sii] as jet~E moves away from its source indicates
a strong decline in extinction towards the observer. The middle panel
in Figure~\ref{cavity-walls} illustrates the decline in the
[\Feii]/[\Sii] ratio.

We conclude that source~Ea is not driving a major molecular outflow as
it emerges from the cloud core, but shows kinematic evidence for
either entrained or sideways splashing gas at blueshifted velocities.



It should be noted that close to, and southwest of, source Ea we also
detect faint redshifted $^{12}$CO emission at $V_{LSR} \sim 16$ to
18~\kms (middle upper panel of Figure~\ref{12co-outflows-mosaic}).  At
these velocities the emission is compact, extending to the southwest
only out to 1\arcsec \/ to 2\arcsec \/ from Ea. We note that VLA
X-band maps of source Ea shows evidence for a stubby extension
perpendicular to the axis of the E-jet, which is unlikely
to be from the circumstellar disk, since it is uncommon to detect
disks at the relatively low frequency of 10~GHz, so it is probably
another bipolar jet, indicating that Ea most likely is a close binary
(Figure~\ref{vla-composite}).


\subsection{Low-velocity Features and Outflow Cavities}\label{subsec:cavities} 

\subsubsection{The Region between Source Ea and Eb}

Between source~Ea and source~Eb there is a redshifted triangular
feature seen at around V$_{lsr}$$\sim$11 to 14 in both $^{12}$CO and
$^{13}$CO (Figures~\ref{12co-outflows-mosaic} and
\ref{13co-outflows-mosaic}). At slightly lower velocities
V$_{lsr}$$\sim$10.0 to 11.6 \kms\ (central two panels in
Figure~\ref{13co-outflows-mosaic}) there appears to be wide-angle,
U-shaped cavity walls opening up from source Eb towards the northwest
at PA $\sim$330\arcdeg.  At slightly higher velocities between 11.6
and 14 \kms, this U-shaped feature disappears.

The triangular redshifted feature northwest of source Ea and its
U-shaped extension (Figure~\ref{cavity-walls}-left) may trace the
receding, far-side of a wide angle cavity excavated over time by
either a wide angle wind or sideways splashing material surrounding
the C-jet ejected by source Ea. Support for this scenario comes from
the detection of both the blue- and red-shifted gas at $^{13}$CO
(Figure~\ref{13co-outflows-mosaic}).

Figure~\ref{cavity-walls}-right illustrates the relationship between the C
and E jets emerging from source Ea and the low-velocity cavity walls
traced by \tco .  At LSR velocities between 10.0 and 11.8 \kms\ (i.e.,
very low redshifted velocities), the \tco\ emission is concentrated in
the center of the ALMA field.  The emission peaks close to source Eb
and shows narrow, curved extensions to the north, south, east and west
of Eb that trace a pair of parabolic structures that open to the
northwest and the southeast. The axis of symmetry of these structures
is close to the orientations of the C and E jets.  The apex of the
south-facing feature coincides with the redshifted CO emission close
to Ea and thus likely traces the walls of the outflow cavity
associated with jet E.  The base (and center) of the northern parabola
is approximately at the position of Eb and its axis is coincident with
that of jet C.  Therefore, this structure likely traces the walls of
the cavity evacuated by the outflow associated with jet C.

The compact MO1 outflow and redshifted SiO emission northwest of and
driven by source Eb has an axis aimed more to the southeast and
northwest, and appears to be unrelated to the cavity walls discussed
above.





\subsubsection{Source Wa}


Near-infrared HST images show a bright compact reflection nebulosity
located about 1~arcsec south-east of source Wa which may trace an
outflow cavity (Figure~\ref{ssv63-h2}).  If so, source Wa may also
contribute to the generation of blueshifted \CO\ emission in the
southern part of the ALMA field.  A filamentary knot complex known as
HH~24B (Herbig 1974) is located a few arc-seconds south of source Wa
which may trace shocks where a wide-angle wind impacts the southern
part of the SSV~63 cloud core, see Figures~\ref{jetJ} ([\Sii]) and
\ref{embedded-S} (H$_2$).  The ALMA \CO\ map of outflow MO2 from source
Wb (Figure~\ref{mo2fig}) shows a wing of redshifted \CO\ emission
south of source Wa.
This may trace the
redshifted side of a wide-angle cavity surrounding the reflection
nebulosity and filamentary \Htwo\ and [\Feii] emission
south of source Wa.

\begin{figure}
\centerline{\includegraphics[angle=0,width=8.3cm]{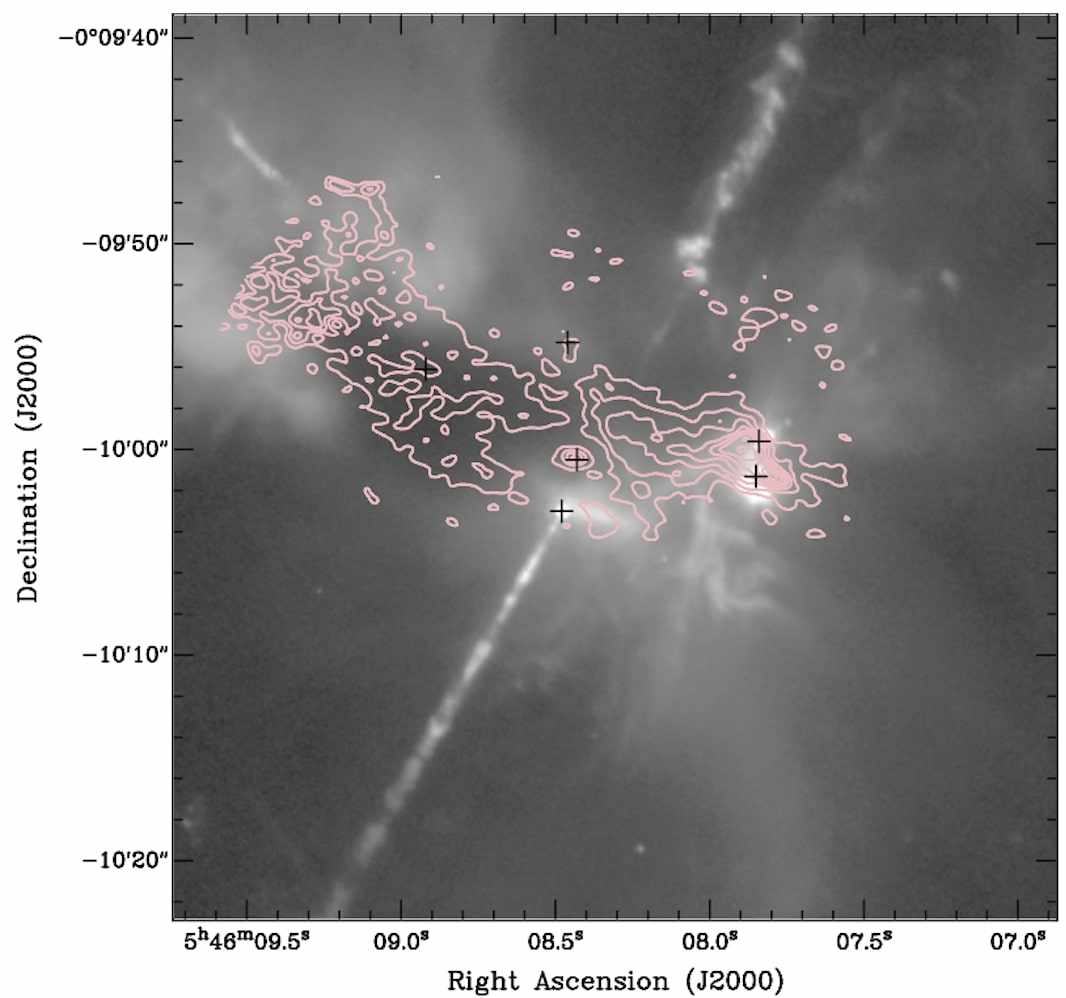}}
\caption{A low-velocity formaldehyde flow stretches towards the
north-east, seemingly following the base of the optical G-jet
emanating from source NE. The contours are integrated over V$_{lsr}$ from
8.6 to 9.7~\kms.  
\label{formaldehyde-flow}} 
\end{figure}

\subsection{Formaldehyde Kinematics}\label{subsec:formaldehyde} 

At low blueshifted velocities (from about 8.6~\kms\ to 9.7 \kms ) the
H$_2$CO emission is concentrated in a 6\arcsec \/ to 7\arcsec\ wide
structure extending from the sources Ea and Eb and to the
east-northeast up to the edge of the ALMA primary beam
(Figure~\ref{formaldehyde-flow}).

Figure~\ref{h2co-mosaic} shows a clear velocity gradient, of about 2.1
km~s$^{-1}$~pc$^{-1}$, along the structure towards the northeast with
decreasing velocity (i.e. greater blueshifted velocities away from the
central cloud velocity) at increasing distances from the field's
center.  This is especially evident in the upper right frame in Figure
\ref{h2co-mosaic}.  The feature shown there appears centered on source
NE and exhibits U-shaped cavities facing away from this source along
the axis of jet G propagating towards the northeast.  At the edge of
the H$_2$CO flow, faint \tco\ emission is detected at about V$_{lsr}$
$\sim$ 9.2 \kms\ which appears to trace the walls of the structure
seen in H$_2$CO.  Faint \co\ emission is detected along the center of
the H$_2$CO flow, at low redshifted velocities (from about 10.2 to
10.7 \kms ). The \co\ emission shows a velocity gradient where we, in
contrast, see higher redshifted velocities the further away from the
center of the field.



The interpretation of this formaldehyde flow is difficult. On
morphological grounds, it appears to be associated with the cavity of
the wide G-jet driven by source NE. The increase in velocity of the
formaldehyde flow with increasing distance (a 'Hubble-flow') from
source~NE could be caused by an explosion in this source. But it could
also simply reflect geometry of the background cavity wall, which
might be curving towards us. The projection into our line-of-sight of
the flow-vectors along such a curve could produce the observed
velocity field.

\subsection{An Infalling Streamer?}\label{subsec:streamer}

At $V_{LSR}$ from 8.2 to 9.8~km~s$^{-1}$ we see a 
filamentary structure, in both the $^{13}$CO and C$^{18}$O maps,
that extends from the field center out to the edge of the field (see
Figure~\ref{streamer-fig}). This structure, referred to as the
streamer, shows a velocity gradient in which the gas at larger radii
have, on average, lower blueshifted velocities compared to the gas
closer to the center of the field (see
Figure~\ref{streamer-pvfig}). This could be interpreted as infall from
the far-side of the SSV~63 core feeding its center.  The streamer is
aimed at source Eb in the center of the SSV~63 cloud core.


As an order-of-magnitude estimate, the total \co\ emission of the
streamer is roughly 5\% of the total \co\ emission seen in the ALMA
data. The mass of the HH~24 core has been measured as
$\sim$2.3~M$_\odot$ by K\"onives et al. (2020) (see
Section~\ref{sec:efficiency}), indicating that the streamer has a mass
of roughly 0.12~\Msol.  However, because ALMA resolves out most of the
extended background emission, this is an upper bound on the mass of
the infalling streamer.

Assuming an infall speed of 2 \kms, the infall time from 7,200
AU, corresponding to the angular radius of the ALMA field-of-view, is
$t_{in}\sim$17,000 years.  Thus, a rough upper limit to the mass accretion 
rate into the center is $\sim 7 \times 10^{-6}$ \Msol yr$^{-1}$.

\begin{figure}
\centerline{\includegraphics[angle=0,width=8.3cm]{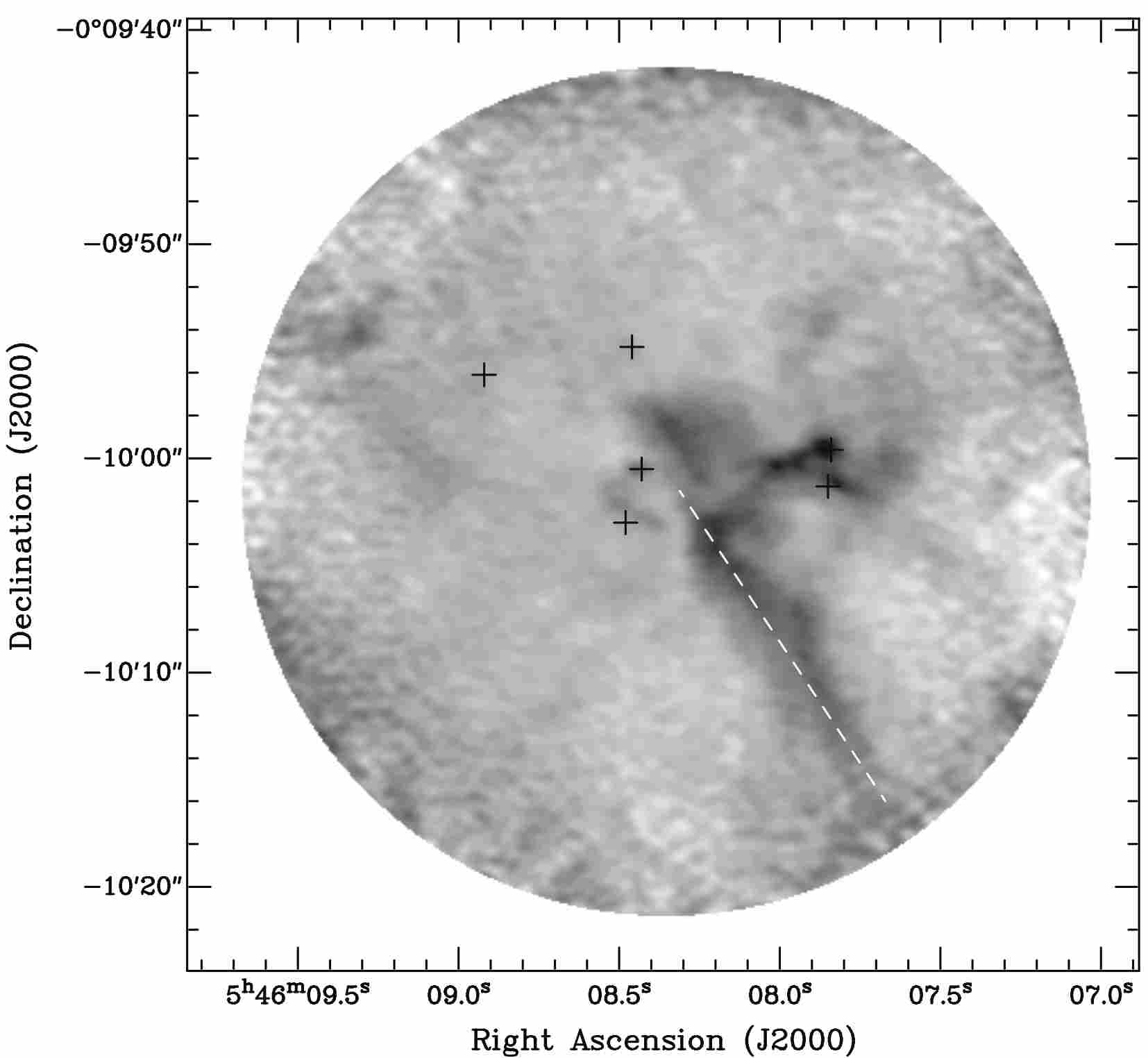}}
\caption{C$^{18}$O emission integrated over
blueshifted velocities shows a streamer feature towards the southwest
of the field. The dashed white line shows the direction along which
the position-velocity diagram in Figure~\ref{streamer-pvfig} is
taken. 
\label{streamer-fig}} 
\end{figure}

\begin{figure}
\centerline{\includegraphics[angle=0,width=8.3cm]{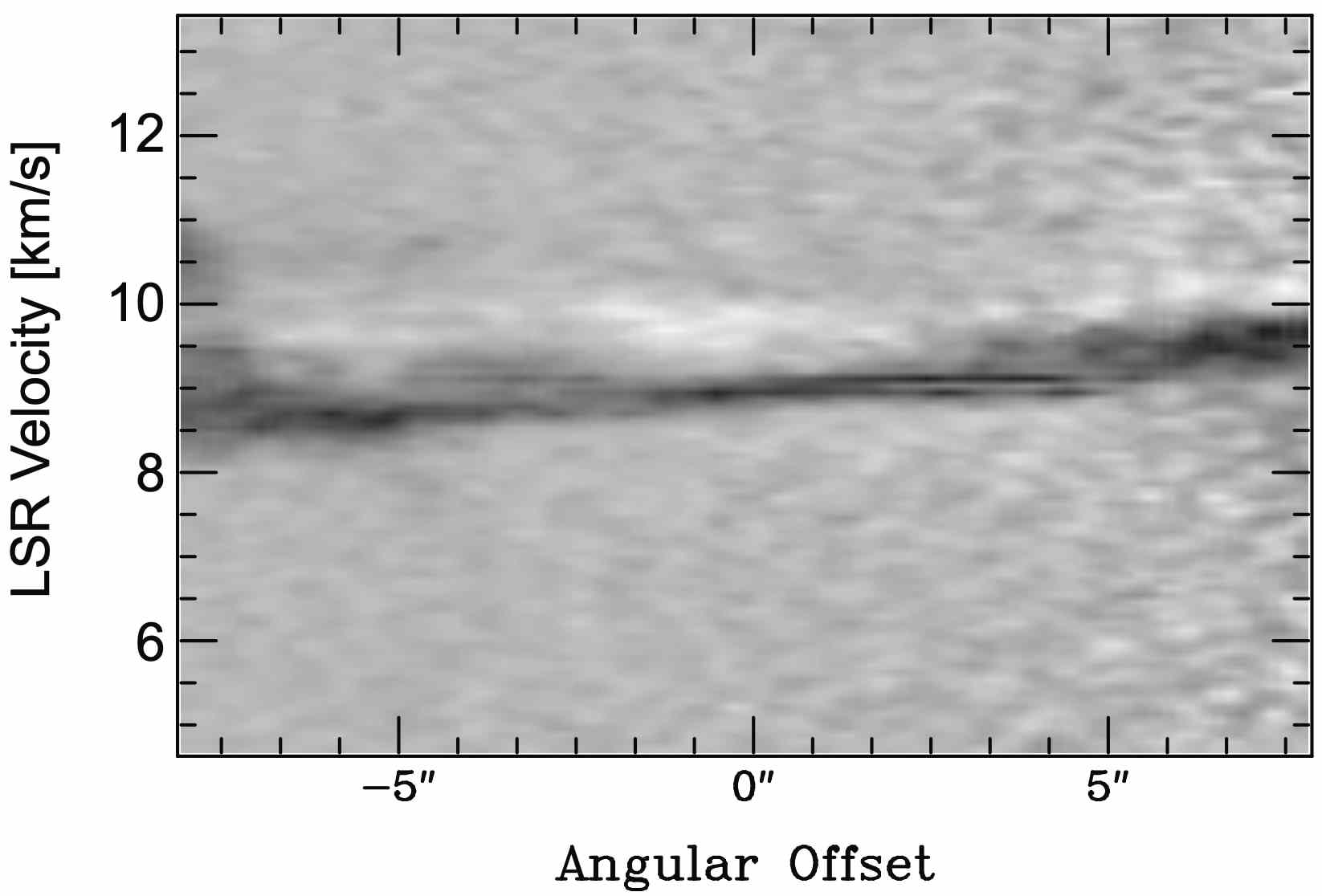}}
\caption{Position-velocity diagram of C$^{18}$O emission along the
major axis of the C$^{18}$O streamer (see Figure~\ref{streamer-fig}).
The center and the edge of the ALMA field is to the left and right,
respectively. 
\label{streamer-pvfig}}
\end{figure}

\subsection{Interpretation of ALMA Data}\label{subsec:interpretation}

The ALMA observations reveal several ultra-compact, bipolar molecular
outflows emerging from YSOs embedded in the SSV~63 cloud core.  The
detected flows emerge from the sources Eb (MO1), Wb (MO2), and N
(MO3).  The CO emission from these flows range in size from
$\sim$2\arcsec\ to $\sim$10\arcsec\ ($\sim$800 to 4,000 AU), one to
two orders-of-magnitude shorter than the chains of HH objects and MHOs
which trace the parsec-scale regions impacted by the jets emerging
from the SSV~63 core. Blue- and redshifted low radial-velocity,
``perturbations" to the \CO\ and \tco\ and the \HtwoCO\ line wings in
the SSV~63 cloud core appear to be linked to outflow activity from
sources Ea, Wa, and NE.  The radial velocities of these \CO\ and \tco\
outflows are one to two orders-of-magnitude slower than the proper
motions and radial velocities of the visual and near-IR wavelength
jets, with detectable molecular emission reaching a maximum V$_{lsr}$
of 15 to 20 \kms\ compared to the V$_{lsr}$ of the core.  As with many
other highly evolved Herbig-Haro outflows, associated \CO\ outflows
are confined to the size-scale of the remnant parent cloud core. These
relatively low-velocity outflow components are likely to be swept-up
gas from the parent cloud by the action of velocity-variable jets or
wide-angle winds that may surround the jets and be confined to outflow
cavity walls.

The association of specific jets with individual sources constrains
the evolutionary stages of the driving YSOs.  The C and E jets are
located at the base of the largest parsec-scale chain of HHs and MHOs
emerging from the SSV~63 core.  This giant flow consists of the MHOs
SSE2-east and SSE2-west located $\sim$14\arcmin\ (1.65 pc) south of source Ea
and the HH 20, 21 and NNW shocks $\sim$13\arcmin\ (1.45 pc) to the north.
Comparison of the ALMA and HST images shows that the northwestern base
of jet E coincides with the position of source Ea to within 0.1\arcsec
.  This implies that the southern lobe of this parsec-scale flow,
powered by the redshifted jet E, emerges from source Ea which has a
mass 1.9 - 2.0 \Msol , the second most massive YSO in the SSV~63 core.
Assuming a steady, average mass accretion rate of $10^{-5}$ \Msol
yr$^{-1}$, it would take $2 \times 10^5$ years to accumulate Ea's
mass, the second most massive YSO in the core.  Thus Ea may be the
oldest or second oldest YSO formed in this core; it continues to drive
active atomic and ionized jets indicating continuing accretion and
stellar growth.

The most massive YSO is source NE that likely powers the G jet and
the associated bow shock located $\sim$90\arcsec\ (0.17 pc) northeast of
the SSV~63 core.  The association of the G jet with source NE is
supported by the orientation of the NE disk, that has a minor axis closely
aligned with this flow.  There is no evidence for a larger,
parsec-scale flow from source NE, indicating that the recent outflow
activity responsible for the G jet and associated shocks followed an
extended period of no outflow activity by source NE prior to the
launch of the G jet.  Assuming a jet speed of 100 \kms , the
dynamical age of the most distant detected bow shock at the head of the G
jet is only $\sim$1,700 years.

\vspace{0.3cm}

\section{DISCUSSION}\label{sec:discussion}

\subsection{The Formation of Jets}\label{subsec:binaries}

In the 70 years since the Herbig-Haro phenomenon was discovered
(Herbig 1950,1951, Haro 1952,1953) the fundamental physical processes
involved have been gradually established (Schwartz 1983, Reipurth \&
Bally 2001), as well as the properties of the molecular outflows that
result from entrainment by the jets of the surrounding molecular
clouds (Bachiller 1996, Bally 2016). There is general agreement that
jets are launched when accreted matter interacts with magnetic fields
within a few AU in the star-disk region, although the specific details
of models vary greatly, see, e.g., Frank et al. (2014) for a
review. In common for all these models is the issue of what triggers
the accretion of matter to the central zone. A number of disk
instability mechanisms have been identified that will lead to
accretion with a concomitant outflow. Reipurth
(2000) postulated that the {\em giant, parsec-scale} HH jets are
driven by disk-instabilities induced by close periastron passages
during the chaotic motions of the components of newborn
non-hierarchical stellar systems, thus force-feeding the jet
engine. This is in contrast to many small jets seen from single stars,
which may result from internal disk instabilities.


\subsection{Breakup of the SSV 63 Multiple System}\label{subsec:breakup}


The SSV~63 stellar group is a prototypical multiple system in a
non-hierarchical configuration. It is an example of the exceedingly
high stellar densities that can be associated with stellar birth: the
stellar density of the HH 24 sources is estimated at about 4 $\times$
10$^5$ pc$^{-3}$, which is a factor of roughly 1000 times the stellar
density in the center of globular clusters.
This naturally leads to powerful dynamical interactions, and
consequently such systems break up on timescales of about 100 crossing
times (e.g., Valtonen \& Mikkola 1991). Numerical simulations show
that half of all break-ups occur during the embedded phase (Reipurth
et al. 2010), lasting about 500,000~yr (Evans et al. 2009), which we
then adopt as the upper limit for the age of the SSV~63 system.

The discovery that a low-mass young object, the borderline brown dwarf
SSV~63~H$\alpha$~5, has been ejected from the SSV~63 multiple system
about 5800~yr ago demonstrates directly the dynamical nature of this
little group of protostars. With an upper limit of, say, 500,000~yr
for the age of the SSV~63 multiple system, it is remarkable that we
find a runaway star precisely during the last $\sim$1\% of the age of
the system. Either this is plain luck, or the ejection of low-mass
members of the system is a more commonly occurring phenomenon. The
top-heavy distribution of masses in SSV~63 might be an indicator that
many other very low mass objects have been ejected during the lifetime
of the system. Our search for runaway stars was limited to a small
area about 10 arcmin around SSV~63. But if an object had been ejected
500,000 yr ago with a velocity of 25~\kms\, then in principle it could
by now have travelled more than 4 degrees. Once Gaia~DR4 is released,
the proper motion uncertainties will be sufficiently low that a
meaningful association with more distant objects can be established.

The escape speed from the SSV~63 system and its core is about
1.5~\kms. Objects ejected with a lower speed will remain loosely
tethered to the system, and will after a while return to the system,
where numerical simulations suggest that they will be ejected again,
until they eventually are kicked out with a velocity higher than the
escape speed. Such almost-escapers can travel substantial distances
before falling back. Given the ejection of H$\alpha$~5 within the very
recent past, it appears likely that there could be a number of other
both escaping and returning bodies that were once members of the
SSV~63 multiple system.  The ejection of low-mass cluster
members has also been observed in regions of high mass star formation
(Orion BN/KL, G\'omez et al. 2008; W49 North, Rodr\'{i}guez et
al. 2020).

The energy for an ejection from an unstable triple system is acquired
by shrinking the separation of two members. Usually the lowest mass
member is ejected, and the two remaining members become bound into an
eccentric binary. But occasionally a low-mass binary is ejected
leaving behind a more massive member. If the triple is part of a
larger multi-body system, the recoil of the remaining binary (or
single) will add to the velocity dispersion of the system, and thus
facilitate further break-up.

It follows that several of the SSV~63 components are likely to be
close binaries. This is then consistent with the observation that a
number of collimated jets are emanating from SSV~63 as a result of the
inspiraling of binaries. Also we note the presence of what appears to
be a quadrupolar radio continuum jet from source Ea, indicating that
Ea is a close binary system. A similar quadrupolar radio morphology was
found for HH~111 (Reipurth et al. 1999).














\subsection{The Fate of the SSV~63 Multiple System}\label{subsec:fate}

The non-hierarchical configuration of the SSV~63 multiple system
implies that the system will inevitably undergo a dynamical
transformation towards a hierarchical configuration, in the process
likely losing several of its present members. There is evidently no
way to predict the details of such a highly stochastic process, but
one can approach the issue in a statistical manner. We have carried
out numerical simulations using the N-body code described in detail by
Reipurth \& Mikkola (2012, 2015), except that a cloud core and
accretion were not included. We model the SSV~63 system in an XYZ
coordinate system, where XY is the plane of the sky, and we have
assumed that the multiple system is as deep along the Z line-of-sight
as it is across the XY-plane, that is, about 6000~~AU. We fix the six
bodies\footnote{The simulations were performed before the seventh
member, N, was discovered with ALMA} at the observed XY positions and
randomly assign Z-values to each of the components in the range
$\pm$3000~AU. We assume that the individual components have randomly
oriented velocity vectors of 1~\kms\ corresponding to the velocity
dispersion in a typical turbulent cloud.  This is supported by the
radial velocity differences of the stars measured by ALMA
(Table~\ref{table:line-fit}).  All bodies are assumed to have equal
masses.  We then run the code 1000~times for 100~Myr and review the
end products at 1, 10, and 100~Myr. The results are the same within
the uncertainties at 1, 10, and 100~Myr. For 1~Myr the values are as
follows: Single bodies: 2919 (69.3\%); Binaries: 958 (22.7\%);
Triples: 319 (7.6\%); Quadruples: 19 (0.4\%); Higher-order systems:
none. The following conclusions can be drawn from these numbers:

(1): Since no system with an order higher than 4 survives, and even
     those are very rare, it follows that the sextuple SSV~63 system is
     almost certainly doomed to disintegrate.

(2): We note that 1000 simulations of six stars should lead to 6000 
    classifications in the above system categories, but the numbers do
    not add up to 6000, i.e. some stars are unaccounted for. While the
    simulations are very precise, in about 10\% of the cases the
    analysis code that classifies the outcome cannot determine
    whether a nearby pair of stars are bound or not. For example two
    stars may be ejected in separate events and move close to each
    other, but it is not clear if the pair is bound or will become
    bound as the result of passing close to a third star. Such cases
    are not counted by the analysis software.

(3):  A comparison between these simulations and observations of
    multiplicity (e.g., Raghavan et al. 2010) shows that in our
    simulations singles are overrepresented and triples are
    underrepresented. This is not surprising since our simulations do
    not include the molecular environment and the related dissipative
    processes that tend to bind pairs of stars into binaries and
    triples.

(4): The virtually unchanged numbers of singles, binaries, triples, and
     quadruples at 1, 10 and 100~Myr shows that the dynamical
     evolution is essentially complete within the first
     million years. It follows that the SSV~63 system is presently in
     a highly dynamical and unstable situation, as expected from its
     multi-component non-hierarchical configuration.

\begin{figure}
\centerline{\includegraphics[angle=0,width=8.3cm]{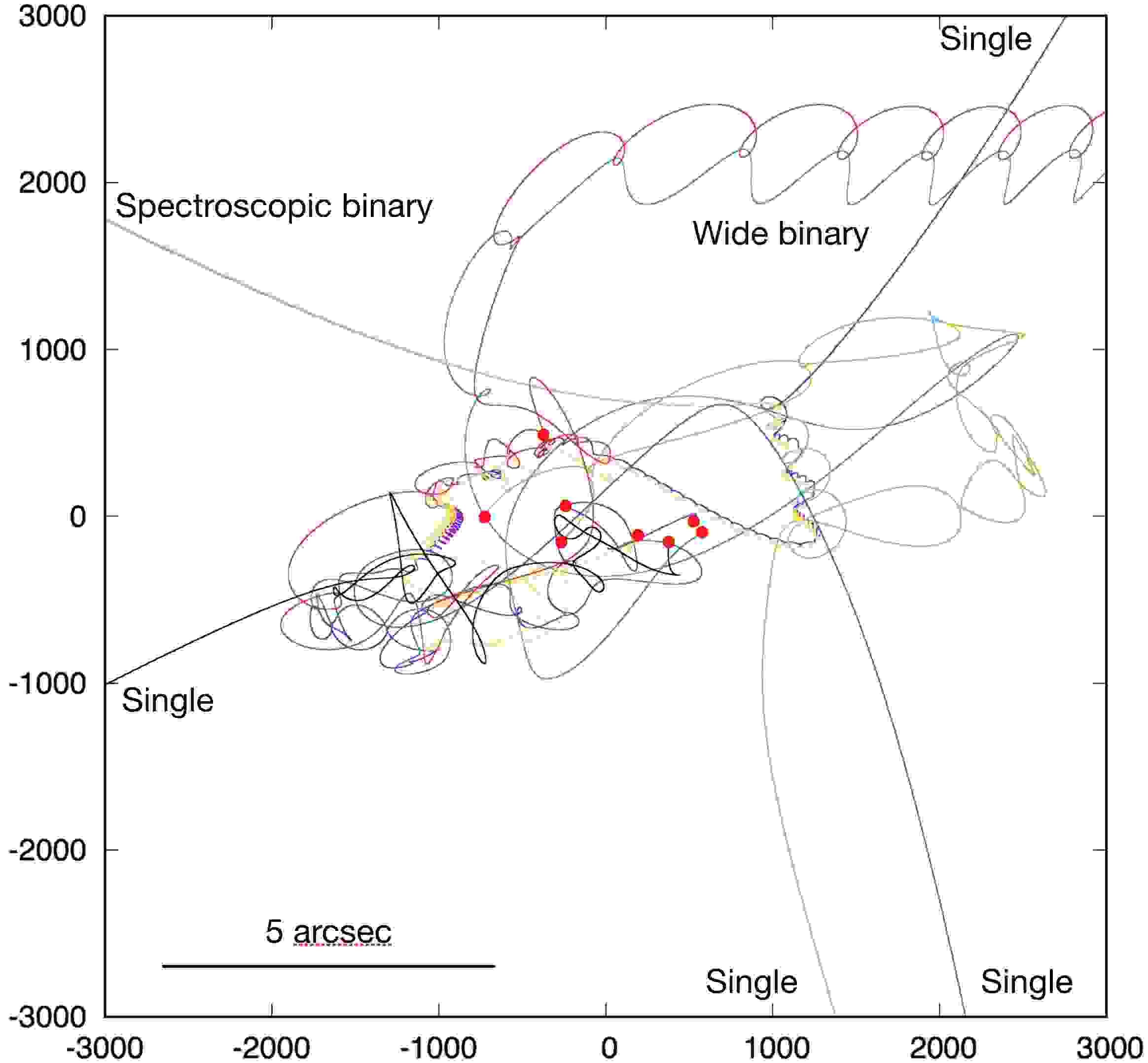}}
\caption{An example of a numerical simulation of an unstable eight-body 
equal-mass system illustrating the chaotic nature of the
interactions. The decay products are single stars as well as a
close and a wide binary. The tickmarks are in AU, and the angular scale
assumes a distance of 400~pc.
\label{eight-body}}
\end{figure}

Similar results are obtained when running the code for higher-order
systems. An example of an 8-body system is shown in
Figure~\ref{eight-body}, in which a system with dimensions comparable
to SSV~63 completely disintegrates within a million years.

SSV~63 is a specific case illustrating the dynamical
evolution of small multiple systems. On a more general level, since
numerical simulations show that about half of all ejections occur
during the embedded phase while stars are still building their masses
(Reipurth et al. 2010), it follows that dynamical interactions in
small multiple systems play an important role in setting the masses of
stars. Early ejections will in some cases lead to the formation of
brown dwarfs (Reipurth \& Clarke 2001), and later ejections at random
times will play an important role in shaping the initial mass function
(e.g., Bate \& Bonnell 2005).  

\subsection{Flybys and Disk Structure} 

It has been known for some time that dynamical interactions in young
binaries can have profound effects on circumstellar disks, as
recognized in the seminal work of Clarke \& Pringle (1993). Similarly,
flybys in clusters can warp and truncate disks (e.g., Pfalzner 2003,
Pfalzner \& Govind 2021). Additionally, small embedded clusters are
subject to ram pressure stripping from their passages through the
ambient medium (Wijnen et al. 2017).  Such effects are particularly
pronounced in small non-hierarchical multiple systems still embedded
in cloud cores, where stars chaotically move around each other on
short time scales.  With modern smoothed particle dynamical
simulations such perturbations can be studied in great detail. Recent
simulations by Cuello et al. (2020a,b) illustrate the various effects
in detail. Among the observable signatures of such dynamical
interactions are spiral arms, disk warping, diffuse halos of material
pulled from disks, and disk truncation. High-resolution observations
with ALMA, like the DSHARP project (Andrews et al. 2018), are able to
detect such features, and they have been seen in several multiple
systems (e.g., Kurtovic et al. 2018).

\begin{figure}
\centerline{\includegraphics[angle=0,width=8cm]{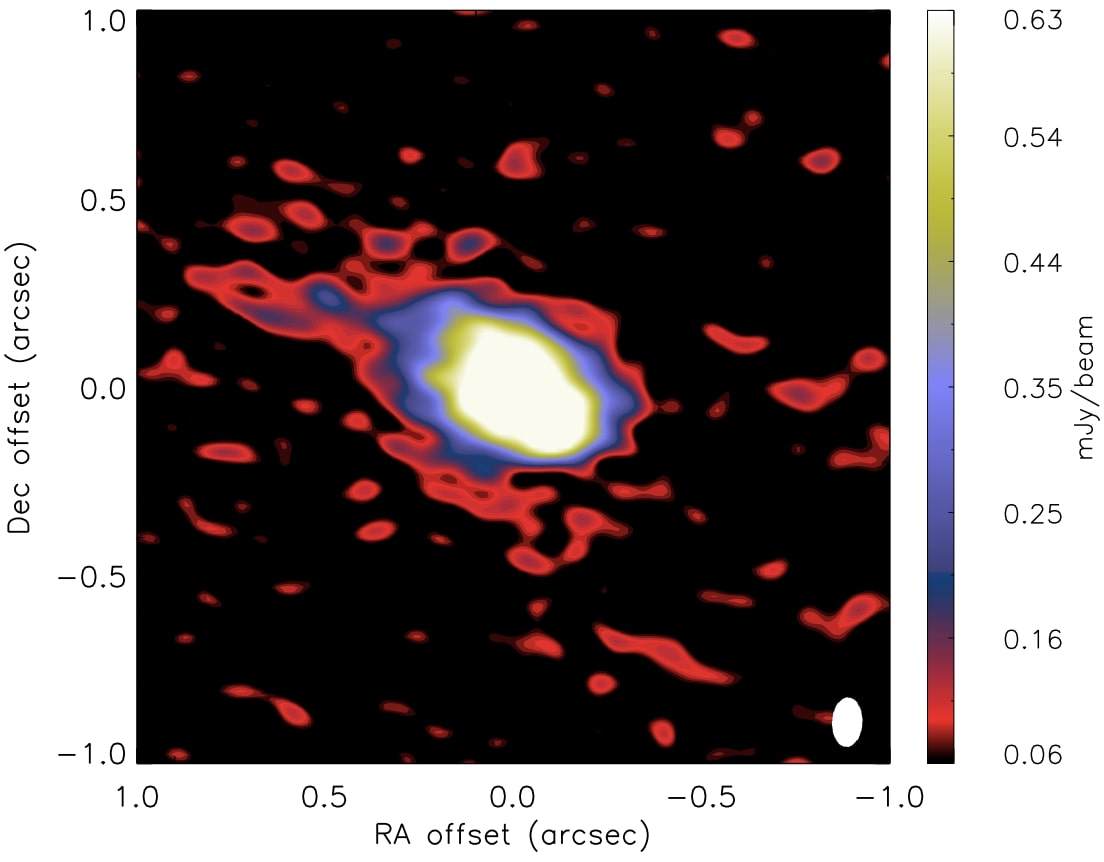}}
\caption{A 1.3~mm map made with ALMA of the Eb source. The disk is
clearly irregular, with an arm protruding to the east. The color scale
starts from 1.5~$\sigma$ [Eb-disk] and goes up to 50\% of the maximum
intensity. 
\label{Eb-disk}} 
\end{figure}


With the 0\farcs12 angular resolution of the extended ALMA
configuration the circumstellar disks of four of the sources in SSV~63
(Ea, Eb, Wb, and NE) are resolved, but at a distance of 400~pc finer
structure of the disks is not discernable. However, disk radii can be
estimated in both dust and C$^{18}$O gas
(Table~\ref{table:alma-disk}).  As is commonly seen, the dust disks
are significantly smaller than the gaseous disks. For NE, Ea, Eb, Wa,
and Wb we find dust radii of 51, 39, 159, 35, and 81~AU.  We can
compare this to the results of Tobin et al. (2020), who used ALMA to
carry out a major 0.87~mm continuum survey of 328 protostars in Orion
with similar angular resolution. For the subset of Class~I
non-multiple sources they find a median dust radius of 37~AU.
However, it should be noted that the majority of protostars observed
by Tobin et al. will arrive on the main sequence as M-dwarfs, whereas
the SSV~63 components will become G, F, and A stars. It is well known
that there is a clear correlation between dust radius and stellar
mass, and Andrews (2020) suggests the relation R$_{mm}$ $\propto$
M$_*^{0.9}$. Thus, it appears that the dust disks in SSV~63 are on
average a factor 2 smaller than expected, which could be a signature
that they have been truncated. However, the radius-stellar mass
relation has significant scatter, and combined with the small number
statistics we cannot be certain that the SSV~63 disks have suffered
dynamically induced truncation.

The disks of sources Ea, Wb, and NE are almost perfectly symmetric,
with no indication of recent perturbations. In contrast, the disk of Eb is
asymmetric, with a diffuse halo or wing stretching
towards the NNE (Figure~\ref{Eb-disk}). The length of this elongation is
almost 0.5 arcsec, that is, about 200~AU in projected
extent. Such an appearance is indicative of a recent
interaction. However, because Eb is deeply embedded,  it
is conceivable that its diffuse appearance is related to an infalling
envelope. Our current data cannot distinguish between these two
possibilities.

%


\vspace{0.3cm}

\section{SUMMARY AND CONCLUSIONS}\label{sec:conclusions}

We have performed a detailed observational study at optical, infrared,
mm and cm wavelengths of the HH~24 jet complex and the multiple system
SSV~63 that drives the various jets in order to better understand the
nature of low- to intermediate-mass star formation in such a small
system. We here summarize the main results:

{\em 1}: The SSV~63 system is embedded in a cloud core, and the known
components are the wide binaries Wa/Wb and Ea/Eb plus the NE source.
All are likely Class~I sources, but both NE and Eb are deeply embedded
and only detected at mid-infrared and longer wavelengths, and may be
borderline Class~0 sources.  Our deep near-IR images have identified
an additional faint source, S, and ALMA maps have discovered another
deeply embedded source, N. Thus the cloud core, which is elliptical
with dimensions of about 5,000 $\times$ 12,500~AU, contains at least 7
sources. The five main sources are all detected by the VLA, and source
Wb has a secondary component.

{\em 2}: The most prominent jet among the outflows is the finely
collimated HH~24E jet.  Multi-epoch HST images show the jet to be very
bright in the [\Feii] 1.64~$\mu$m transition, to have a transverse
velocity of around 250~\kms\ away from the driving source Ea, and
expand with an opening angle of $\sim$2.6$^\circ$. Spectra show the
jet to be redshifted, and to have an angle of $\sim$35$^\circ$ to the
plane of the sky. Our VLA maps show a bipolar radio continuum jet from
source Ea along the axis of the E-jet, with a smaller bipolar jet at
right angles, indicating that source~Ea is an unresolved binary.

{\em 3}: The counter jet HH~24C displays a chaotic jumble of knots,
likely the result of it having burrowed through the cloud core. High
tangential velocities of about 300~\kms\ combined with a radial
velocity around -200~\kms\ indicates that the jet is moving towards us
at an angle of about 35$^\circ$ to the plane of the sky. A major new
knot appeared at visual wavelengths from behind a cloud edge between
2006 and 2014.

{\em 4}: The HH~24G jet has an unusual morphology, with fragments of a
collimated jet surrounded by a tubular cavity with a diameter of
$\sim$5000~AU and walls outlined by shocks. Its driving source is
SSV~63~NE. Near the base of the jet is a bright and highly variable
reflection nebulosity, indicating motion of shadowing material close
to the source.

{\em 5}: At large distances from HH~24, a group of HH objects,
including HH~19, 20, and 21, is found to the NW; another, HH~27, is
found to the SE. Proper motion measurements confirm previous
suggestions that HH~19 and HH~27 form distant bow shocks from the
faint jet HH~24J driven by the Wb source. The total extent of this
giant bipolar flow is 1.39~pc in projection.

{\em 6}: The group of objects HH~20 and 21 form part of a giant
fractured working surface driven by the HH~24C jet. We have searched
for further distant bow shocks along the well-defined flow axis and
found an object, HH~24NNW, 
1.45~pc from source~Ea and along its flow axis. To the SE we have found
several distant bow shocks, at distances of 0.98 and 1.67~pc from
source~Ea. The total extent of the E/C jet pair is thus 3.2~pc in
projection, or 3.8~pc at an inclination of 35$^\circ$ to the plane of
the sky. 

{\em 7}: The deeply embedded Class~0 VLA source HH~24-MMS, located 
$\sim$40~arcsec south of SSV~63, is shifted by 0.8~arcsec (320~AU)
from its location observed in 2000. Possible explanations include variability
in a binary, motion of the source, or dust heated through a
lighthouse effect, none of which are without problems. Our H$_2$
images reveal an extended series of shocks from the nearby Class~0
source HOPS~317.

{\em 8}: The brightest shock in the HH~24 complex, HH~24A, is
structurally and kinematically complex, with knots on its eastern side
moving along the axis of the E-jet, while the central bright part is
essentially stationary, and may represent a shock in the counterflow
from the nearby source HOPS~317.

{\em 9}: We have searched for additional YSOs near SSV~63, and have
found five H$\alpha$-emission stars and brown dwarfs in the vicinity
of SSV~63, with spectral types between M3.5 and M7. They are
1.5 to 2~arcmin from SSV~63, far outside the dense molecular core. 
Proper motions from Gaia show that one of these, SSV~63~H$\alpha$~5,
moves straight away from the embedded sources with a tangential
velocity of 26~\kms. The object has a spectral type of M5.5, and is a
borderline brown dwarf. H$\alpha$~5 was very close to the Class~0/I NE
protostellar source about 5800~yr ago, and we assume NE is the source
from which H$\alpha$~5 was dynamically ejected. Such an ejection
requires that either NE or H$\alpha$~5 must be a close binary. If
H$\alpha$~5 was ejected from a protostellar system it follows that it
is itself protostellar, and hence it falls into the category of
orphaned protostars (Reipurth et al. 2010). None of the other four
H$\alpha$ emission stars have significant motions, and their origin is
unclear. Among these, H$\alpha$~1 drives a faint highly collimated
jet, HH~1200, and H$\alpha$~2 is a young M7 brown dwarf.

{\em 10}: Our $^{12}$CO observations with ALMA have revealed a few
small molecular outflows.  A bipolar one, labeled MO1, is centered on
the deeply embedded source Eb and is perpendicular to the well-defined disk
axis. The flow has a total extent of only about 2~arcsec, and is the
only one that is also associated with SiO emission.  Another bipolar
flow, MO2, lies along the axis of the jet~HH~24J driven by source Wb.
The third bipolar outflow, MO3, is associated with the mm continuum
source N. Surprisingly, the major bipolar E/C jet from source Ea does
not show evidence of a molecular outflow, although some low-velocity
emission may be associated with gas flowing along cavity walls.

{\em 11}: A peculiar formaldehyde flow, 6$''$-7$''$ wide centered on
source NE, is detected at low blueshifted velocities partly along the
wide G-jet. Its velocity is increasing with distance from NE, and
could be caused by an explosion, or be a flow gliding along a curved
background cavity wall.

{\em 12}: ALMA detects a large filamentary structure in $^{13}$CO and
C$^{18}$O extending from the edge of the field to its center with a
slight 1.6 \kms\ gradient. This may be interpreted as a streamer of
infalling material for which we estimate a rough upper limit to the
mass feeding the core of $\sim$ 7 $\times$
10$^{-6}$~M$_\odot$yr$^{-1}$. Thus star formation in the core may be
continously fed with fresh gas.

{\em 13}: We have derived stellar masses of the four sources Ea, Eb,
Wb, and NE assuming Keplerian rotation of their disks detected with
ALMA. The masses are 2.0, 1.3, 0.9, and 2.1~M$_\odot$, with estimated
uncertainties of about 0.1~M$_\odot$. The masses of Ea and NE indicate
that they are proto-Herbig~Ae stars. Eb and Wb have masses on the high
end of T~Tauri stars, but since both stars are heavily extincted and
detectable only at mid-IR wavelengths, they may still gain a
significant amount of gas.

{\em 14}: The five dominant sources, Ea, Eb, Wa, Wb, and NE, display
circumstellar disks in the ALMA observations, with major axes oriented
almost precisely perpendicular to the prominent jets they drive. For
four of the sources, Ea, Eb, Wb, and NE, disk radii are derived for
the gas and the dust. On average, they are about a factor 2 smaller
than inferred from a disk radius-stellar mass correlation. This might
be the result of truncation in this dynamically active system, but due
to the large scatter of the correlation and the small number of
sources observed, a firm conclusion is premature.  The disk of Eb
is irregular with a larger eastern lobe that might be the result of a
close encounter with another of the sources.

{\em 15}: We determine a mass of $\sim$3.3~M$_\odot$ for the cloud
core in which the SSV~63 multiple system resides based on the
850~$\mu$m data of Kirk et al. (2016a,b).  A lower limit to the total
stellar mass of the multiple system is roughly 7~M$_\odot$. A filamentary
structure in the region which may be an infalling streamer of gas,
suggests that the core may be continously forming stars as its gas
content is replenished.

{\em 16}: SSV~63 is an excellent example of a protostellar multiple
system of at least 7 embedded sources and one low-mass runaway
borderline brown dwarf. With a non-hierarchical configuration, the
system is unstable with the stars moving chaotically among each
other. This will eventually lead to the breakup of the system, in the
process ejecting a number of the members, preferentially those with
lowest mass.  Numerical simulations indicate that the system will
almost completely disintegrate within less than 1 million years.  As
the stars are ejected from their feeding zones their masses are set,
and dynamical interactions in small protostellar multiple systems are
thus an important factor in defining the initial mass function.



\section{Acknowledgements}

We thank an anonymous referee for an insightful report, which improved
this paper.  We also thank Helen Kirk for providing
Figure~\ref{fig850micron}, G\"oran Sandell for help with the Herschel
data, and Isabel Baraffe for advice on the BHAC15 models.
B.R. acknowledges support by NASA through grant HST-GO-13485.
J.B. acknowledges support by the National Science Foundation through
grant~AST-1910393.
H.-W.Y. acknowledges support from Ministry of Science and Technology
(MOST) in Taiwan through the grant MOST 110-2628-M-001-003-MY3 and
from the Academia Sinica Career Development Award (AS-CDA-111-M03).
H.G.A. acknowledges support from the National Science Foundation award
AST-1714710. 
L.F.R. acknowledges the financial support of DGAPA (UNAM) IN105617,
IN101418, N110618 and IN112417 and CONACyT 238631 and 280775.
A.C.R. acknowledges support by DGAPA (UNAM) grant 
IG100422.
This paper makes use of the following ALMA data:
ADS/JAO.ALMA\#2018.1.01194.S. ALMA is a partnership of ESO
(representing its member states), NSF (USA) and NINS (Japan), together
with NRC (Canada), MOST and ASIAA (Taiwan), and KASI (Republic of
Korea), in cooperation with the Republic of Chile. The Joint ALMA
Observatory is operated by ESO, AUI/NRAO and NAOJ. The National Radio
Astronomy Observatory is a facility of the National Science Foundation
operated under cooperative agreement by Associated Universities, Inc.
Based in part on data collected at the Subaru Telescope, which is
operated by the National Astronomical Observatory of Japan (NAOJ).
Thanks are due to the Subaru staff, in particular Miki Ishii and
Hisanori Furusawa for excellent
and dedicated support during the observations.  We are grateful to
Nobunari Kashikawa for permission to use his [\Sii] filter.
Based in part on observations  (GN-2010A-Q-10, GN-2013B-Q-77)
obtained at the international Gemini Observatory, a program of NSF's NOIRLab, which is managed by the Association of Universities for Research in Astronomy (AURA) under a cooperative agreement with the National Science Foundation on behalf of the Gemini Observatory partnership: the National Science Foundation (United States), National Research Council (Canada), Agencia Nacional de Investigaci\'{o}n y Desarrollo (Chile), Ministerio de Ciencia, Tecnolog\'{i}a e Innovaci\'{o}n (Argentina), Minist\'{e}rio da Ci\^{e}ncia, Tecnologia, Inova\c{c}\~{o}es e Comunica\c{c}\~{o}es (Brazil), and Korea Astronomy and Space Science Institute (Republic of Korea).
 We are thankful to Richard McDermid for help with the Gemini Phase II submission.
This research is based in part on observations made with the NASA/ESA
Hubble Space Telescope obtained from the Space Telescope Science
Institute, which is operated by the Association of Universities for
Research in Astronomy, Inc., under NASA contract NAS 5-26555.
The VLA observations were part of our project 19A-012, made with the
NSF's Karl G. Jansky Very Large Array (VLA) of the National
Radio Astronomy Observatory, which is a facility of the National Science
Foundation operated under cooperative agreement by Associated
Universities, Inc.
Observations were obtained with the Apache Point Observatory 3.5-meter
telescope, which is owned and operated by the Astrophysical Research
Corporation. We thank the APO Observing Specialists for their
assistance during the observations.
This work is based in part on observations made with the Spitzer Space
Telescope, which is operated by the Jet Propulsion Laboratory,
California Institute of Technology under a contract with NASA,
and by Herschel, which is an ESA space observatory with science instruments provided by European-led Principal Investigator consortia and with important praticipation from NASA.
This publication makes use of data products from the Two Micron All
Sky Survey, which is a joint project of the University of
Massachusetts and the Infrared Processing and Analysis
Center/California Institute of Technology, funded by the National
Aeronautics and Space Administration and the National Science
Foundation.
Based on observations collected at the European Organisation for Astronomical 
Research in the Southern Hemisphere and extracted from the ESO archives.
This material is based upon work supported by the National
Aeronautics and Space Administration through the NASA Astrobiology
Institute under Cooperative Agreement No. NNA09DA77A issued through
the Office of Space Science.
This research has made use of the SIMBAD database,
operated at CDS, Strasbourg, France, and of NASA's Astrophysics Data System
Bibliographic Services.

Facilities: Gemini (GMOS,GNIRS), Spitzer, Subaru (SuprimeCam,IRCS),
Herschel, ALMA, VLA, HST (WFC3,ACS), Apache Point Observatory 3.5m
(ARCES, ARCTIC, NICFPS), VLT (NACO)



\end{document}